\documentclass[11pt, a4paper]{article}

\usepackage{amsmath}
\makeatletter
\newcommand{\mathleft}{\@fleqntrue\@mathmargin0pt}
\usepackage{mathrsfs}

\usepackage{graphicx}
\usepackage{caption}
\usepackage{subcaption}
\usepackage{float}
\usepackage{geometry}

\usepackage{hyperref}
\usepackage{cleveref}
\hypersetup{colorlinks=true, allcolors = cyan,}

\usepackage{natbib}
\setcitestyle{authoryear,open={(},close={)}}
\graphicspath{ {./Pics/} }
\usepackage{epstopdf}

\newcommand\Tstrut{\rule{0pt}{2.6ex}}
\newcommand\Bstrut{\rule[-0.9ex]{0pt}{0pt}}
\newcommand{\TBstrut}{\Tstrut\Bstrut}
\begin{document}
	\newgeometry{left=2.54cm,right=2.54cm,top=2.54cm,bottom=2.54cm}
	\title{Credit Spread Approximation and Improvement using Random Forest Regression}
	\author{Mathieu Mercadier $^{a,b,}$\footnote{Corresponding author at: +335 55 32 81 88.\newline E-mail addresses: mathieu.mercadier@unilim.fr (M. Mercadier), jp.lardy@jplc.fr (J.P. Lardy).\newline We are grateful to A. Tarazi, A. Sauviat, L. Lepetit and R. Tacneng for their advice and revisions, to S. Koulidiati for early tests, and to C. Lardy for proof-reading this article.} , Jean-Pierre Lardy $^{b}$}
	\maketitle
	\noindent
	\footnotesize
	a: Universit\'e de Limoges, LAPE, 5 rue F\'elix Ebou\'e, 87031 Limoges Cedex, France\\
	b: JPLC SASU, 54 avenue de la R\'evolution, 87000 Limoges, France\\
	\large
	\begin{center}
	\textbf{Draft. Not edited.\\Please do not circulate.}
	\end{center}
	\normalsize
	\noindent\rule{\textwidth}{1pt}
	\textbf{Abstract}\\
	Credit Default Swap (CDS) levels provide a market appreciation of companies' default risk. These derivatives are not always available, creating a need for CDS approximations. This paper offers a simple, global and transparent CDS structural approximation, which contrasts with more complex and proprietary approximations currently in use. This Equity-to-Credit formula (E2C), inspired by CreditGrades, obtains better CDS approximations, according to empirical analyses based on a large sample spanning 2016-2018. A random forest regression run with this E2C formula and selected additional financial data results in an 87.3\% out-of-sample accuracy in CDS approximations. The transparency property of this algorithm confirms the predominance of the E2C estimate, and the impact of companies' debt rating and size, in predicting their CDS.
	\noindent\rule{\textwidth}{1pt}
	\newpage
	\section{Introduction}\label{sec:Intro}
	The ability to accurately apprehend companies' credit risk is at the core of financial professionals' concerns and an extensive branch of literature is dedicated to this matter
	\citep[e.g.][]{Merton1974,Black1976,Vasicek1987,Finger2002}. Since the mid-nineties, the uncertainty regarding an obligor's capacity to fulfill its contractual duty can be insured by a derivative, the Credit Default Swap. However, not all firms have actively traded CDS, leading to the development of estimations. \cite{Merton1974} was the first to define a relationship between a company's probability of default and its capital structure. He used the option pricing framework to relate the three balance sheet segments, i.e. asset, debt and equity. \cite{Black1976} added the possibility for a default to occur prior to maturity. But numerous approximations were developed in the private sector when these derivatives started to be traded. For obvious reasons, many of these models are highly proprietary and are not accessible to researchers.
	
	Despite being developed by practitioners, the CreditGrades model \citep{Finger2002} is available for use to everyone. This formula relies on companies' asset values dynamic modeled with a diffusion process and a default barrier that can be crossed prior to maturity as in \cite{Black1976}. Overall, academics have focused on strengthening the definitions of some variables of the CreditGrades model to improve its accuracy. For instance \cite{Zhou2001} takes into account market changes using a jump-diffusion process. \cite{Sepp2006} makes the model even more complex assuming that the firm's asset value process follows a double-exponential jump-diffusion and that the variance of the firm's assets return is stochastic. Another extension of the CreditGrades model, using implied (and not historical) volatilities to take into account both equities volatilities and leverage, is proposed by \cite{Stamicar2006}. More recently, \cite{Escobar2012} have offered a multivariate extension of the CreditGrades model under the assumption of stochastic variance and correlation among the companies' assets. Credit Default Swap is a topic of interest in much recent operational research literature \citep{Guarin2011,Tomohiro2014,Cont2016,Chalamandaris2018,Koutmos2018,Irresberger2018}, but in practice, models for traded CDS spreads tend not to be popular among portfolio managers. Although the aforementioned models provide very close approximations \citep[e.g.][]{Imbierowicz2008}, they remain prohibitively complex.
	
	In this paper, our aim is to develop an intuitive, accurate and open-access instrument to approximate CDS spreads, intended for use by credit market practitioners or as a reliable proxy for researchers. We first provide an elementary formula, called Equity-to-Credit (E2C henceforth), comparable to the CreditGrades model in terms of accuracy. We then further improve the precision of this straightforward formula using machine learning tools, which have become more prevalent in the financial sector. 
	
	The first contribution of this paper is the development of a concise and broad-based approximation assessing credit spreads. The Equity-to-Credit formula which we develop is a pared-down elementary equation inspired by RiskMetrics Group's CreditGrades model \citep{Finger2002}. The formula relies on two main mathematical concepts: the reflection principle and Gauss's inequality. The probability of default is based on the reflection principle and its symmetrical property. Its upper bound is specified with Gauss's inequality which is a refinement of Chebyshev's inequality, converted to a conservative equality following Roy's ``principle of safety first'' \citep{Roy1952}. We obtain a simple and transparent credit approximation which only depends on leverage, equity volatility and standard debt recovery parameters, assessed on a conservative setting and CreditGrades parameterization.

	We then use the CreditGrades model as a comparative basis to empirically assess the E2C formula's reliability in the second part of this paper. More specifically, both models' results are confronted to the actual CDS. The E2C formula proves to be more correlated and a better regressor than the alternative model. In addition, a comparison is performed gathering data by buckets of senior unsecured debt rating and industrial sectors. The analysis is done with medians and truncated means, both proving in favor of the E2C formula. Consistently with the existing literature, our 2016-2018 sample confirms that structural models generally underpredict actual CDS spreads. \cite{Rodrigues2011} reach the same conclusions for three structural models including the CreditGrades model. Moreover, we notice that both the E2C and the CreditGrades models yield lesser underestimations for riskier companies' spreads, in line with \cite{Teixeira2007}. Notably, the E2C returns closer estimates for the financial sector where CreditGrades, and Merton models more generally, are traditionally poor.
	
	By construction, some parameters influencing credit spreads are missing from structural models such as our E2C. Therefore, our second contribution in this paper is an improvement of our CDS approximation accuracy, taking into account selected complementary features. We suggest the use of a supervised learning algorithm which allows the treatment of a multivariate universe, composed of the E2C and a limited number of chosen variables. The E2C formula being only based on equity information, a credit derivatives index is set as a time-specific independent variable. Company-specific variables are then added, such as credit ratings, size, industrial sector, and geographical location.
	
	Its intelligibility and its ability to handle reasonable sample sizes were the main reasons driving our choice of the random forest regression algorithm \citep{Breiman2001}. The purpose is to average multiple randomly bootstrapped decision trees \citep[``bagging'', in][]{Breiman1996} built with subsets of the features randomly chosen at each node. Once the hyper-parameters are tuned, the algorithm is parallelized to decrease computation time. The main result is an 87.3\% out-of-sample accuracy which emphasizes the power of this machine learning algorithm and our database to approximate CDS spreads.
	
	In addition, this algorithm's choices are easily accessible through the transparency of its decisions at each split, displaying the feature it chooses in order to reduce the error. This allows us to evaluate each feature's contribution in predicting the CDS spreads, using two methods. First, we measure the improvement brought by a specific selected variable at each node. Then, the importance of the variable is measured by how much it decreases the model accuracy when this feature is randomly permuted. This two-pronged evaluation confirms that our E2C formula is the main variable selected by the decision trees when approximating the CDS. Additionally, it underlines the next best source of improvements as the company's credit rating and its size.
	
	The remainder of this paper is organized as follows. In the next section, we present the main steps to build our E2C formula. We then confirm that the accuracy provided by the E2C is at least as good as its closest parent's. Finally, the approximation accuracy is improved using random forest regressions on the E2C and other independent variables.
	\section{The E2C Formula}
	In this section, we present the main steps to build the E2C formula, leading to an elementary CDS approximation equation. The stock price $S_t$ is defined as an additive stochastic process, such that:
	\begin{align*}
	S_t = S_0 + \delta z_t \sqrt{t}
	\end{align*}
	We assume that the stock price $S_t$ varies by a random amount of standard deviation $\delta$ per time unit. Where $\delta$ stands for a level of volatility and $z_t$ a symmetric standard random variable: $z_t$ has zero mean and unit variance but is not necessarily assumed a normal distribution. Hence, the expected value and the variance of the stock price process are E[$S_t$] = $S_0$ and Var($S_t$) = $\delta^2 t$.

\begin{figure}[H]
		\begin{center}
			\vspace*{0.5cm}\hspace*{0.0cm}\includegraphics[width=10cm, height=5cm]{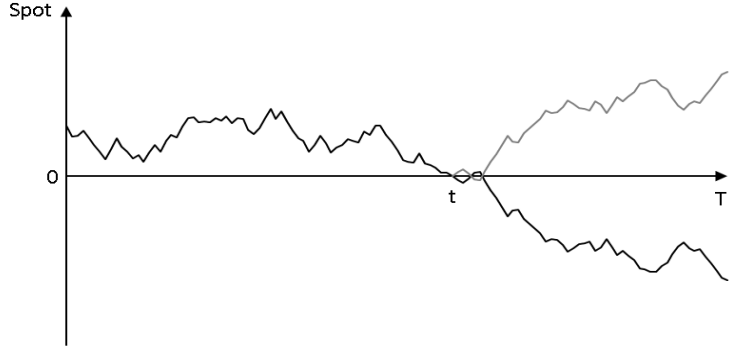}\\
			\caption{Reflection Principle}
			\label{fig:1}
		\end{center}
	\end{figure}

	Once the prices' path model is defined, we derive the probability for a bankruptcy to occur. In this kind of model, the failure happens if the stock price reaches a certain level. For obvious economic reasons, it is set at zero. Therefore, we look for the probability that a stock price reaches zero at any time $t$ prior to a certain maturity $T$.
	\begin{align*}
	{\rm I\!P}(Default \mid t \leq T) = {\rm I\!P}(\min S_t \leq 0 \mid t \leq T)
	\end{align*}
	Knowing that the process starts at $S_0 > 0$, the probability of reaching $S_t = 0$ at any time $0 < t < T$ is measured. According to the reflection principle \citep[][see \Cref{fig:1} below]{Levy1940}, if the zero value is reached before $T$, each possible path ending positively has a symmetric counterpart ending negatively from this point until $T$. In other words, if the drift is null, the number of all paths reaching zero is exactly twice the number of paths terminating below zero.\\
	\vspace*{-1.3cm}
	\noindent
	Hence,
	\begin{align*}
	{\rm I\!P}(\min S_t \leq 0 \mid t \leq T) & = 2{\rm I\!P}(S_T \leq 0) &\\
	& = 2{\rm I\!P}(S_0 + \delta z_T \sqrt{T} \leq 0) &\\
	& = 2{\rm I\!P}(z_T \leq \dfrac{-S_0}{\delta \sqrt{T}}) &\\
	\end{align*}
	According to $z_t$ symmetry:
	\begin{align} \label{eq:1}
	{\rm I\!P}(\min S_t \leq 0 \mid t \leq T) = {\rm I\!P}(\mid z_T \mid \geq \dfrac{S_0}{\delta \sqrt{T}})
	\end{align}
	\noindent
	The idea above is closely related to the barrier concept introduced by \cite{Black1976} which assesses the first time a firm's asset value drops below a certain time-dependent barrier.

	To define this probability boundary and for simplicity, we note $X$ a random variable corresponding to $z_T$. The boundary could be estimated with Chebyshev's well-known over-conservative inequality \citeyearpar{Chebyshev1867}, ${\rm I\!P}(|X - \mu| \geq k\sigma) \leq \dfrac{1}{k^2}$. However, a lesser-known refinement developed by \cite{Gauss1821} \citep[or more recently, in][]{Perks1947} allows to sharpen Chebyshev's boundary by a $\dfrac{4}{9}$ factor on the condition that $X$ is unimodal with a mode $\mu$. This assumption is correct regarding our model with $E[X] = \mu= 0$. Applying Gauss's inequality to our main equation \ref{eq:1}, we get:
	\begin{align*}
	{\rm I\!P}(\mid z_T \mid \geq \dfrac{S_0}{\delta \sqrt{T}}) & \leq \dfrac{4 \delta^2 T}{9 S_0^2}
	\end{align*}
	For simplicity and economic sense, we get rid of the inequality operator based on the paper of \cite{Roy1952}. Following his ``principle of safety first'' and a conservative objective, we maximize our upper bound. It has a financial sense therefore to conclude: \\
	\begin{align*}
	{\rm I\!P}(Default \mid t \leq T) & = \dfrac{4 \delta^2 T}{9 S_0^2}
	\end{align*}
	Dividing this probability by the time horizon, we define the hazard rate $h$.
	\begin{align} \label{eq:2}
	h = \dfrac{{\rm I\!P}(Default \mid t \leq T)}{T} = \dfrac{4 \delta^2}{9S_0^2}
	\end{align}
	Using R as the asset specific recovery rate, we model the credit approximation as:
	\begin{align} \label{eq:3}
	C = (1 - R) \cdot h = (1 - R) \cdot \dfrac{4 \delta^2}{9S_0^2}
	\end{align}
	Next, we derive a closed-form expression for the volatility parameter $\delta$ based on the CreditGrades assumptions, and focusing on the downside evolution of the stock price. In fact, in this asymmetric approach which is presented in \nameref{sec:A}, the above equation \ref{eq:3} is still valid with a ``downside delta'' $\delta_{S_T \leq S_0}$, and is justified by reconsidering the Gauss inequality on only one side of the distribution.

	Much research has been done on the link between bankruptcies, assets, debts and equities, notably by \cite{Merton1974} and \cite{Black1976}. A company defaults if a certain level of insolvency is reached. In line with the assumption of the CreditGrades approach, we apprehend this level, or default barrier, as the remaining amount of the firm's assets in the case of default, corresponding to the recovery value received by the debt holders. We note this amount $\bar{L}D$, where $\bar{L}$ is the average recovery on the debt and $D$ the firm's debt-per-share.

	To keep the simplest expression consistent with the definition of $\bar{L}$ (such that the asset value $V=\bar{L}D$ when $S=0$), we write $V = S + \bar{L}D$. With $\sigma_S$ being the equity volatility, the equity and asset volatilities are related as follows:
	\begin{align*}
	\delta=\sigma_S S = \sigma_V V = \sigma_V (S + \bar{L}D)
	\end{align*}
	Then, as such $\delta$ is not constant, we anchor it by emphasizing its downside boundary values:
	\begin{itemize}
     \item At $t=0$, $V = S_0 + \bar{L}D$, therefore,
    \begin{align*}
    \delta_{t=0} = \sigma_V (S_0 + \bar{L}D)
    \end{align*}
    \item If the default occurs, $S=0$ thus, $V = \bar{L}D$, hence,
    \begin{align*}
    \delta_{S=0} = \sigma_V \bar{L}D
    \end{align*}
	\end{itemize}
	And we retain for $\delta$ the average of these values, using the geometric average as we are dealing with magnitudes. In addition, it is indifferent whether the average is squared or the squared values are averaged.\\\\
	Hence,
	\begin{align*}
	\delta_{S_T \leq S_0}^2 = \sigma_V^2 (S_0 + \bar{L}D)\bar{L}D
	\end{align*}
	And, as
	\begin{align*}
	\sigma_V = \sigma_S \dfrac{S_0}{(S_0 + \bar{L}D)}
	\end{align*}
	We get:
	\begin{align*}
	\delta_{S_T \leq S_0}^2 = \sigma_S^2 S_0^2 \dfrac{\bar{L}D}{(S_0 + \bar{L}D)}
	\end{align*}
	where $\dfrac{\bar{L}D}{S_0 + \bar{L}D}$ is the ratio ``debt to enterprise market value'' of the company, also known as the Market-Adjusted Debt (MAD) ratio. The final credit approximation\footnote{The hazard rate is null when the debt is null, which is another property from the use of a geometric average.} is computed replacing this result in equation \ref{eq:3}.
	\begin{align} \label{eq:4}	
	C = (1 - R) \cdot \dfrac{4}{9} \cdot \dfrac{\bar{L}D}{S_0 + \bar{L}D} \cdot \sigma_{S_0}^2
	\end{align}
	The above formula is thereafter referred to as the Equity-to-Credit (E2C) formula. The E2C is computed using both market and balance sheet inputs \textendash \ on the one hand the current stock price and its volatility, and on the other hand the company's debt-per-share. The debt-per-share $D$ is computed following the CreditGrades method, further detailed in \nameref{sec:A}. With regard to the volatilities, for each company, both historical and implied ones are extracted for different maturities. The historical volatilities are computed over 30, 60, 120, 200, 260 and 360 days and the implied volatilities correspond to 3, 6, 12, 18 and 24 months options maturities. After a few tests, we retain the implied volatilities for strikes of put options ``out of the money'' by 0.5 standard deviation. For robustness, the E2C's volatility parameter is defined as the median of all the above available volatilities. In general, recovery rates of traded CDS are set at 0.4 corresponding to the developed market average. However, being conservative, we set it at 0.3, as some practitioners do and as it is commonly attributed to emerging markets. After some empirical tests, the RiskMetrics team chose an $\bar{L}$ of 0.5 for their model. Hence, knowing that the CreditGrades parameterization is relevant for our equation, we calibrate our model with $R = 0.3$ and $\bar{L} = 0.5$.
	
	The equation presented in this section is very intuitive as the variables' impact is consistent with market specialists' expectations. This credit spread approximation is an increasing function of the debt and the corresponding stock volatility. On the other end, when the share value increases, the spread value decreases. 
	\section{E2C and CreditGrades Models Comparison}
	\subsection{Data and Softwares}\label{sec:Data}
	Our panel data sample is a blend of market and fundamental reports information over 308 listed companies from 27 countries (of which 12 are gathered in a single Eurozone group)\footnote{Country (number of Firms per Country) list: Australia (14), Canada (5), Denmark (2), Eurozone (78), Great Britain (22), Hong Kong (6), India (3), Japan (22), Malaysia (1), New Zealand (2), Norway (2), Singapore (1), South Korea (8), Sweden (3), United States (139).}. Each company belongs to one of the following ten major sectors: basic material, communications, consumer cyclical, consumer non-cyclical, energy, financial, industrial and utilities\footnote{Technology and diversified sectors are not displayed, they only respectively count 8 and 2 companies on average.}. The data set spans weekly, every Friday over three years (i.e. 155 dates), from 2016-02-03 until 2018-12-28, and is based entirely on Bloomberg L. P. data.

	Data preparation and handling is entirely conducted in Python 3.6 (Python Software Foundation, \citeyear{Python2016}), relying on the packages ``numpy'' by Van Der \cite{Walt2011}, ``pandas'' by \cite{McKinney2010} and for visual outputs on ``matplotlib'' by \cite{hunter2007}. Linear regressions are performed with Stata 13.0 \citeyearpar{Stata2013}. Moreover, we use ``sci-kit learn'' by \cite{Pedregosa2011} for the random forest and the ``randomForest'' R package by \cite{Liaw2002} to confirm the results.
	\subsection{Basic Analysis}\label{sec:BasicAnalysis}
	Based on the data sample defined above, we compare the two CDS spreads approximations obtained by our E2C formula, i.e. equation \ref{eq:4}, and the CreditGrades version defined by RiskMetrics Group. This version requires the use of the Cumulative Normal distribution and logarithm functions, and an additional parameter $\lambda$ which stands for the standard deviation of the global recovery rate ($\lambda = 0.3$ according to the RiskMetrics paper, \citeyear{Finger2002}). To convert the CreditGrades survival probabilities to spreads, we define the maturity $t = 5$ years. All other parameters are set as those defined previously for the E2C. We analyze the 5-year CDS against the E2C equation defined above, and against the CreditGrades estimate $(1 - R) \cdot h_{CG}$, where $h_{CG}$ is the hazard rate implied by CreditGrades (described in \nameref{sec:B}).

	According to the descriptive tables available in \nameref{sec:B}, the E2C and CreditGrades outputs are relatively close to one another and more importantly to the 5-year CDS. Both approximations have a lower median and are slightly more volatile than the actual CDS. A comparative table of the quality of the CDS approximations at a country level produced by each of these models is available in \Cref{tab:4}. Moreover, the E2C's density function shape is closer to the CDS, as it is more right-skewed and leptokurtic than the one obtained with CreditGrades. We also notice, for all curves, a higher heterogeneity between companies than within. Finally, we compute the averages of each firm's correlation over time and of each date's correlation over all firms. In both correlation tables, E2C is highly correlated to CreditGrades with a coefficient around 90\%. Above all, E2C is more correlated to the 5-year CDS than CreditGrades between companies and through time. Across the board, our approximation shows conclusive results.

	We further study the predictivity of these equations with econometric techniques. According to the appropriate testing\footnote{In both cases, panel (fixed-effects) models are chosen over pooled OLS. The null hypothesis of homoscedasticity of the Breusch and Pagan Lagrangian multiplier test is rejected. Then, we perform a test of overidentifying restrictions (similar to Hausman test, with $H0: cov(X_{i,t}, v_i)=0$) on our model which includes time-variant variables. The result indicates that the fixed-effects model should be preferred over the random-effects model.}, we handle both regressions (CDS onto E2C and onto CreditGrades) with fixed-effects models. Thus, both company-specific characteristics and time must be taken into account. In the summarized results below (\Cref{tab:1}), we highlight the $R^2$, and in particular the $R^2-within$ along with the fixed-effects model.
	\begin{table}[H]
		\begin{center}
			{
				\vspace*{-0.0cm}
				\centering
				\def\sym#1{\ifmmode^{#1}\else\(^{#1}\)\fi}
				\begin{tabular}{l*{2}{c}}
					\hline\hline
					&\multicolumn{1}{c}{CDS\_5y}&\multicolumn{1}{c}{CDS\_5y}\\
					\hline
					E2C         &       0.637\sym{***}&                     \\
					&    (203.81)         &                     \\
					CrdGrd      &                     &       0.671\sym{***}\\
					&                     &    (162.29)         \\
					constant      &       58.16\sym{***}&       44.41\sym{***}\\
					&     (103.7)         &     (61.26)         \\
					\hline
					\(N\)       &       47476         &       47476        \\
					$R^{2}$-within  &       0.468         &       0.358         \\
					$R^{2}$-between        &       0.796         &       0.618         \\
					$R^{2}$-overall        &       0.692         &       0.543         \\
					\hline\hline
					\multicolumn{3}{l}{\footnotesize \textit{t} statistics in parentheses}\\
					[-0.5em]
					\multicolumn{3}{l}{\footnotesize \sym{*} \(p<0.05\), \sym{**} \(p<0.01\), \sym{***} \(p<0.001\)}\\
				\end{tabular}
			}	
		\end{center}
		\vspace*{-0.4cm}
		\caption{Goodness-of-Fit of fixed effect model using E2C and CreditGrades}
		\label{tab:1}
	\end{table}
	With an $R^2$ over 10 points of percentage above the CreditGrades regression, the E2C formula is shown to approximate the CDS with globally more reliability than CreditGrades.
	\subsection{Rating and Sectorial Analyses}\label{sec:TruncMean}
	Finally, we focus on two universes in which firms' repartition is overall well balanced: unsecured senior debt ratings and industrial sectors. The E2C results are compared, along with those of CreditGrades, to the actual CDS spreads using two methods. The basic averages being exposed to outliers, we firstly choose to compare their medians. The second method compares the truncated means, where averages are computed after having dropped the extreme 10\% top and bottom points. For the sake of visualization, we summarize our results below in \Cref{fig:2,fig:3}.
	\subsubsection{Debt Rating Comparison}
	Our dataset is composed of senior unsecured debt ratings provided by Standard \& Poor's and Moody's. Unfortunately, their rating scales are different, and their ratings for specific firms and dates might be different. This issue is overcome through the following guidelines: if both agencies provide the same grades \citep[based on the comparison scale in][]{Santos2008}, the corresponding grade is chosen; if only one agency gives a grade, this grade is selected; and if the grades are different, still being conservative, the worst one is kept. Our sample includes no AAA grades and too few AA, therefore we group them in a global A bucket. Furthermore, too little data for firms rated below B is available in our sample, so we remove them for this comparison.
	\begin{figure}[H]
		\caption{Comparisons between original 5y CDS, E2C and CreditGrades approximations in terms of senior unsecured debt rating}
		\label{fig:2}
		\centering
		\advance\leftskip-3cm
		\advance\rightskip-3cm
		\vspace*{-0.4cm}\begin{subfigure}[b]{0.65\textwidth}
			\includegraphics[width=\textwidth]{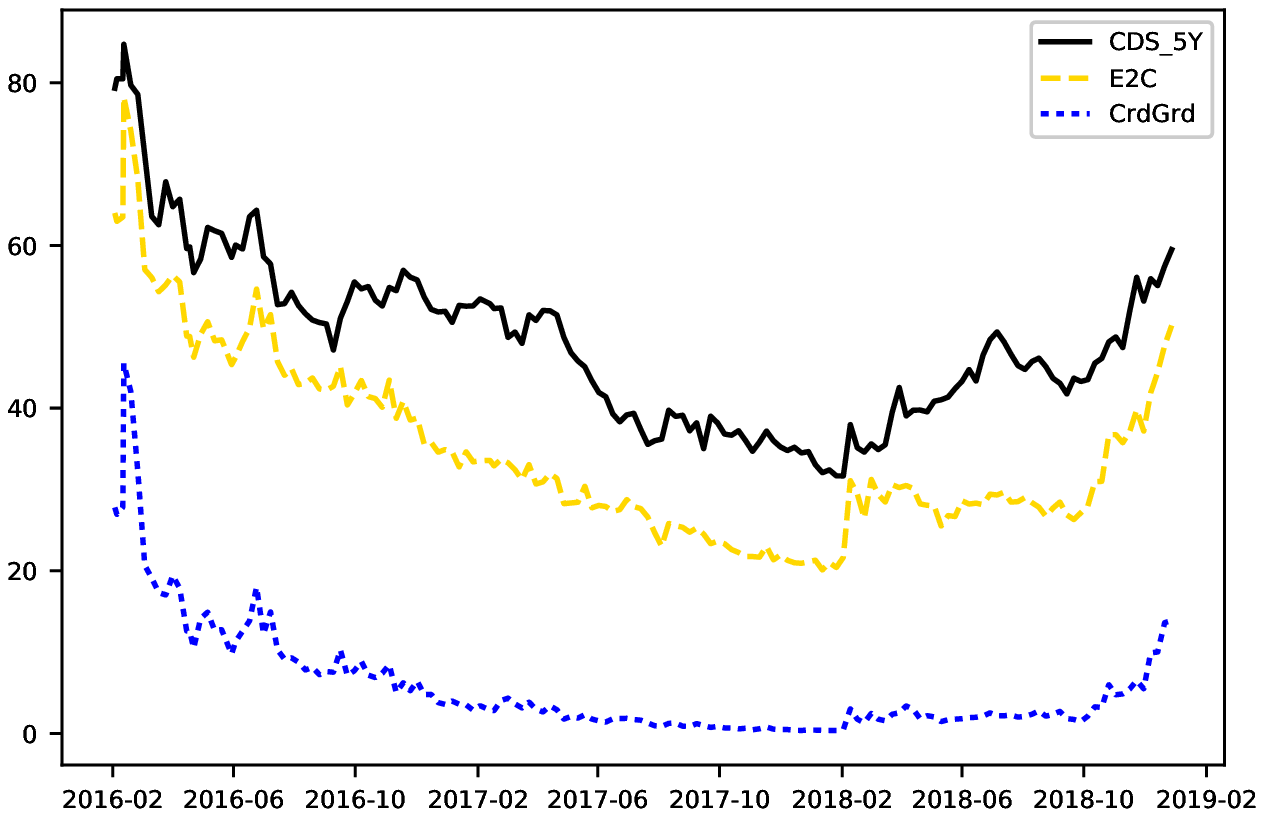}
			\vspace*{-1.0cm}
			\caption*{A: Median}
		\end{subfigure}
		\hspace*{-0.0cm}
		\vspace*{-0.4cm}\begin{subfigure}[b]{0.65\textwidth}
			\includegraphics[width=\textwidth]{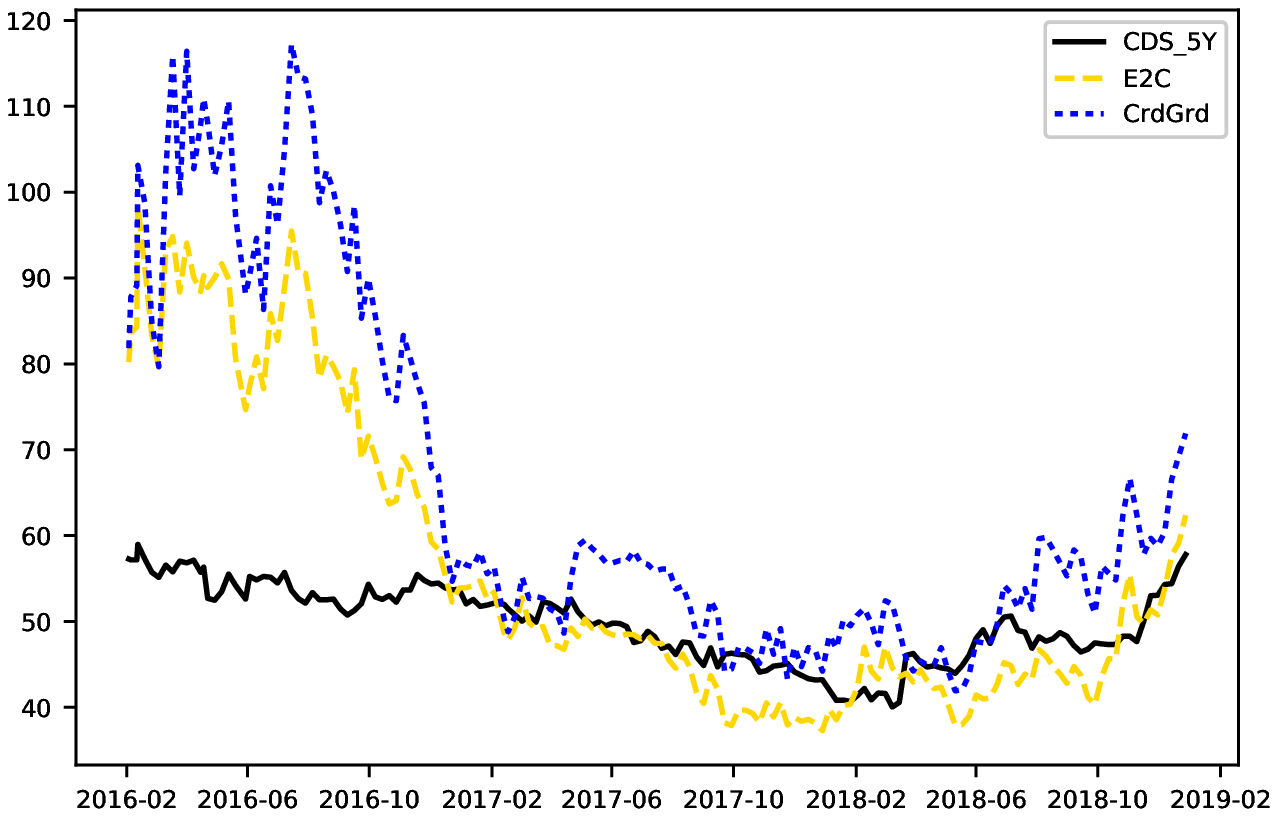}
			\vspace*{-1.0cm}
			\caption*{A: Truncated Mean}
		\end{subfigure}
	\end{figure}
	\vspace*{-0.6cm}
	\begin{figure}[H]\ContinuedFloat
		\centering
		\advance\leftskip-3cm
		\advance\rightskip-3cm	
		\begin{subfigure}[b]{0.65\textwidth}
			\includegraphics[width=\textwidth]{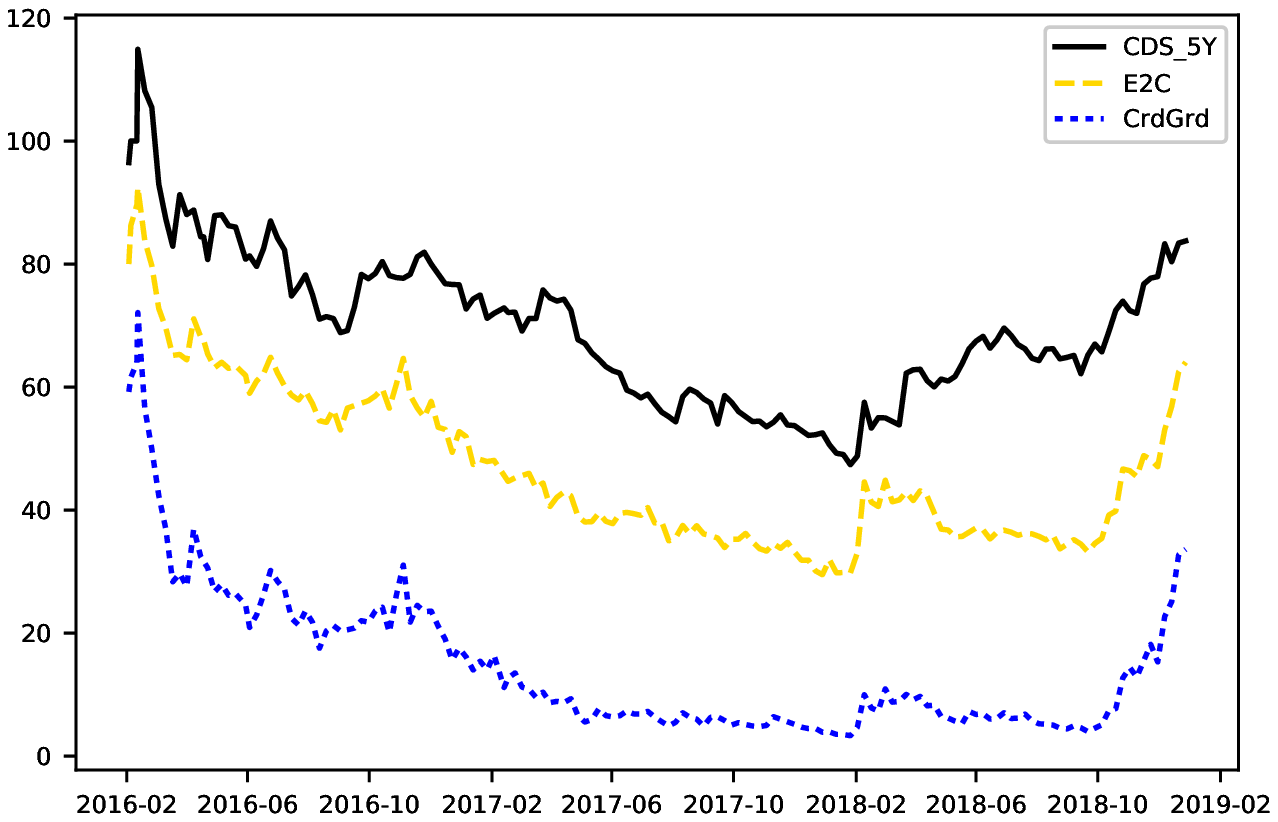}
			\vspace*{-1.0cm}
			\caption*{BBB: Median}
		\end{subfigure}
		\hspace*{-0.0cm}
		\begin{subfigure}[b]{0.65\textwidth}
			\includegraphics[width=\textwidth]{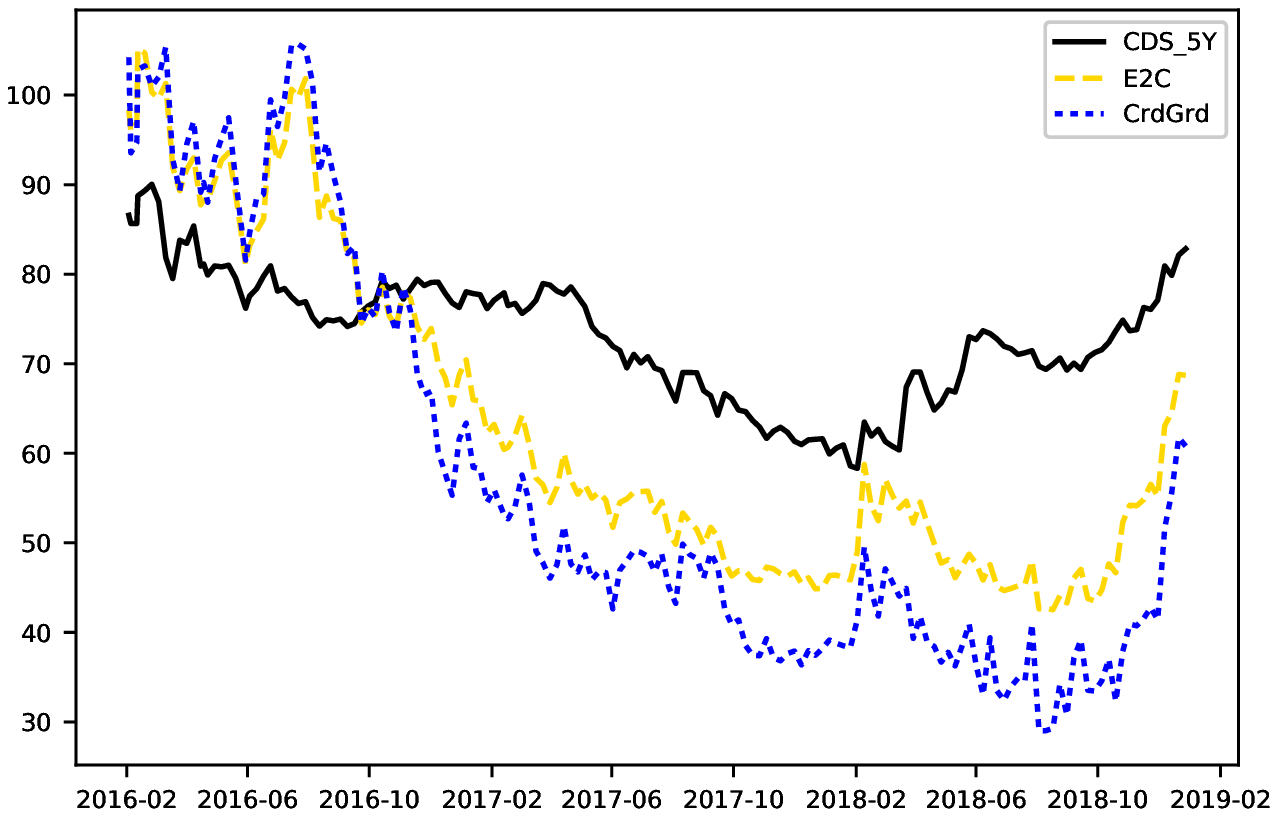}
			\vspace*{-1.0cm}
			\caption*{BBB: Truncated Mean}
		\end{subfigure}
	\end{figure}
	\vspace*{-0.6cm}
	\begin{figure}[H]\ContinuedFloat
		\centering
		\advance\leftskip-3cm
		\advance\rightskip-3cm	
		\begin{subfigure}[b]{0.65\textwidth}
			\includegraphics[width=\textwidth]{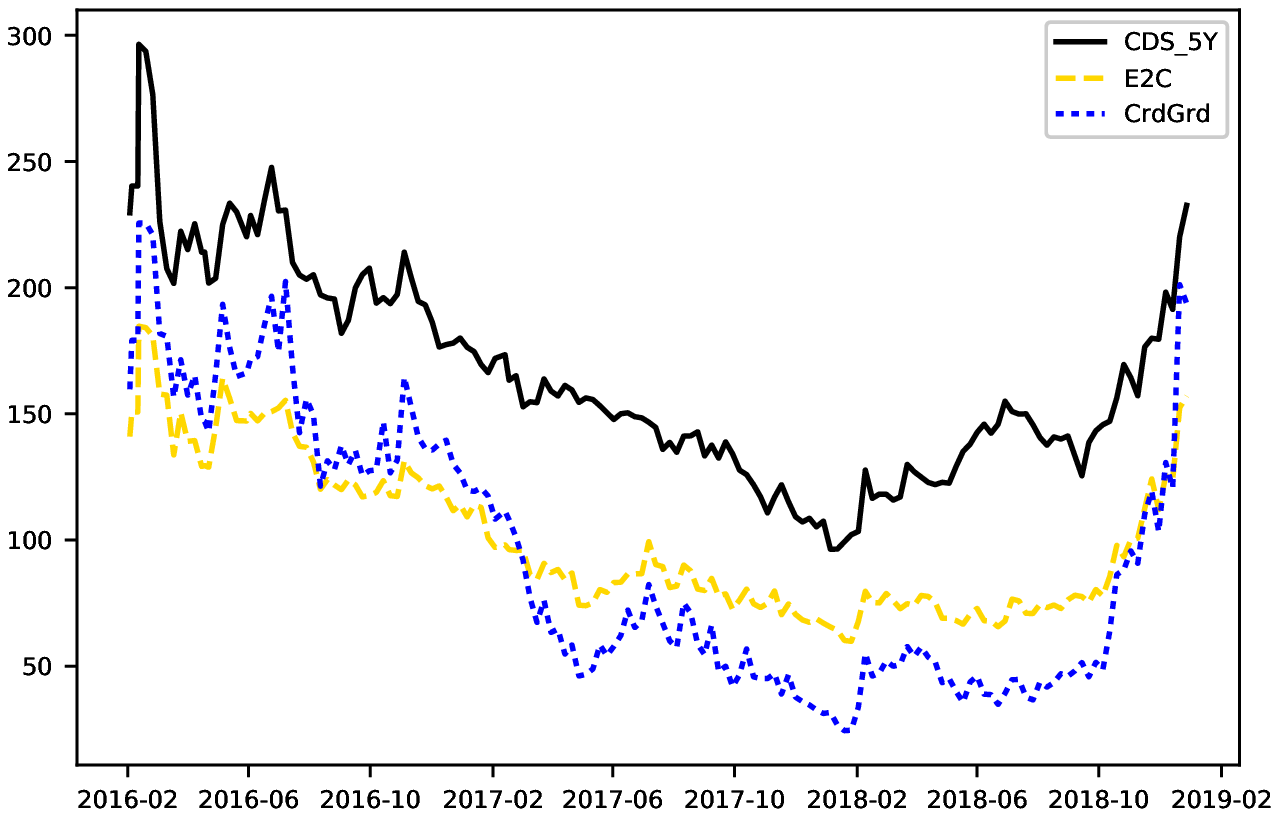}
			\vspace*{-1.0cm}
			\caption*{BB: Median}
		\end{subfigure}
		\hspace*{-0.0cm}
		\begin{subfigure}[b]{0.65\textwidth}
			\includegraphics[width=\textwidth]{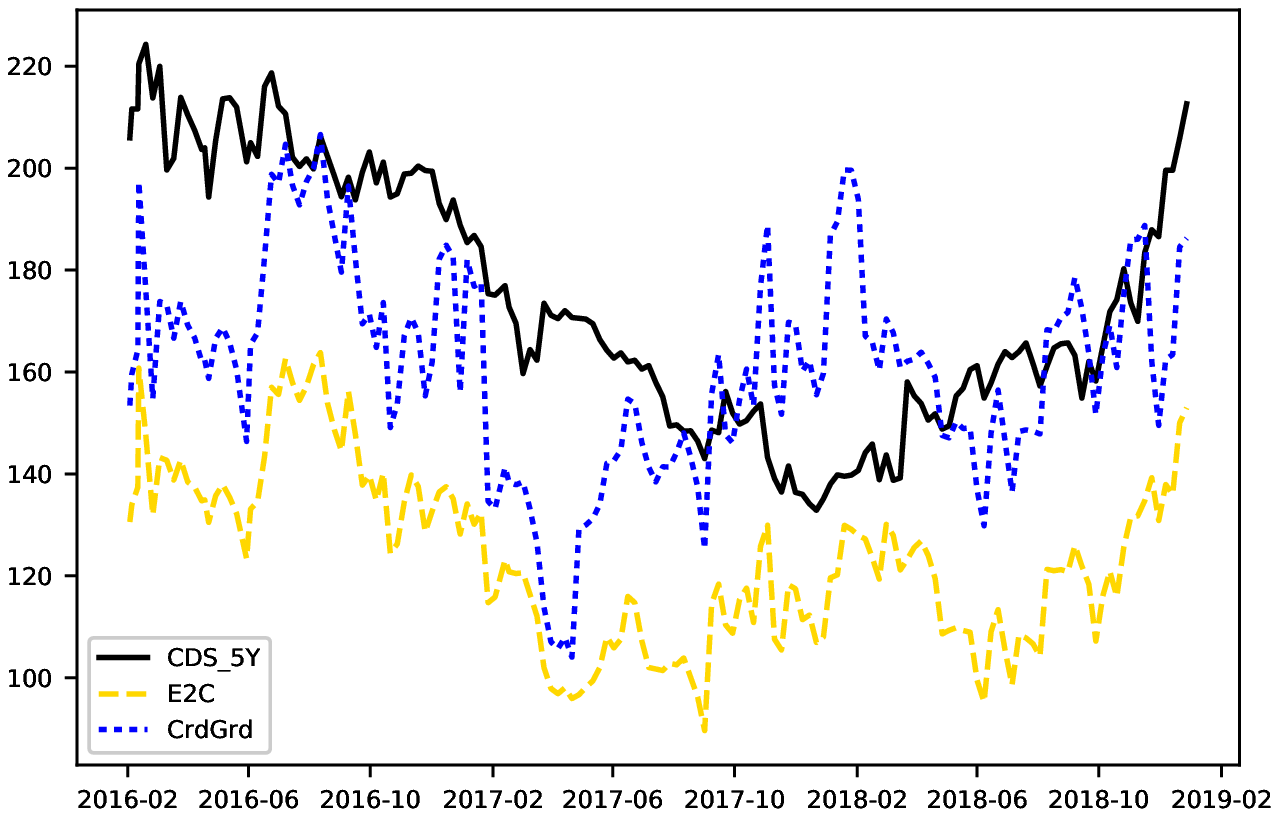}
			\vspace*{-1.0cm}
			\caption*{BB: Truncated Mean}
		\end{subfigure}
	\end{figure}
	\vspace*{-0.6cm}
	\begin{figure}[H]\ContinuedFloat
		\centering
		\advance\leftskip-3cm
		\advance\rightskip-3cm
		\begin{subfigure}[b]{0.65\textwidth}
			\includegraphics[width=\textwidth]{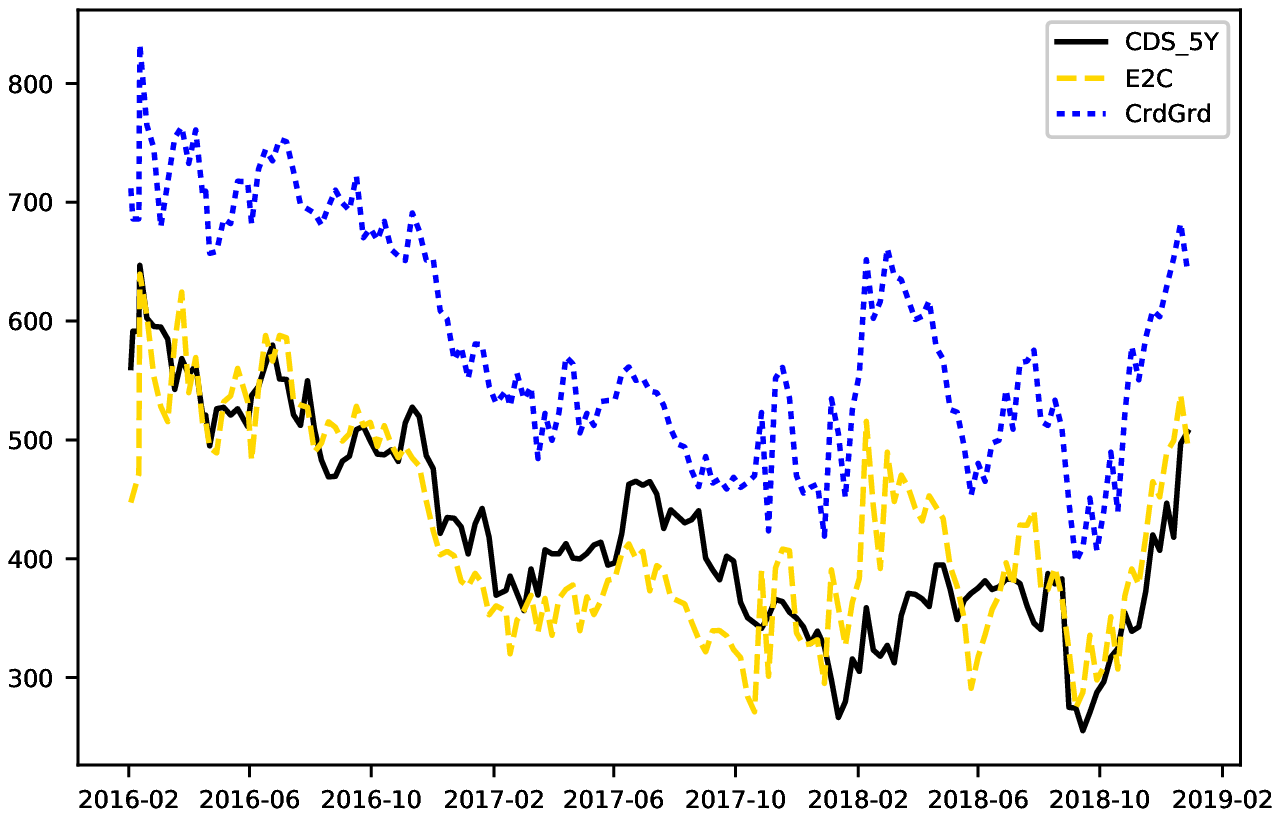}
			\vspace*{-1.0cm}
			\caption*{B: Median}
		\end{subfigure}
		\hspace*{-0.0cm}
		\begin{subfigure}[b]{0.65\textwidth}
			\includegraphics[width=\textwidth]{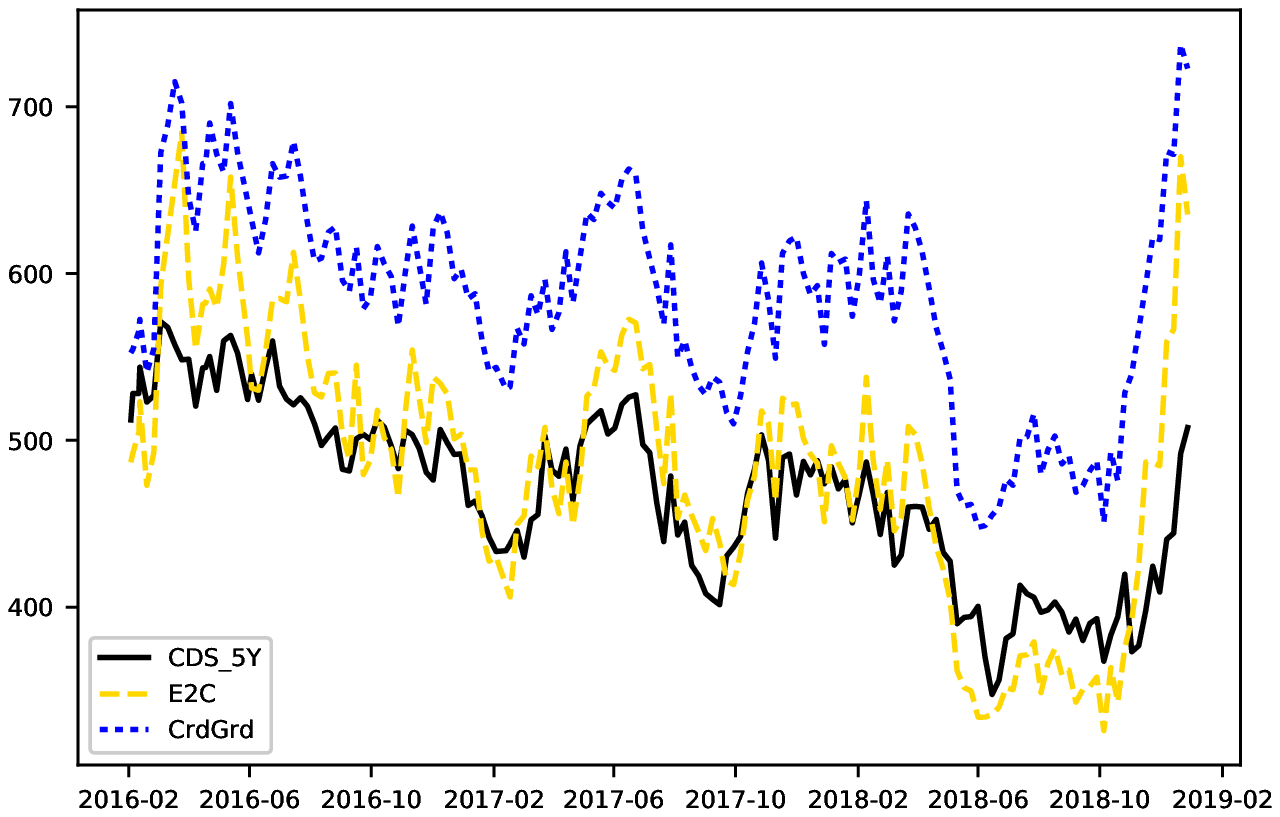}
			\vspace*{-1.0cm}
			\caption*{B: Truncated Mean}
		\end{subfigure}
		\vspace*{-0.25cm}
	\end{figure}
	In all cases, the E2C formula provides a closer median to the CDS than CreditGrades, with the exception of the double-B grade for which E2C and CreditGrades medians are almost alike. Regarding the truncated mean, E2C is indubitably better for the single B and comparable for single A and triple-B.
	These positive results are strengthened by the measures of accuracy shown in \nameref{sec:B}, \Cref{tab:7}. 
	Additionally, we find that these structural models do not underestimate spreads for riskier companies \citep[cf.][]{Teixeira2007}.
	\subsubsection{Industrial Sector Comparison}
	The firms are classified among ten major sectors (introduced in the above data section, \ref{sec:Data}). The sectoral comparisons are displayed below based on the same metrics.
	\begin{figure}[H]
		\caption{Comparisons between original 5y CDS, E2C and CreditGrades approximations in terms of industrial sectors}
		\label{fig:3}
		\centering
		\advance\leftskip-3cm
		\advance\rightskip-3cm
		\vspace*{-0.4cm}
		\begin{subfigure}[b]{0.65\textwidth}
			\includegraphics[width=\textwidth]{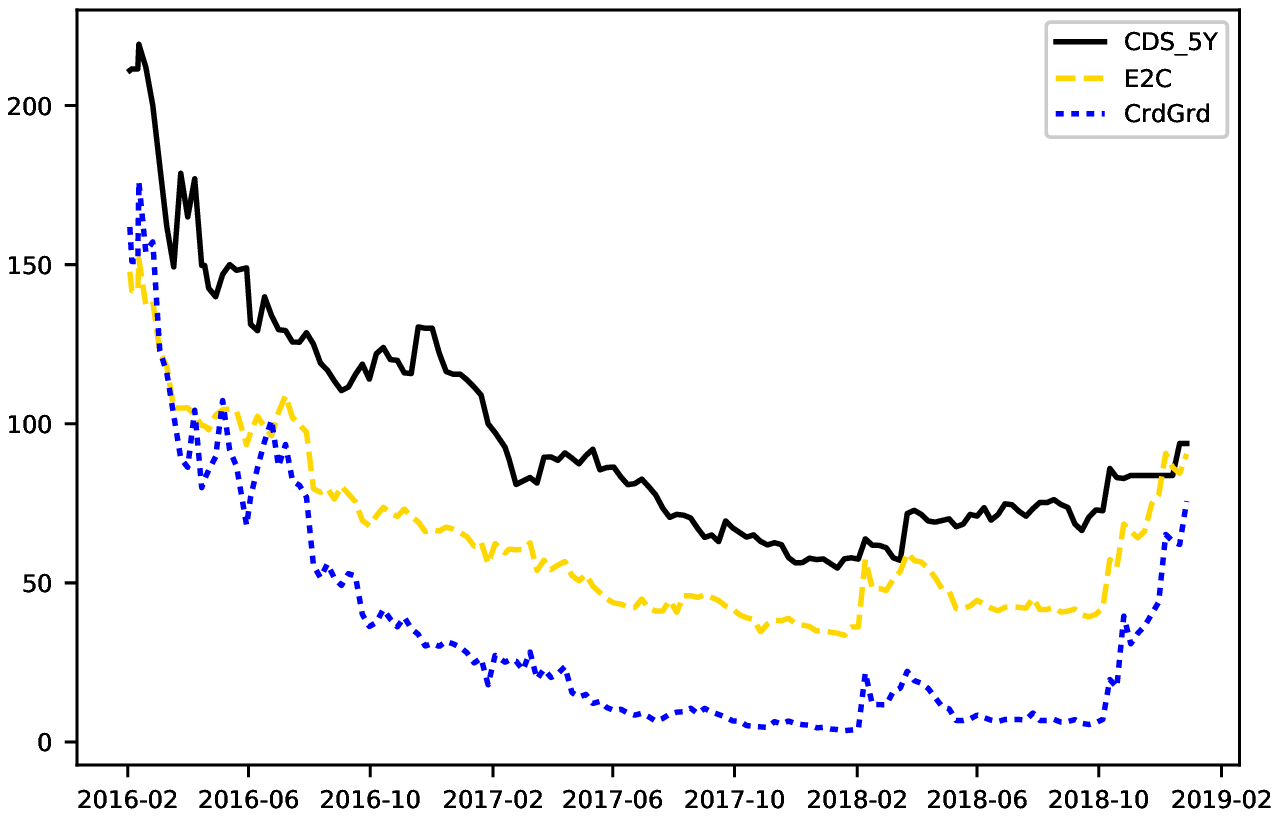}
			\vspace*{-1.0cm}
			\caption*{Basic Materials: Median}
		\end{subfigure}
		\hspace*{-0.0cm}
		\begin{subfigure}[b]{0.65\textwidth}
			\includegraphics[width=\textwidth]{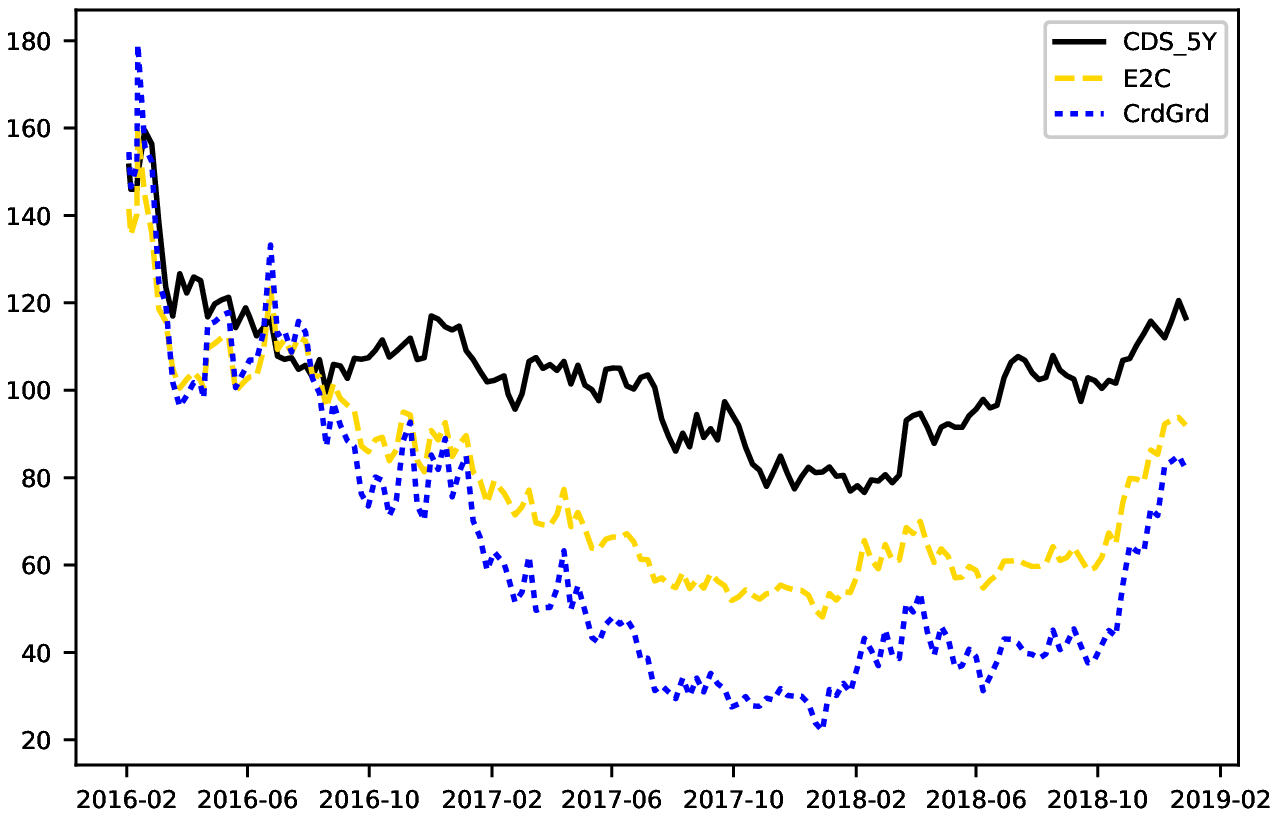}
			\vspace*{-1.0cm}
			\caption*{Basic Materials: Truncated Mean}
		\end{subfigure}
	\end{figure}
	\vspace*{-0.6cm}
	\begin{figure}[H]\ContinuedFloat
		\centering
		\advance\leftskip-3cm
		\advance\rightskip-3cm	
		\begin{subfigure}[b]{0.65\textwidth}
			\includegraphics[width=\textwidth]{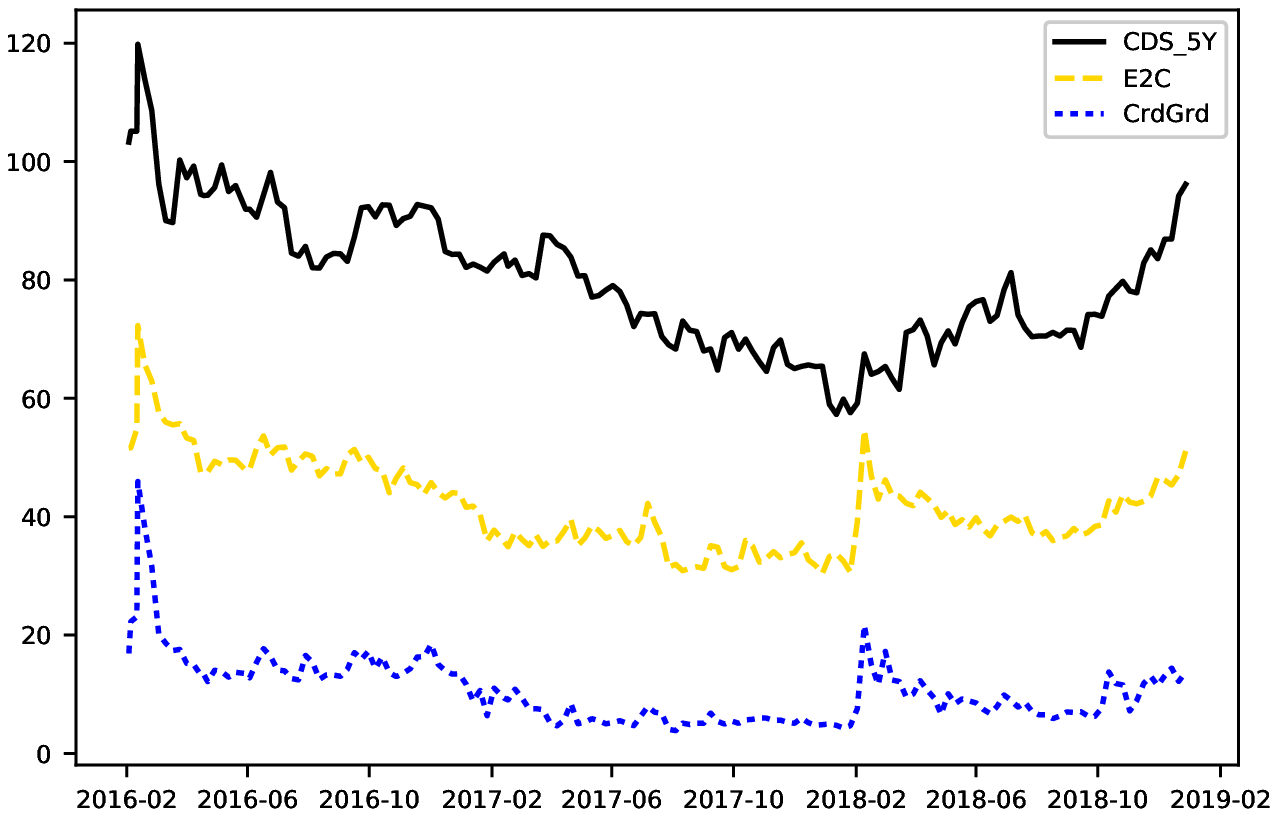}
			\vspace*{-1.0cm}
			\caption*{Communications: Median}
		\end{subfigure}
		\hspace*{-0.0cm}
		\begin{subfigure}[b]{0.65\textwidth}
			\includegraphics[width=\textwidth]{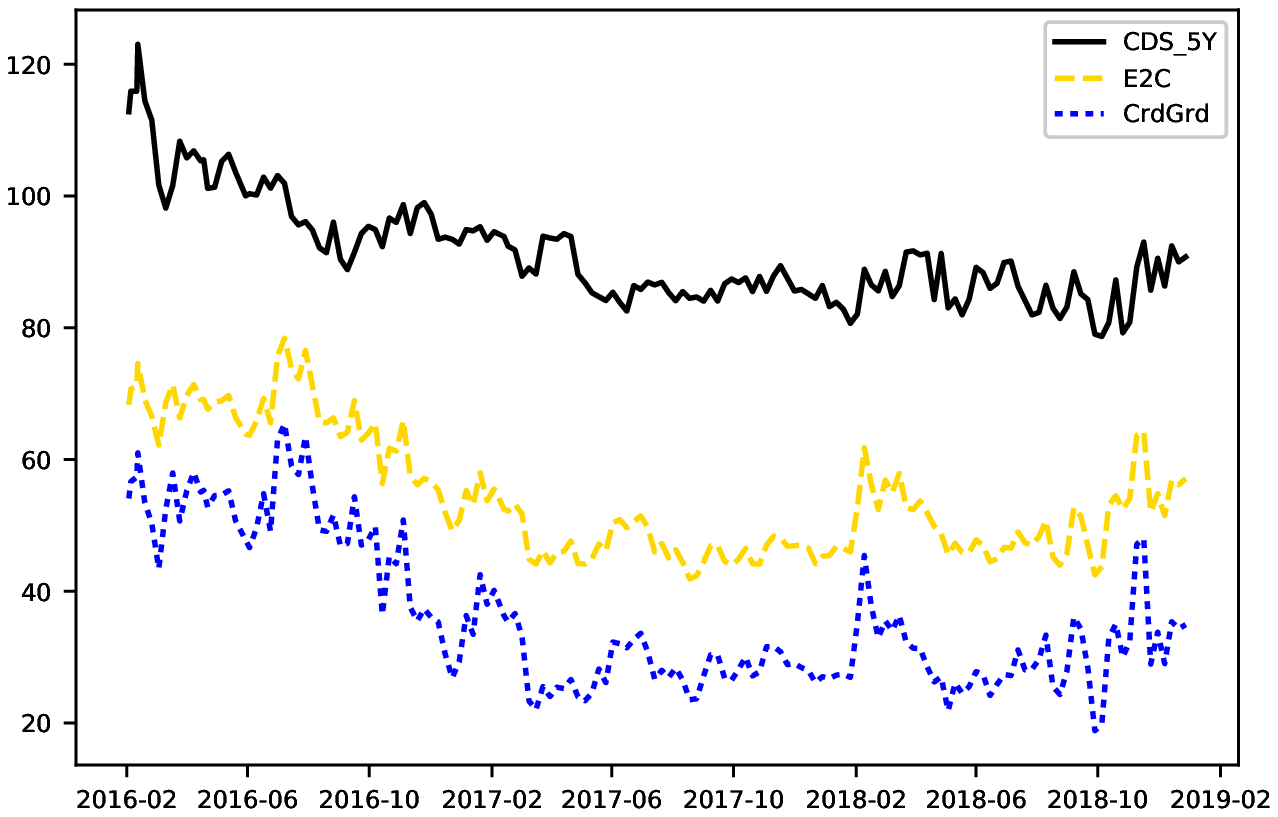}
			\vspace*{-1.0cm}
			\caption*{Communications: Truncated Mean}
		\end{subfigure}
	\end{figure}
	\vspace*{-0.6cm}
	\begin{figure}[H]\ContinuedFloat
		\centering
		\advance\leftskip-3cm
		\advance\rightskip-3cm	
		\begin{subfigure}[b]{0.65\textwidth}
			\includegraphics[width=\textwidth]{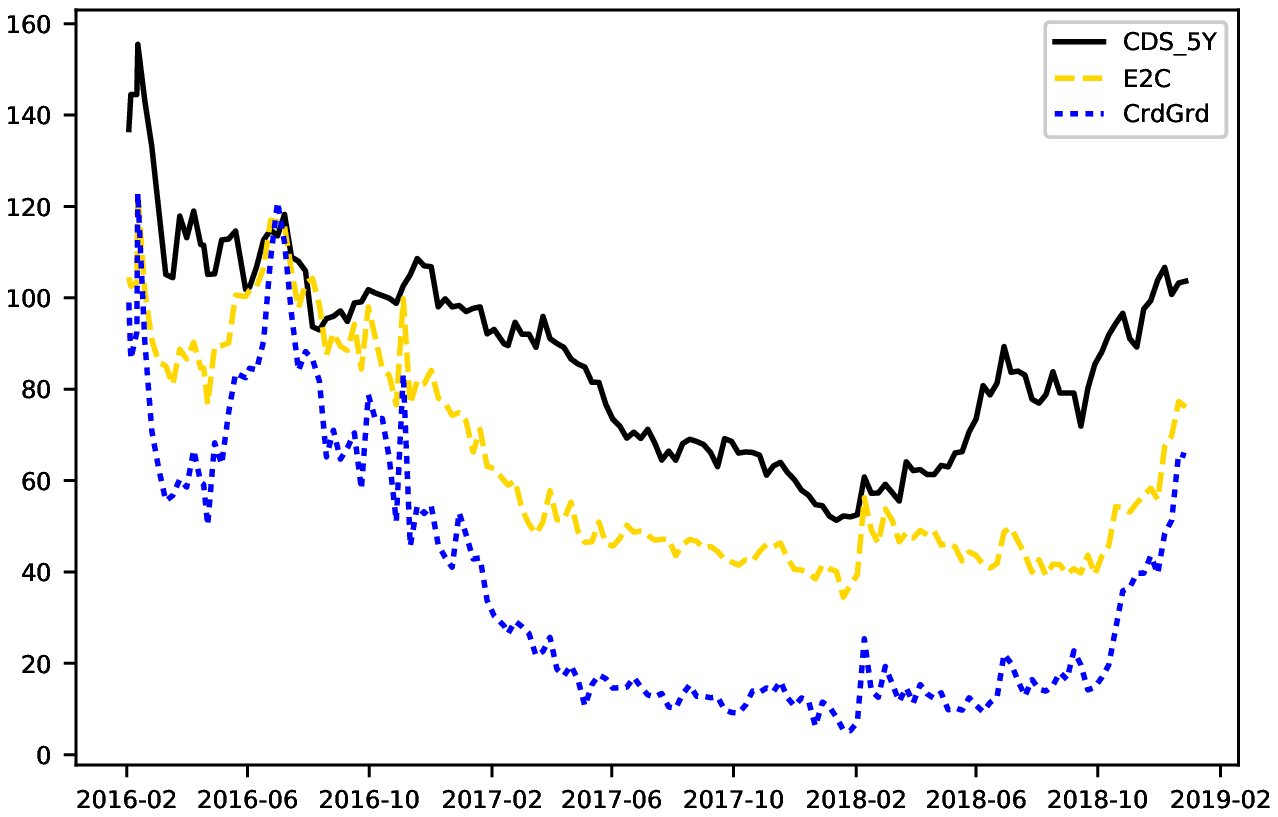}
			\vspace*{-1.0cm}
			\caption*{Consumer, Cyclical: Median}
		\end{subfigure}
		\hspace*{-0.0cm}
		\begin{subfigure}[b]{0.65\textwidth}
			\includegraphics[width=\textwidth]{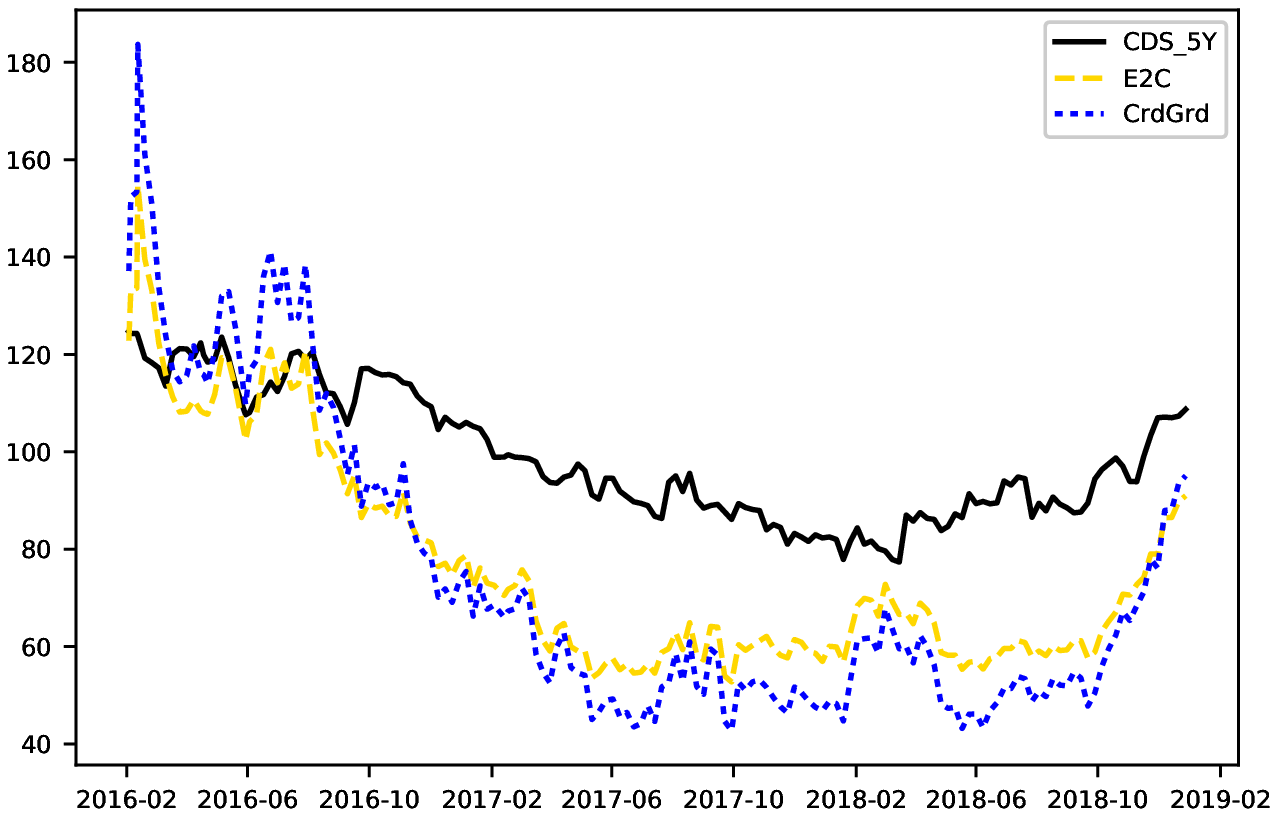}
			\vspace*{-1.0cm}
			\caption*{Consumer, Cyclical: Truncated Mean}
		\end{subfigure}
	\end{figure}
	\vspace*{-0.6cm}
	\begin{figure}[H]\ContinuedFloat
		\centering
		\advance\leftskip-3cm
		\advance\rightskip-3cm
		\begin{subfigure}[b]{0.65\textwidth}
			\includegraphics[width=\textwidth]{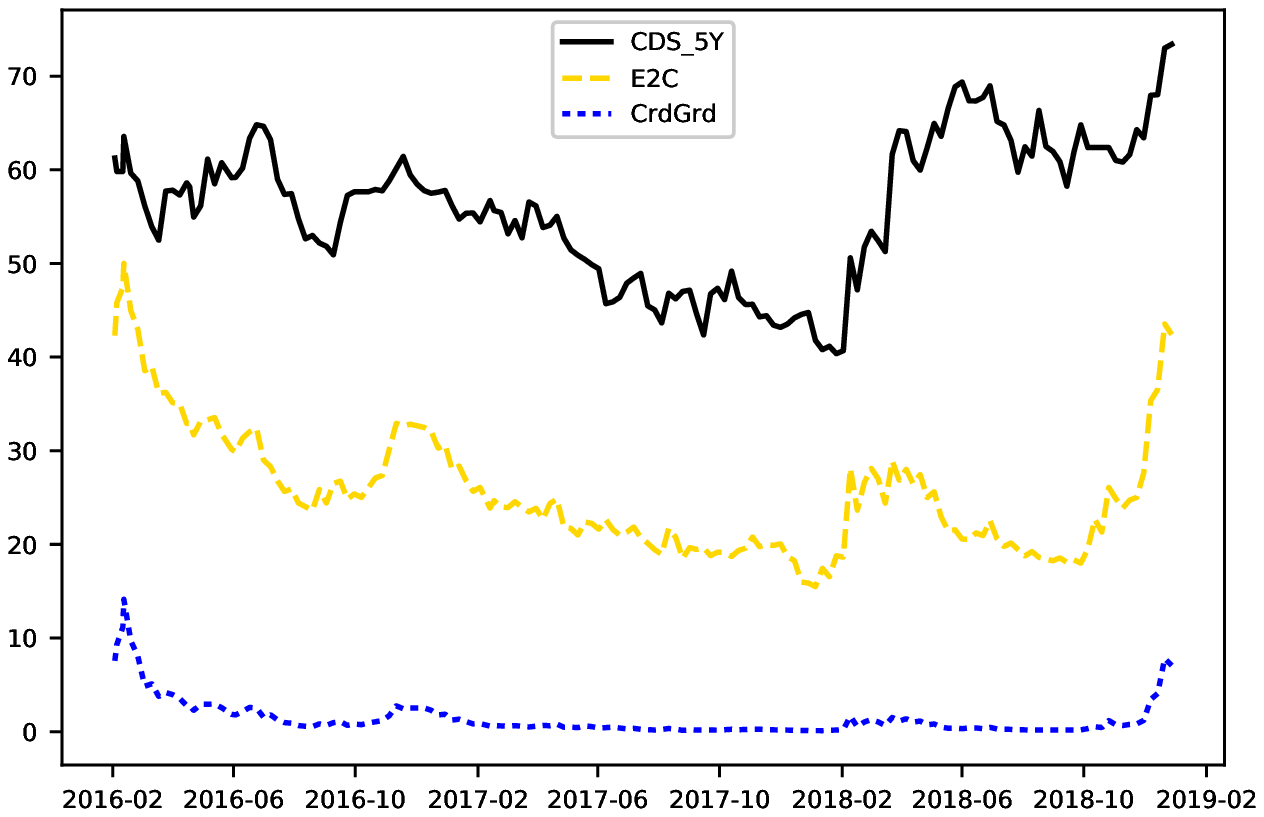}
			\vspace*{-1.0cm}
			\caption*{Consumer, Non-Cyclical: Median}
		\end{subfigure}
		\hspace*{-0.0cm}
		\begin{subfigure}[b]{0.65\textwidth}
			\includegraphics[width=\textwidth]{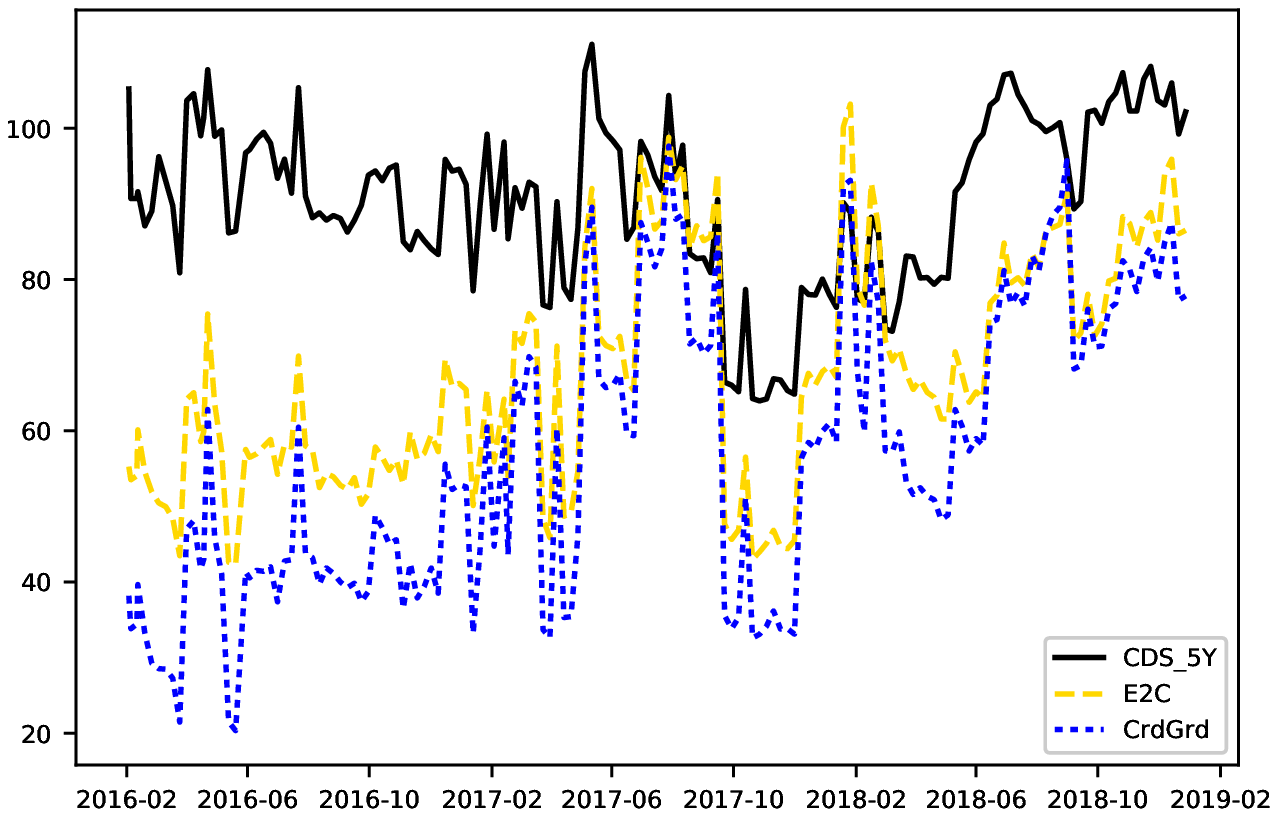}
			\vspace*{-1.0cm}
			\caption*{Consumer, Non-Cyclical: Truncated Mean}
		\end{subfigure}
	\end{figure}
	\vspace*{-0.6cm}
	\begin{figure}[H]\ContinuedFloat
		\centering
		\advance\leftskip-3cm
		\advance\rightskip-3cm	
		\begin{subfigure}[b]{0.65\textwidth}
			\includegraphics[width=\textwidth]{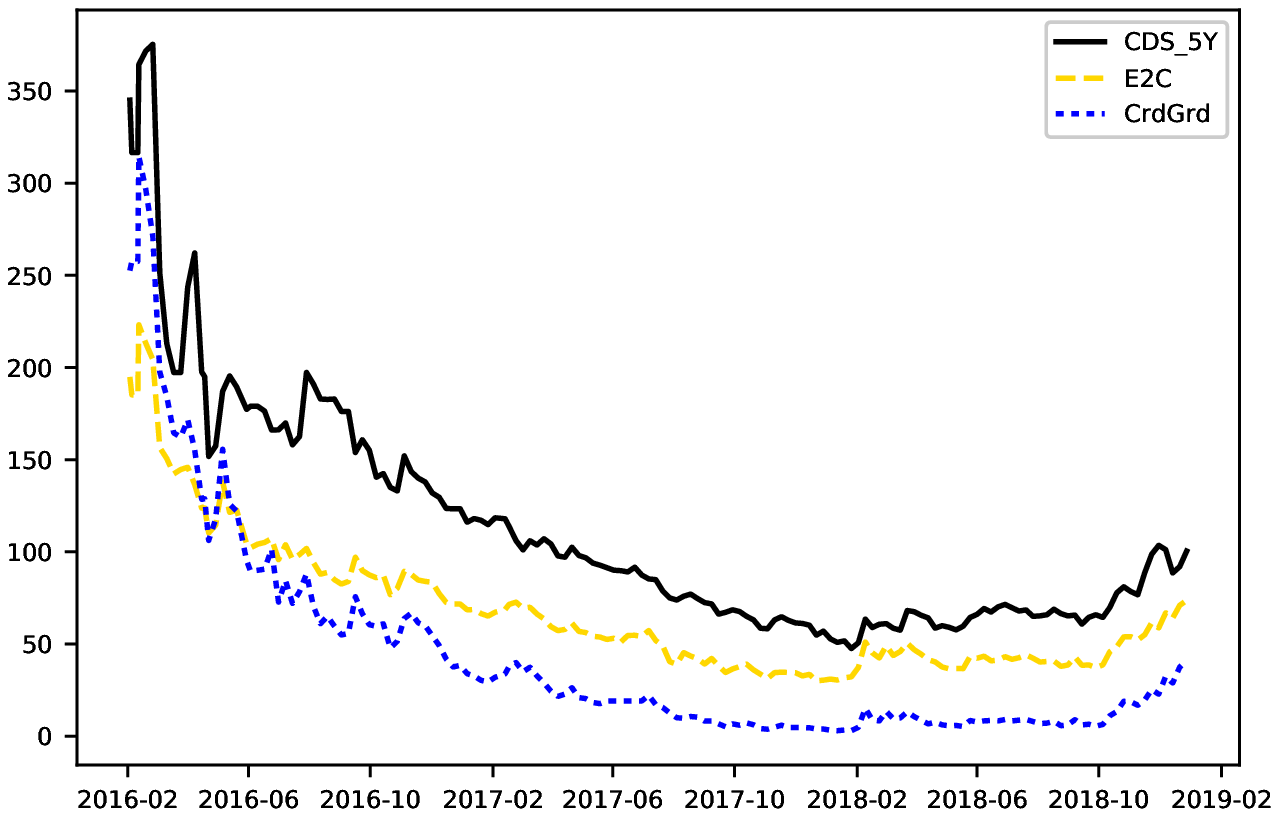}
			\vspace*{-1.0cm}
			\caption*{Energy: Median}
		\end{subfigure}
		\hspace*{-0.0cm}
		\begin{subfigure}[b]{0.65\textwidth}
			\includegraphics[width=\textwidth]{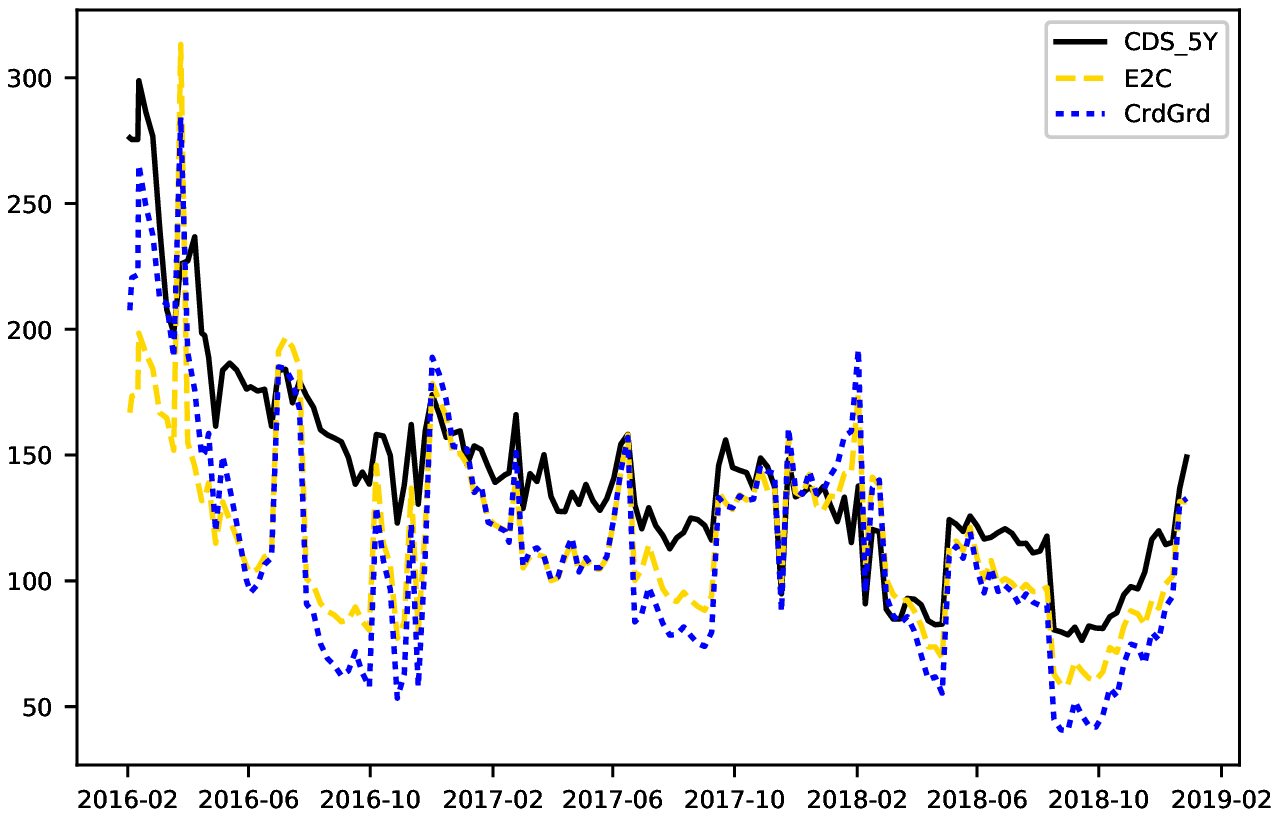}
			\vspace*{-1.0cm}
			\caption*{Energy: Truncated Mean}
		\end{subfigure}
	\end{figure}
	\vspace*{-0.6cm}
	\begin{figure}[H]\ContinuedFloat
		\centering
		\advance\leftskip-3cm
		\advance\rightskip-3cm	
		\begin{subfigure}[b]{0.65\textwidth}
			\includegraphics[width=\textwidth]{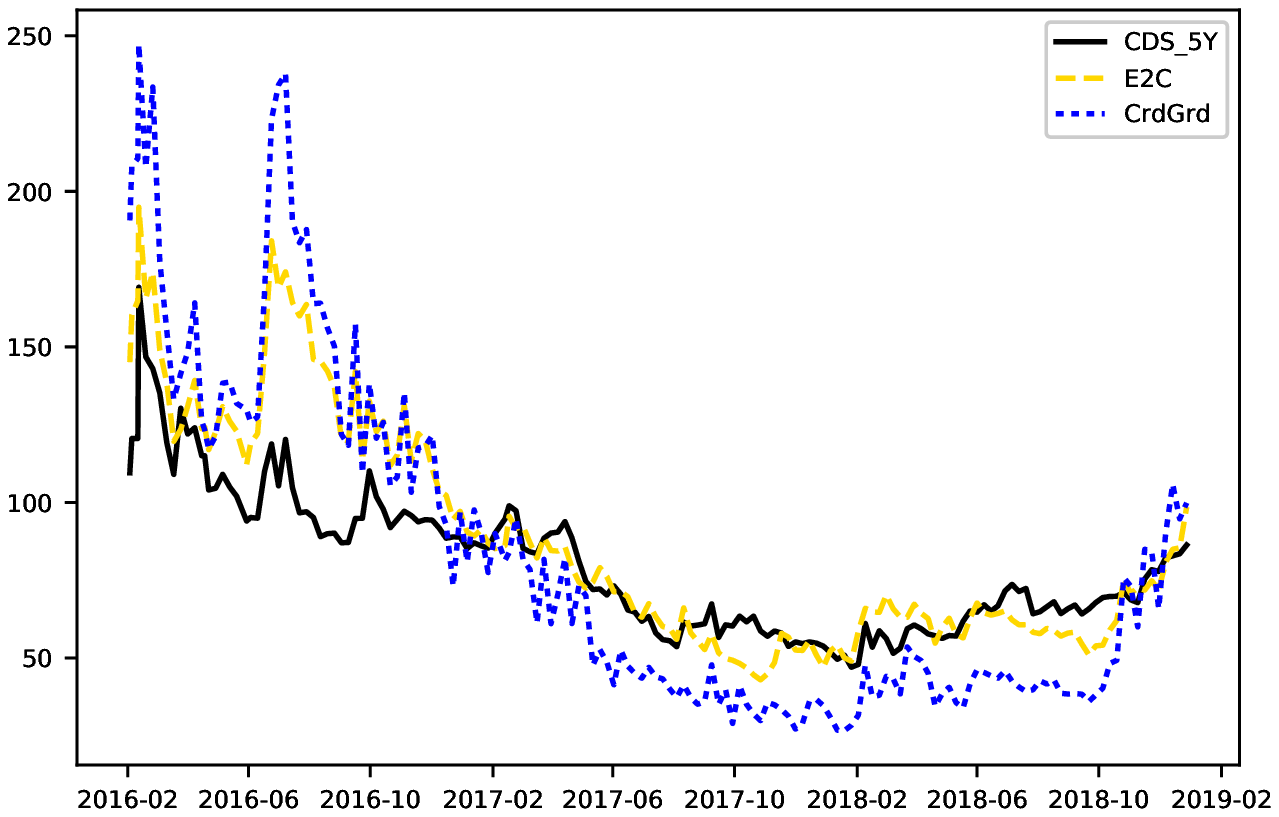}
			\vspace*{-1.0cm}
			\caption*{Financial: Median}
		\end{subfigure}
		\hspace*{-0.0cm}
		\begin{subfigure}[b]{0.65\textwidth}
			\includegraphics[width=\textwidth]{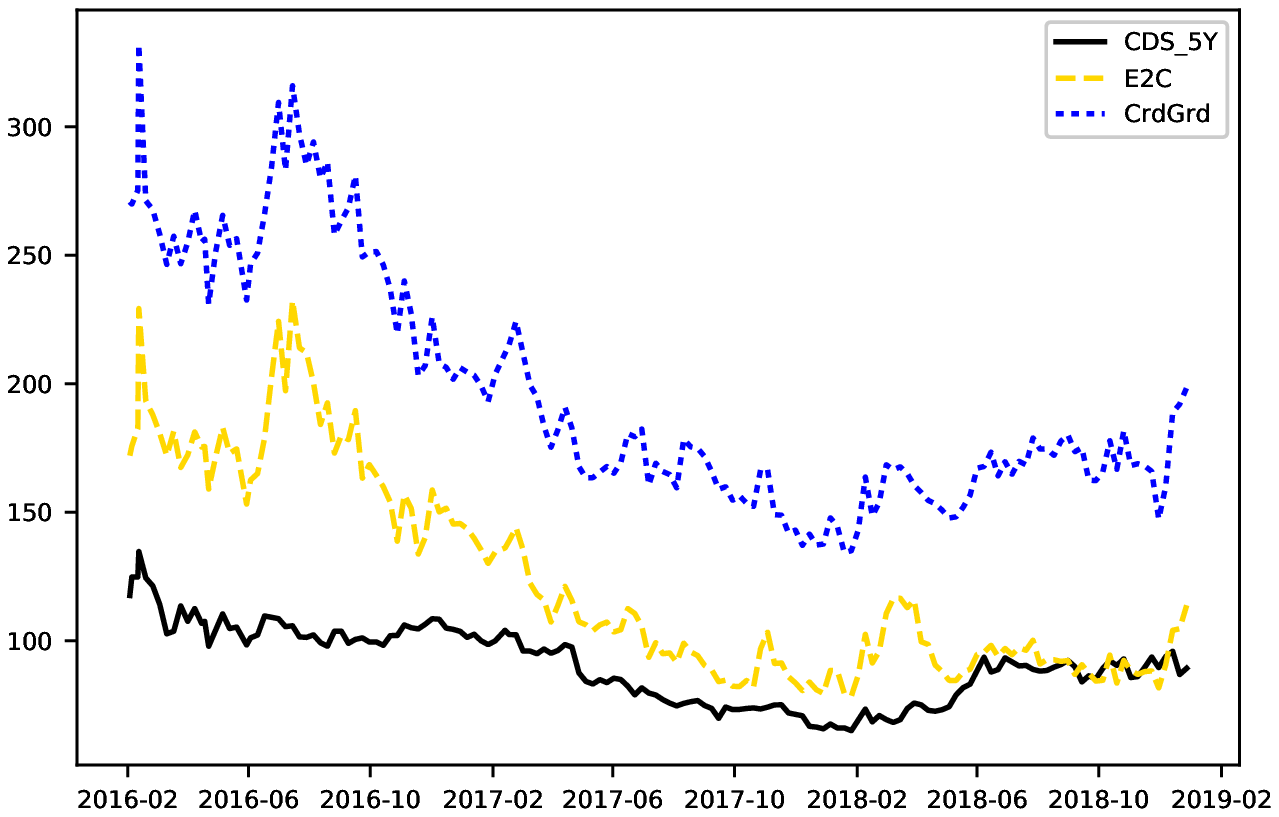}
			\vspace*{-1.0cm}
			\caption*{Financial: Truncated Mean}
		\end{subfigure}
	\end{figure}
	\vspace*{-0.6cm}
	\begin{figure}[H]\ContinuedFloat
		\centering
		\advance\leftskip-3cm
		\advance\rightskip-3cm
		\begin{subfigure}[b]{0.65\textwidth}
			\includegraphics[width=\textwidth]{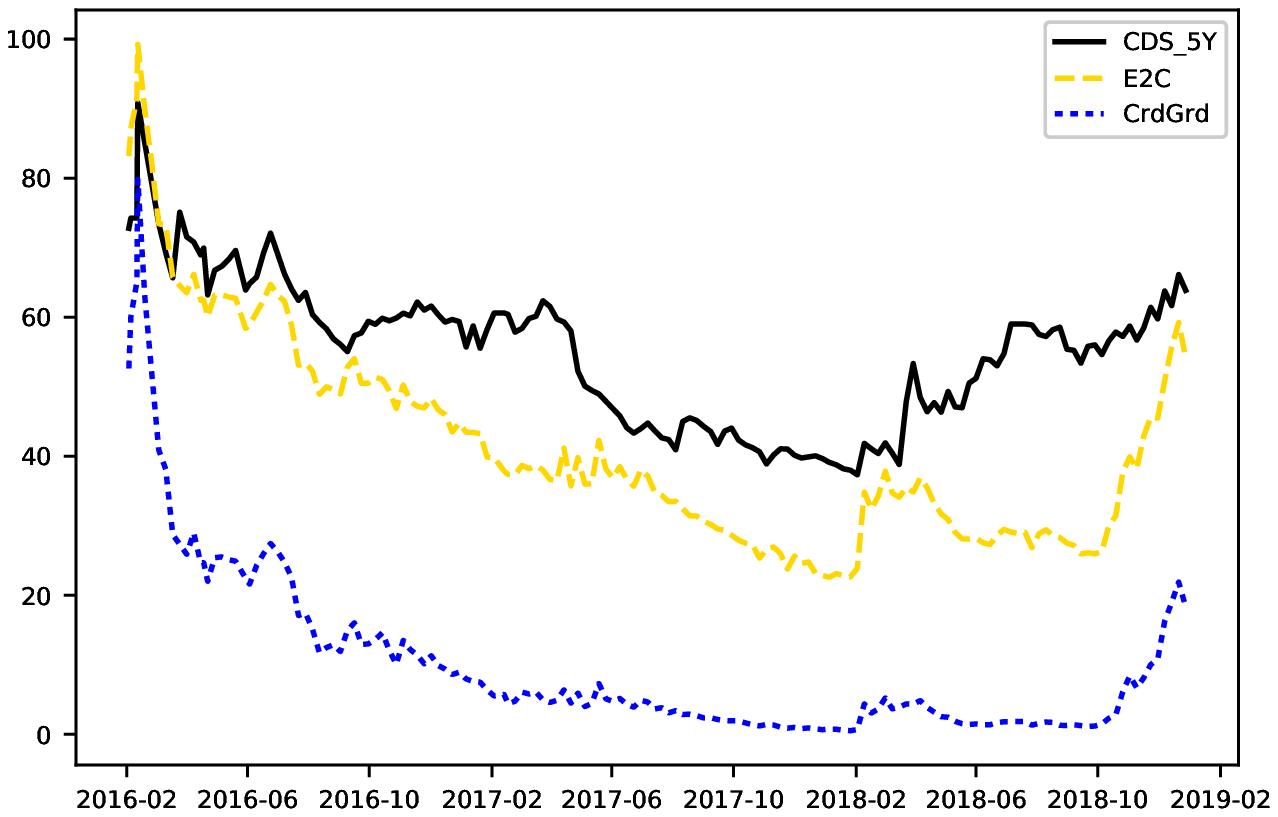}
			\vspace*{-1.0cm}
			\caption*{Industrial: Median}
		\end{subfigure}
		\hspace*{-0.0cm}
		\begin{subfigure}[b]{0.65\textwidth}
			\includegraphics[width=\textwidth]{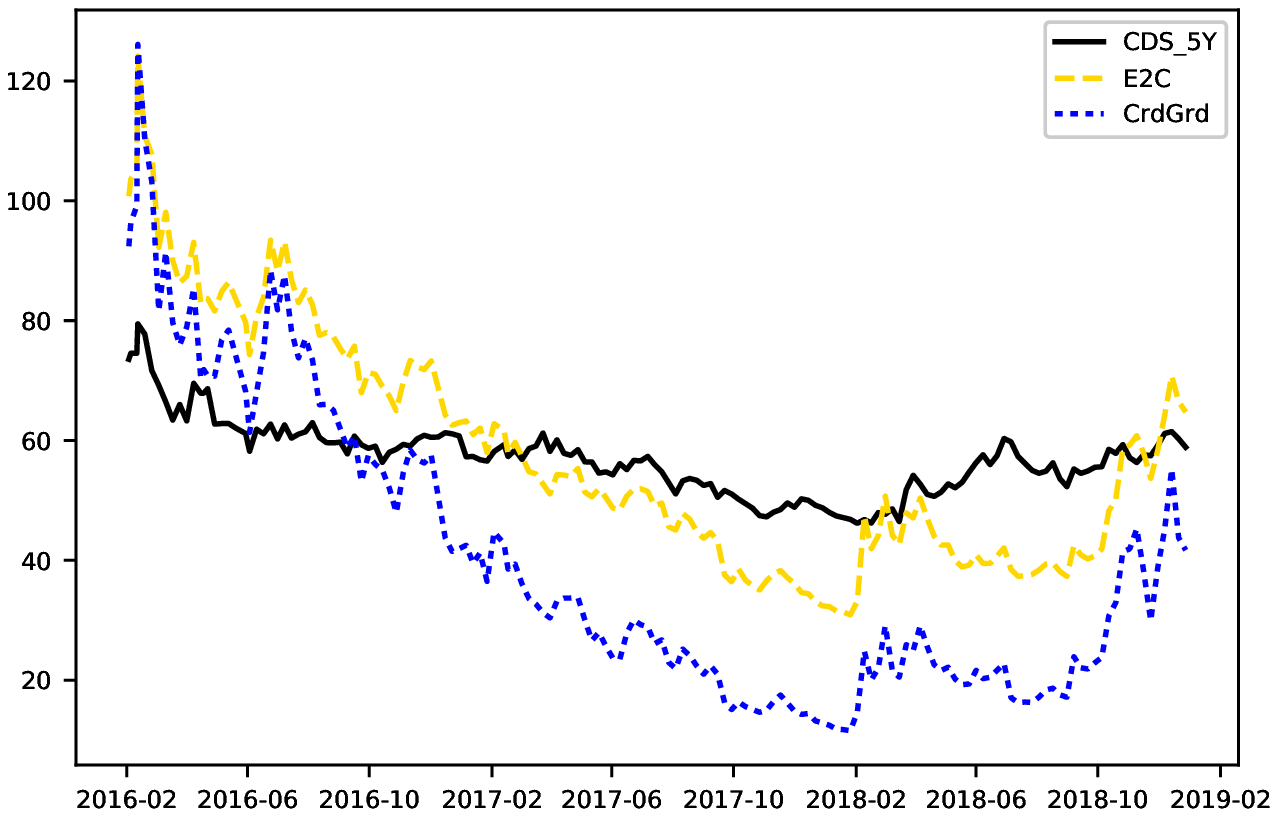}
			\vspace*{-1.0cm}
			\caption*{Industrial: Truncated Mean}
		\end{subfigure}
	\end{figure}
	\vspace*{-0.6cm}
	\begin{figure}[H]\ContinuedFloat
		\centering
		\advance\leftskip-3cm
		\advance\rightskip-3cm	
		\begin{subfigure}[b]{0.65\textwidth}
			\includegraphics[width=\textwidth]{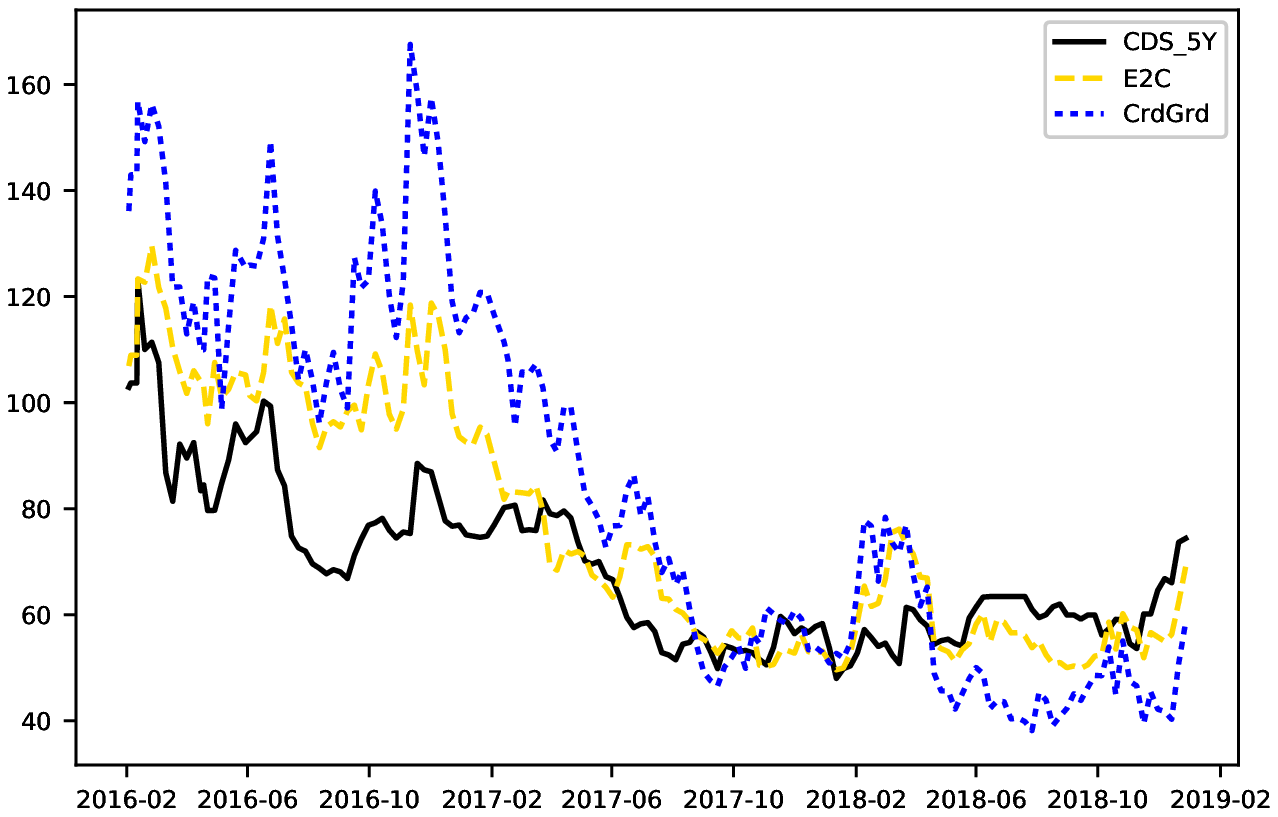}
			\vspace*{-1.0cm}
			\caption*{Utilities: Median}
		\end{subfigure}
		\hspace*{-0.0cm}
		\begin{subfigure}[b]{0.65\textwidth}
			\includegraphics[width=\textwidth]{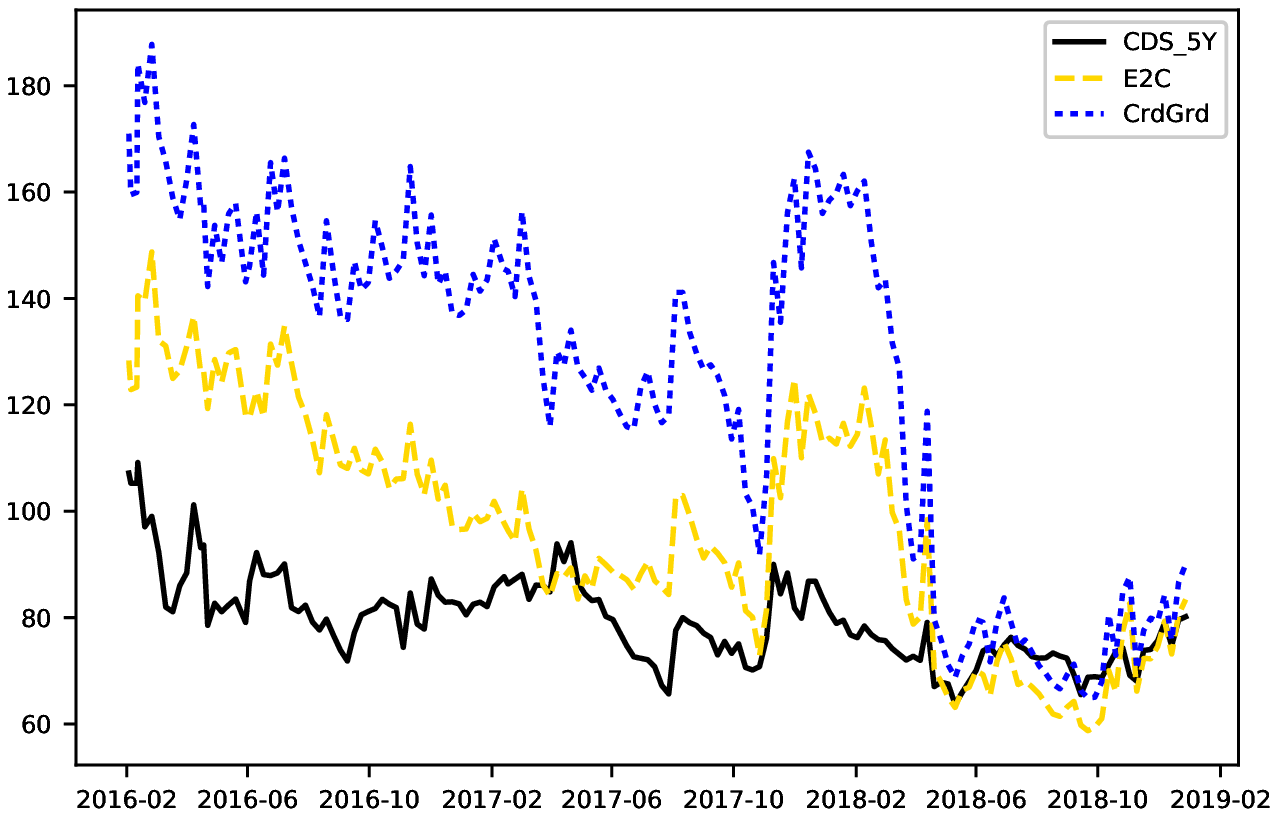}
			\vspace*{-1.0cm}
			\caption*{Utilities: Truncated Mean}
		\end{subfigure}
	\end{figure}
	 Except for utilities and financial, we reach the same conclusions as \cite{Rodrigues2011}\footnote{Their study includes the CreditGrades model.}, \cite{Lardic1999} and \cite{Eom2004} who emphasized that structural models generally underpredict the observed credit spreads, at least for low levels of risks. Indeed, it is hard to define and calibrate a meaningful process for the firms' asset values \citep{Schoenbucher2003}. Those are generally modeled following continuous diffusion paths, and thus are prevented from reaching the default barrier in very short time-spans, as may conceivably occur in reality. Zhou's work \citeyearpar{Zhou1997,Zhou2001a} overcame this issue by introducing a jump-diffusion process, at the cost of further complexity (cf. Section \ref{sec:Intro}). In addition, credit spreads bear residual agency and covenant risks, which are not shared by equity.
	 Regarding the median, it is clear that in every case, the E2C formula approximates the CDS spreads at least as well as CreditGrades. The E2C approximations are clearly superior for basic materials, communications, consumer cyclical and non-cyclical, energy and industrial. Almost identical conclusions are reached computing the truncated mean. E2C results are unquestionably better for basic materials, communications, industrial, utilities and financial. The latter is particularly interesting as it is a key sector where CreditGrades, and more generally ``Merton'' models, are traditionally poor.
	These encouraging charts are confirmed by the measures of accuracy shown in \Cref{tab:8} of \nameref{sec:B}. We now know, at least with these samples, that our simple estimation is at least as good as CreditGrades in almost all of the above cases.
	
	After the positive statistics of section \ref{sec:BasicAnalysis}, the above detailed analysis in terms of rating and sector confirms the satisfactory level of accuracy of the E2C formula with regards to the CreditGrades model. We therefore provide credit market practitioners with an equation which is slightly better than its closest parent, the CreditGrades model, while surpassing it in terms of intuitiveness and clarity.
	
	The period studied in this paper (2016-2018) was overall characterized by low volatilities and highly expansive monetary policies, limiting our assessment of the E2C formula's reaction to changes in the volatility regime. However, we performed an OLS analysis at dates of higher volatility within our data sample. We chose to match the lower bound of the 3-month VIX volatility index\footnote{Which is steadier than the VIX index.} during the 2007-2008 financial crisis, which was at 20\%; a threshold\footnote{As robustness check, the volatility analysis was performed for $1.5\sigma$ above the mean and at the 75\% quantile, obtaining the same results.} exceeded 20 times out of our 155 dates (i.e. 6078 observations). The results available in \Cref{tab:9} (\nameref{sec:B}) show a stable accuracy for E2C even when the market is more volatile\footnote{\Cref{tab:9} (\nameref{sec:B}) provides corresponding results for CreditGrades as well.}.
	\section{Approximation Improved by Random Forests}
	\subsection{Sample Construction}
	The previous positive results give us ground to use and test our approximation further. More precisely, we look to improve our CDS approximation using the simple E2C formula and some additional input variables, and the actual CDS spreads are thereafter used as labels to train supervised learning algorithms. As usual, the original sample is split between training and test sets; and the results are checked with traditional error-squared metrics.

	We suppose that the major part of the CDS approximation should be explained by the E2C formula, which is based on equity information. But CDS are evolving on a different market and thus are not only company-dependent but also correlated to the overall credit market. Therefore, we add the most liquid Investment Grade CDS index, the IG CDX\footnote{As robustness check, the Itraxx \& HY CDX indexes and the VIX 3-month volatility index were set as independent variables; various combinations led to similar results.}, which evolves through time but is not company-dependent. The default likelihood of a debtor is qualitatively and quantitatively assessed by a credit rating agency and disclosed with a credit rating grade. Obviously, the grade level should have an impact on the corresponding CDS spread. Companies' size should also matter, being related to a level of importance and liquidity and linked to a stability concept, thus being integrated in CDS levels. This latter independent variable is measured by the market capitalization. It is worth noting that debt ratings and size are company-specific but not strongly time-dependent. Finally, companies' industrial sectors and geographical locations\footnote{The currencies stands as proxy for geographical locations.} are included as easily accessible additional variables which satisfy the same criteria.
	
	Before running our algorithms, we perform light features engineering. According to \cite{LeCun2012}, ``to efficiently perform a basic machine learning algorithm, it is fundamental to pre-process the data, reduce dimensions and extract hand-crafted, domain specific features''. First, points with one or more missing data are removed (8\% of our data sample), leaving 47,476 observations. Incidentally, random forests are generally robust to numerical instabilities, so there is no need to remove multicollinearity and features with little variance. There is also no need for monotone functions such as centering, scaling or Box-Cox transformations, to which random forests are invariant. The second step in feature engineering concerns the handling of three non-numeric polytomous variables, i.e. countries, industrial sectors, and debt ratings. We perform a label encoding for the unsecured debt ratings, which are an ordinal variable. However, countries and industrial sectors are plain non-ordinal categorical values, which we handle with one-hot encoding\footnote{As usual, for each binary variable group, the dummies with the fewest observations (here, ``CTRY\_MYR'' and ``INDUSTRIES\_DIVERSIFIED'') are removed.} (if countries and industrial sectors were converted through label encoding, artificial relationships between the values would be enforced, whereas creating dummy variables dilutes the features' strength). After a short presentation of our metrics, we run the algorithms on 4 independent variables and 23 dummy variables.

	The in-sample set is built by randomly removing 20\% of the companies and 20\% of the dates. This method leaves us with an out-of-sample standing for 36\% of the whole dataset. The approximation accuracy is assessed using a well-known econometric measure, the R-squared\footnote{$R^2(y, \hat{y}) = 1 - \dfrac{\sum_{i=1}^{N} (y_i - \hat{y}_i)^2}{\sum_{i=1}^{N} (y_i - \bar{y})^2}$}. Other distance-related metrics, such as the relative absolute and squared errors, deliver the same conclusions.
	\subsection{Random Forest Regressions}
	\subsubsection{Algorithm Description}
	An efficient algorithm minimizes both the bias and the variance. Generally, decision trees deliver a low bias but are, by construction, prone to overfit. Therefore, ensemble methods \citep{Opitz1999} running multiple trees are appropriate for relevant predictions. Specifically, we choose a meta-algorithm called bootstrap aggregating \citep[or ``bagging'', in][]{Breiman1996} which builds multiple trees from a training set's sample, randomly drawn with replacement. Bagging contributes to reducing the variance and avoiding overfitting. In the process known as random forest algorithm, developed by \cite{Breiman2001}, this property is reinforced by the random selection, at each node, of a subset of the features used for splitting. Earlier work on random forests had also been launched at Bell Laboratories by \cite{Ho1995,Ho1998}. This random subset might slightly increase the bias along with the decrease in variance (bias-variance tradeoff).
	
	Random forests have recently been used within the fields of economics and finance \citep{Nyman2016,Khaidem2017,Tanaka2016,Behr2017,Yeh2012}. The validity of the use of random forests among a wide choice of learning algorithms has been studied, and confirmed, by \cite{Fernandez-Delgado2014}. \cite{Brummelhuis2017} further verify this use in the context of CDS proxy construction.
	
	Even though recent machine learning research is more oriented toward deep learning algorithms \citep{Badrinarayanan2017,Sun2018,Sangineto2018}, recent papers in applied machine learning have shown that the outputs of deep learning and ensemble based methods are comparable, at least for reasonable sample sizes. An argument in favor of deep learning is its marginally better accuracy \citep{Ahmad2017} although random forests have in some cases outperformed neural networks \citep{Liu2013,RodriguezGaliano2015,Krauss2017}. However, our own priority in this paper is to keep our methods as simple and above all as understandable as possible, leading us to strongly lean in favor of random forests. Beyond performing an internal cross-validation, random forests require less parameter tuning \citep{Ahmad2017}. As far as transparency goes, random forests are also preferable to neural networks, which are often considered to be black boxes \citep[although see][for a recent improvement]{ShwartzZiv2017}.
	
\begin{figure}[H]
		\caption*{Figures 4-6: Goodness-of-Fit in function of the number of trees, features and nodes.}
		\centering
		\advance\leftskip-3cm
		\advance\rightskip-3cm
		\vspace*{-0.0cm}
		\hspace*{0.5cm}
		\begin{subfigure}[b]{0.430\textwidth}
			\includegraphics[width=\textwidth]{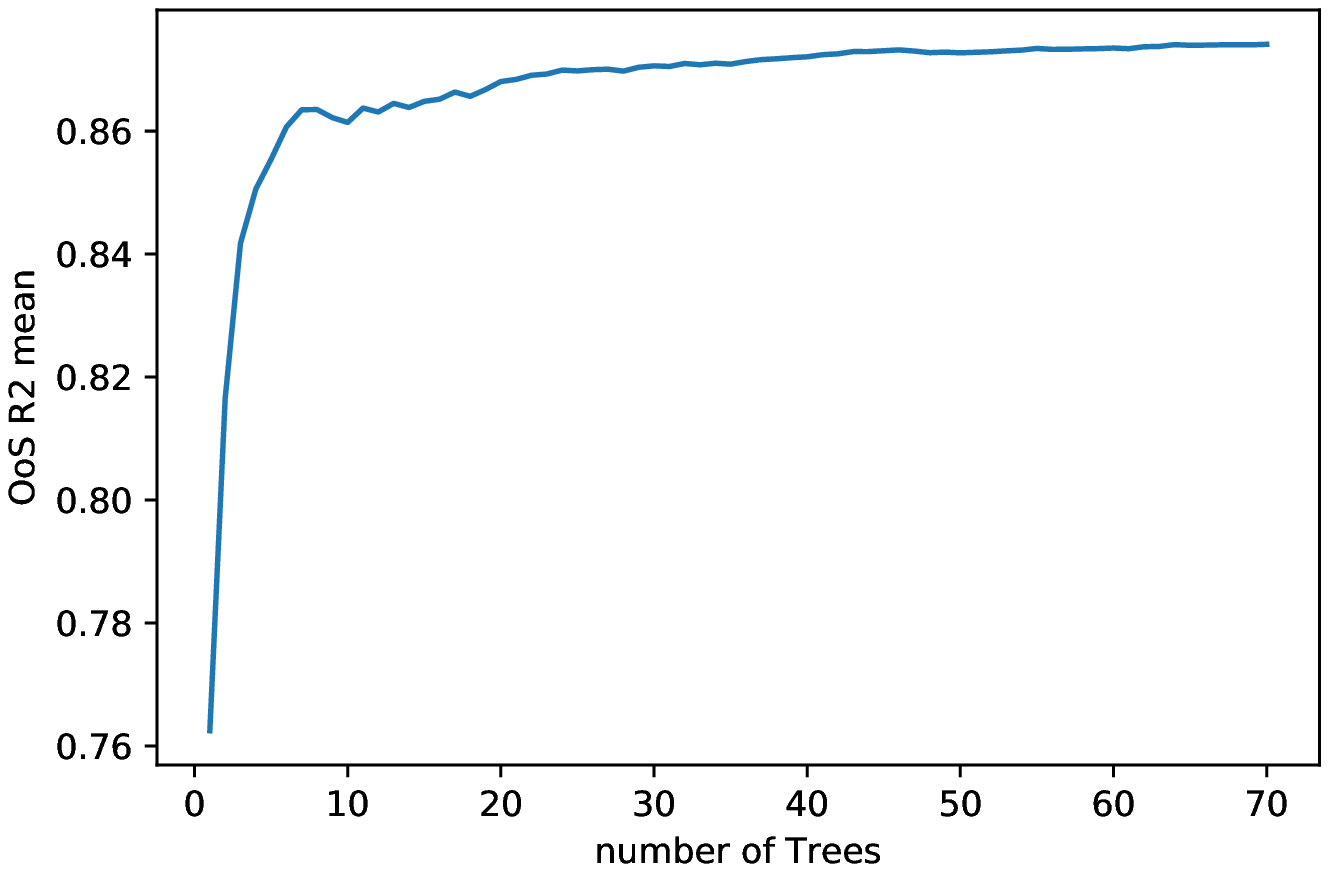}
			\caption*{OoS: mean}
		\end{subfigure}
		\hspace*{-0.5cm}
		\begin{subfigure}[b]{0.430\textwidth}
			\includegraphics[width=\textwidth]{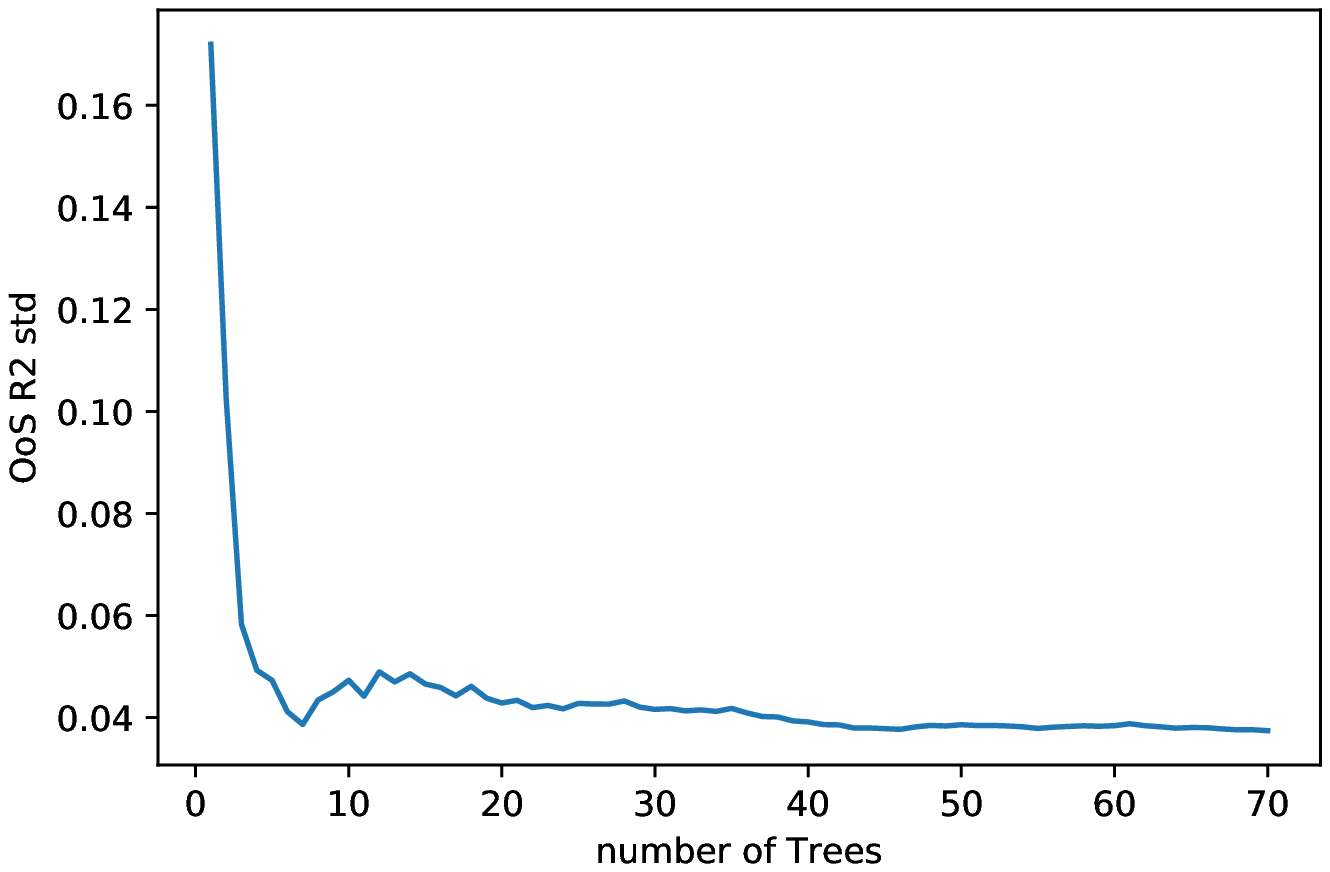}
			\caption*{OoS: std}
		\end{subfigure}
		\hspace*{-0.5cm}
		\begin{subfigure}[b]{0.430\textwidth}
			\includegraphics[width=\textwidth]{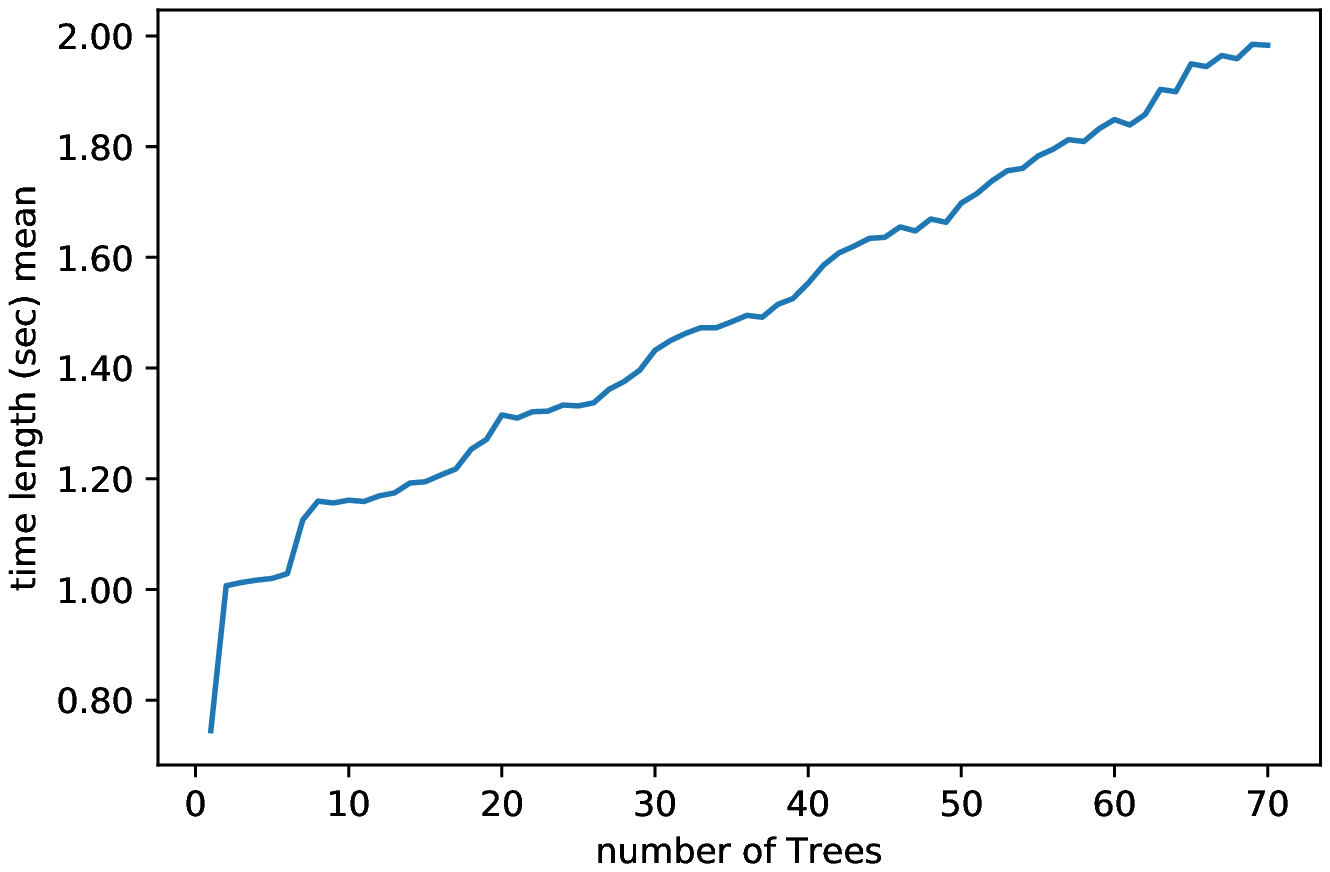}
			\caption*{Time: mean}
		\end{subfigure}
		\vspace*{-0.5cm}
		\caption{Number of trees stands for the number of bootstrapped random trees}
		\label{fig:4}
	\end{figure}

	While random forests are generally run with unpruned classification and regression trees \citep[CARTs, in][]{Breiman1984}, limiting the trees' depth is an option. This has no negative impact in terms of overfitting, and has the positive result of contributing to faster computation. Consequently, we set a maximum depth to the trees, and choose to run them in parallel for further speed. The following is a brief explanation of the random forest algorithm, based on the notations of \cite{Hastie2009}. At step 1(b), we shortly present a decision trees algorithm.\\
	\hrule
	\begin{enumerate}\label{rf:method}
		\item for b= 1 to B:
		\begin{enumerate}
			\item draw a bootstrap subset from the original dataset.
			\item grow a decision tree $T_b$ using the following process:
			\begin{enumerate}  
				\item use a greedy algorithm seeking the splitting feature $j$ among a randomly selected subset $m$ of the $p$ features ($m \leq p$) and split point $s$ that solve:

				\hspace*{-2.2cm}\vbox{\footnotesize\begin{align*}
				&\min_{j, s}\left[\min_{c_1, c_2}\left(\sum_{i\in R_1(j,s)}(y_i-c_1)^2 + \sum_{i\in R_2(j,s)}(y_i-c_2)^2\right)\right]\\\\
				&where\ R_1(j,s) = \{i\mid x_{i,j}\leq s\}\\
				&and\ R_2(j,s) = \{i\mid x_{i,j}> s\}\\
				&c_l=avg\left(y_i\mid i\in R_l(j,s)\right)
				\end{align*}}
				\item once the best fit found, partition the data in two regions ($R_1$, $R_2$) \item repeat (i) on each resulting region until a leaf node in all the branches of the tree is found, or the maximum depth is reached. 
			\end{enumerate}
		\end{enumerate}
		\item output the ensemble of trees $\{T_b\}_1^B$.
	\end{enumerate}
	Thereafter, the prediction on a new point $x$ from the OS is made using the function:
	\begin{align*}
	\hat{f}(x)=\dfrac{1}{B}\sum_{b=1}^BT_b(x).
	\end{align*}
	\hrule
	\noindent\\
	Random forests require to set at least two hyper-parameters. First of all, we determine the number $B$ of random trees we want the algorithm to run. While more trees reduce the variance, we choose $B$ as usual in function of accuracy constrained by the time of computations. \Cref{fig:4} summarizes various criteria supporting our decision.

\begin{figure}[H]
		\centering
		\advance\leftskip-3cm
		\advance\rightskip-3cm
		\hspace*{0.5cm}
		\begin{subfigure}[b]{0.430\textwidth}
			\includegraphics[width=\textwidth]{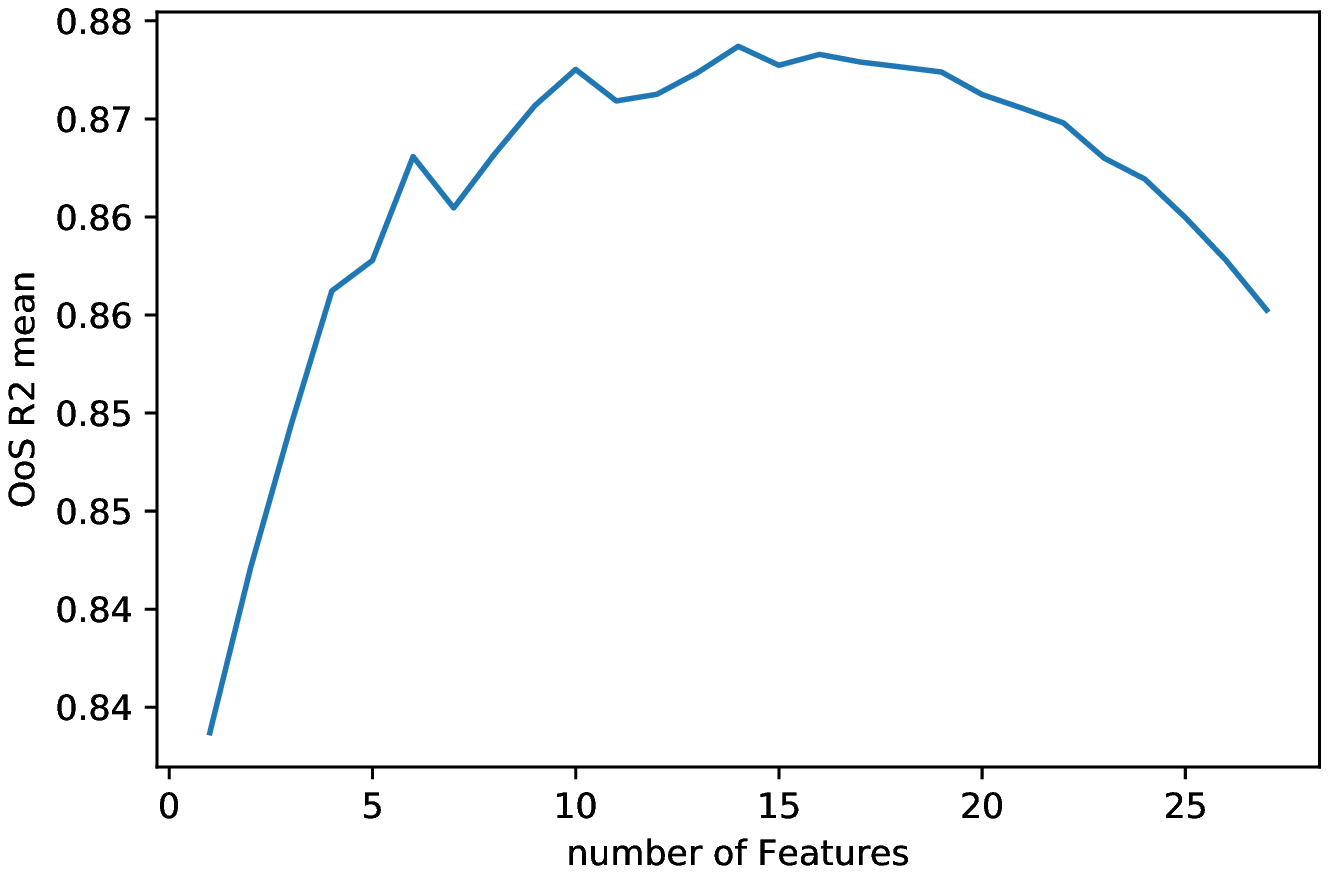}
			\caption*{OoS: mean}
		\end{subfigure}
		\hspace*{-0.5cm}
		\begin{subfigure}[b]{0.430\textwidth}
			\includegraphics[width=\textwidth]{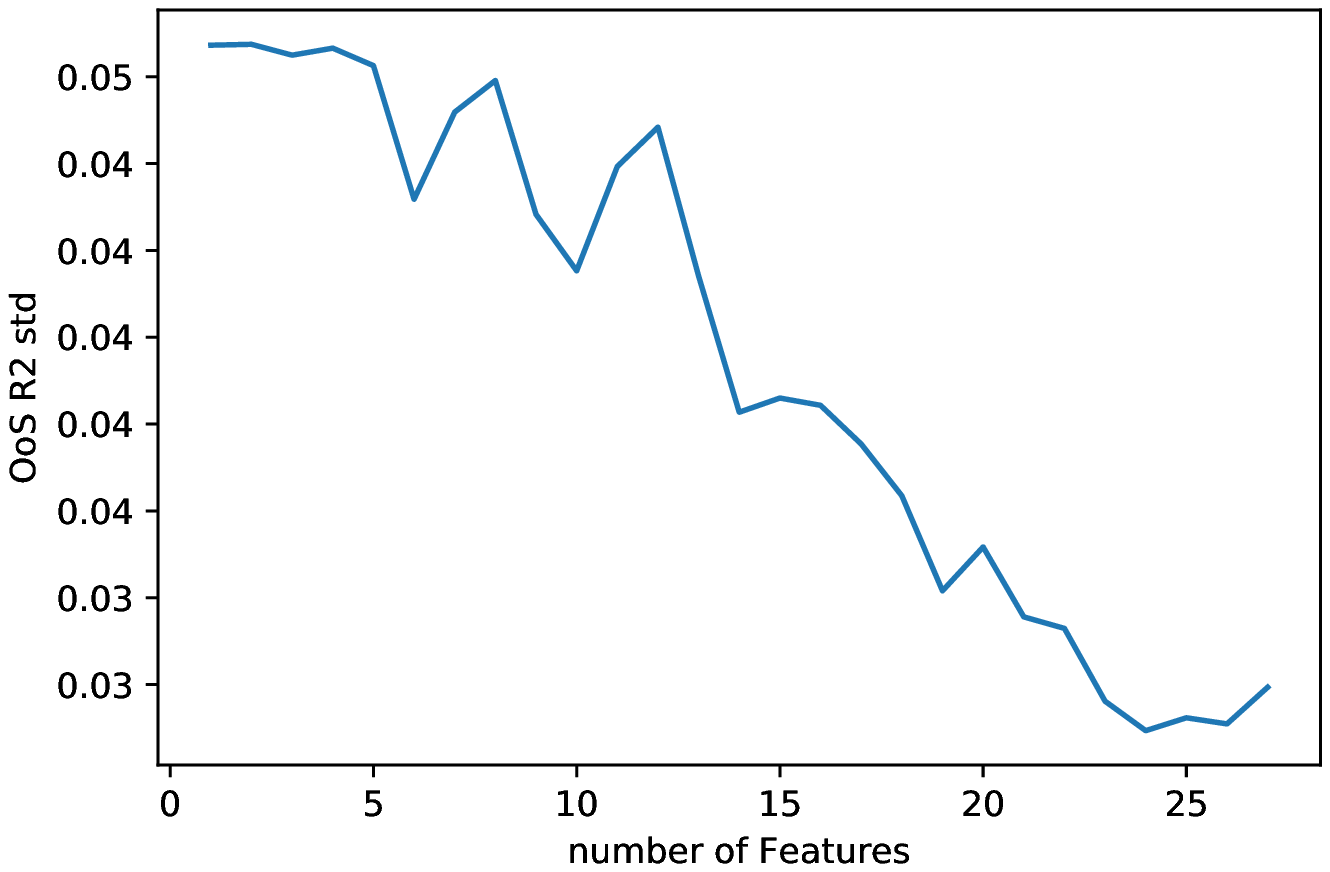}
			\caption*{OoS: std}
		\end{subfigure}
		\hspace*{-0.5cm}
		\begin{subfigure}[b]{0.430\textwidth}
			\includegraphics[width=\textwidth]{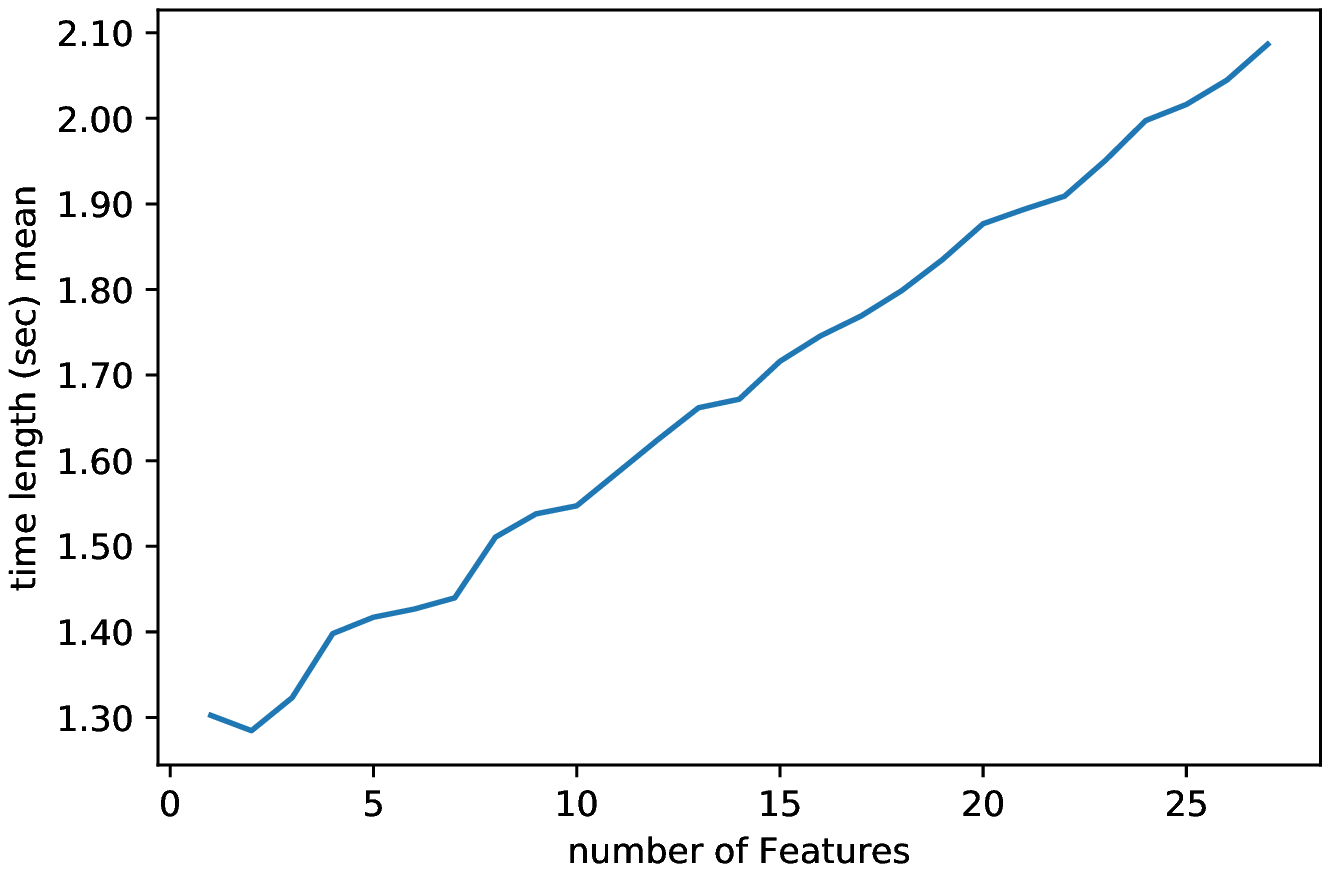}
			\caption*{Time: mean}
		\end{subfigure}
		\vspace*{-0.5cm}
		\caption{Number of features stands for the number of features randomly composing the subset of features tested as splits at each nodes}
		\label{fig:5}
	\end{figure}

	Thereafter the chosen number of random trees is set at 50. Secondly, we set the $m$ number of features randomly selected at each node. According to \cite{Breiman2001}, having categorical and dummy variables, $m$ should be set at two or three times $int(log2(p) + 1)$. This parameter is thus set at 15, a number $m$ additionally confirmed by \Cref{fig:5}, maximizing the averaged accuracy while minimizing its variance.

\begin{figure}[H]
		\centering
		\advance\leftskip-3cm
		\advance\rightskip-3cm
		\hspace*{0.5cm}
		\begin{subfigure}[b]{0.430\textwidth}
			\includegraphics[width=\textwidth]{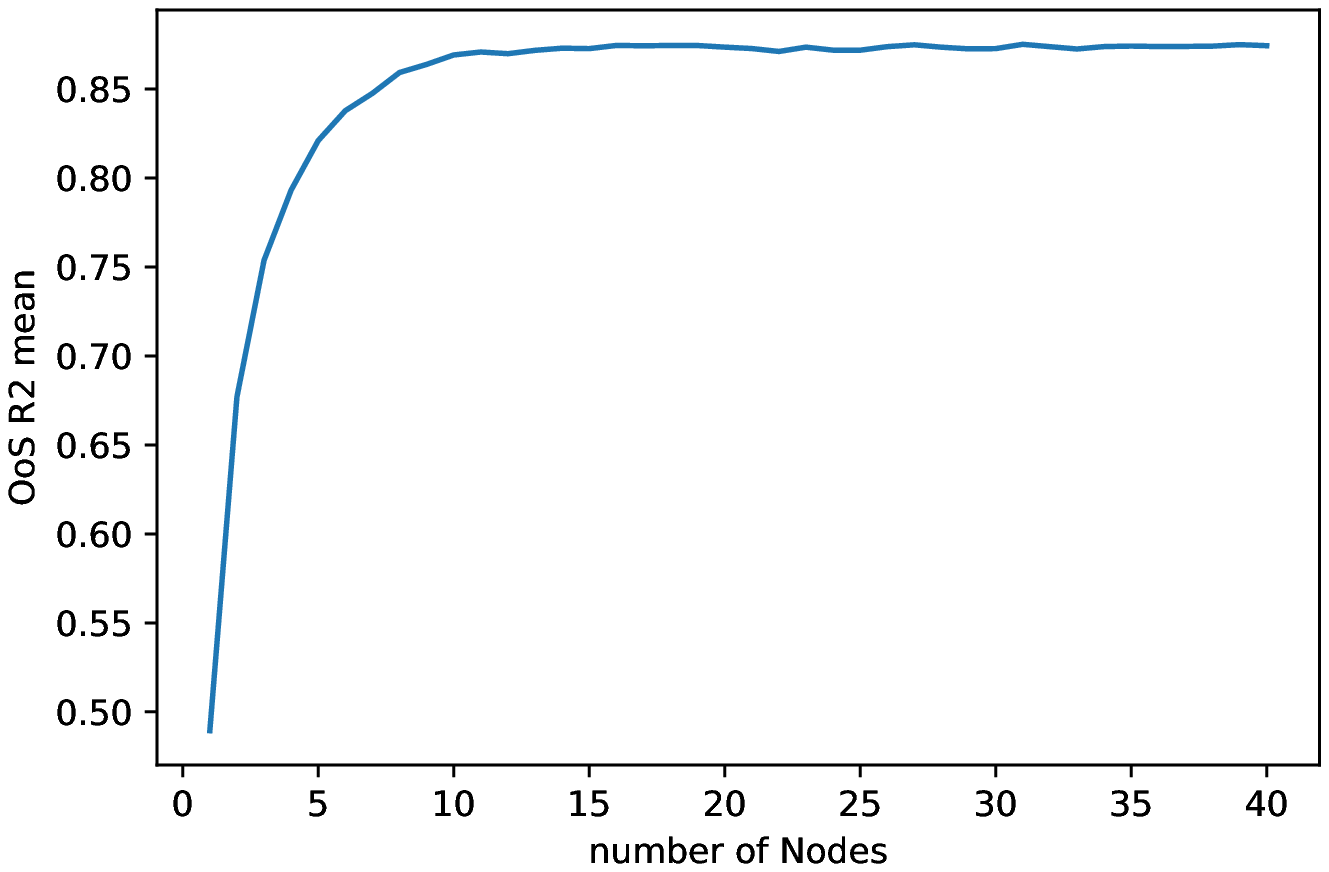}
			\caption*{OoS: mean}
		\end{subfigure}
		\hspace*{-0.5cm}
		\begin{subfigure}[b]{0.430\textwidth}
			\includegraphics[width=\textwidth]{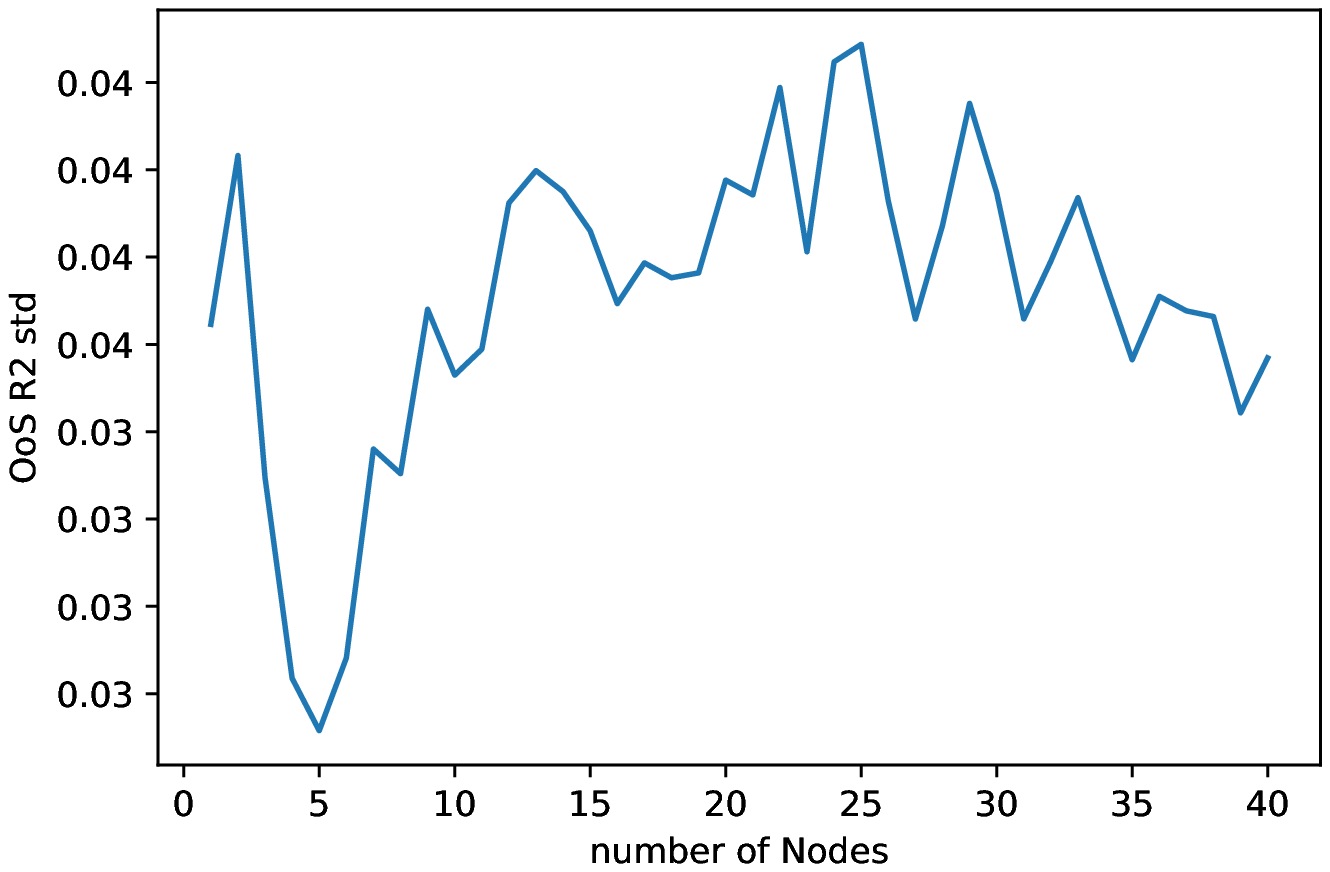}
			\caption*{OoS: std}
		\end{subfigure}
		\hspace*{-0.5cm}
		\begin{subfigure}[b]{0.430\textwidth}
			\includegraphics[width=\textwidth]{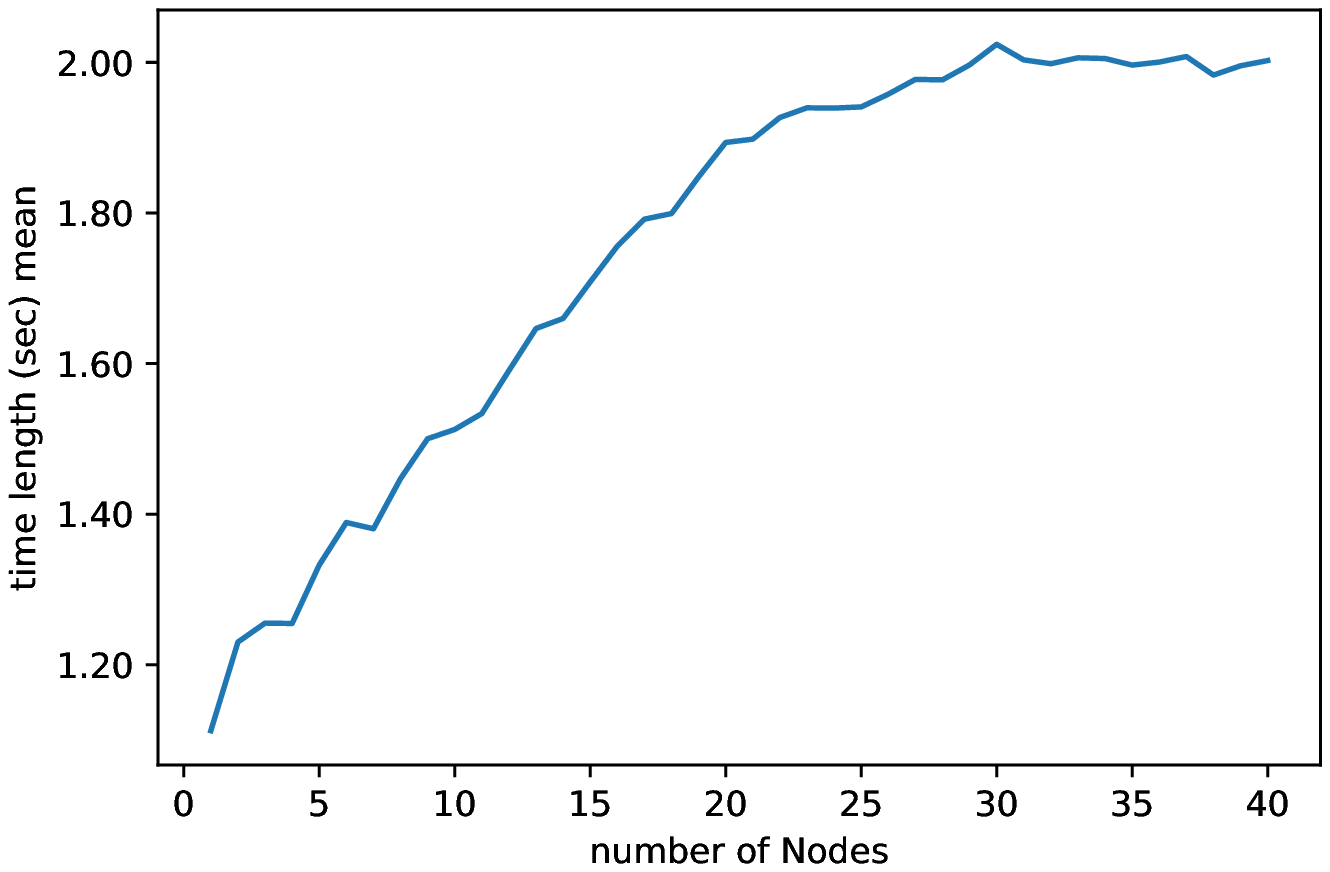}
			\caption*{Time: mean}
		\end{subfigure}
		\vspace*{-0.5cm}
		\caption{Number of nodes defines the trees' depth}
		\label{fig:6}
	\end{figure}

	Based on \Cref{fig:6} and on the concept that pruning can correct potential remaining overfitting, we notice that beyond a sufficient number of nodes the accuracy remains the same. In addition, it leads to quicker computations. Thus we expand the trees' depth until 15 nodes.
	\noindent
	\vspace*{-1.6cm}
	\vspace*{-0.5cm}
	\noindent
	In a nutshell, we launch 50 arbitrary decision trees, each one randomly checking 15 features at each node and having a 15 nodes depth. With this parameters setting, we end up with an 87.3\% out-of-sample $R^2$, see \Cref{tab:2}.\footnote{As a comparison, we obtained close results using a gradient boosting regression \citep{Friedman2001} \textendash \ another algorithm based on decision trees. For the sake of brevity, we do not report these results here. They are, however, available upon request to the authors.}
	\begin{table}[H]
		\centering
		\begin{tabular}[t]{|c|c|c|} 
			\hline
			& IS Mean (Std) & OoS Mean (Std)\TBstrut\\ 
			\hline
			$R^2$ & 99.2\% (0.06\%) & 87.3\% (3.86\%)\Tstrut\\ 
			\hline
		\end{tabular}
		\caption{Goodness-of-Fit of averaged random forest regressions}
		\label{tab:2}
	\end{table}
	These out-of-sample results are illustrated for 10 companies chosen randomly, see \Cref{fig:7}. On these graphs, the black points stand for the 5y CDS benchmarks. The yellow line stands for the sole E2C on the left-hand side and the predicted results using random forest on all features on the right-hand side.
	\begin{figure}[H]
		\caption{Goodness-of-Fit comparison between the sole E2C approximations and the multivariate set-up handled with random forests.}
		\label{fig:7}
		\centering
		\advance\leftskip-3cm
		\advance\rightskip-3cm
		\begin{subfigure}[b]{0.6\textwidth}
			\includegraphics[width=\textwidth]{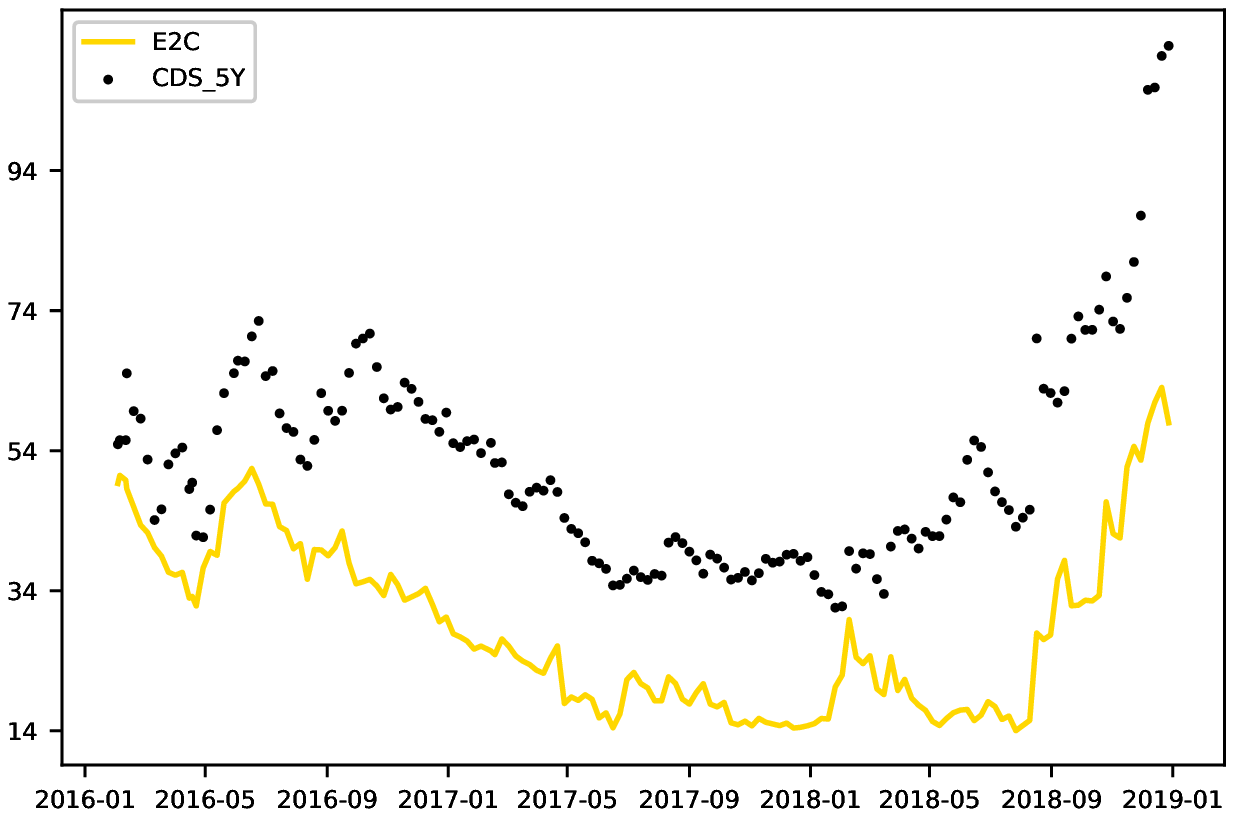}
			\vspace*{-1.0cm}
			\caption*{Bayer}
		\end{subfigure}
		\hspace*{-0.0cm}
		\begin{subfigure}[b]{0.6\textwidth}
			\includegraphics[width=\textwidth]{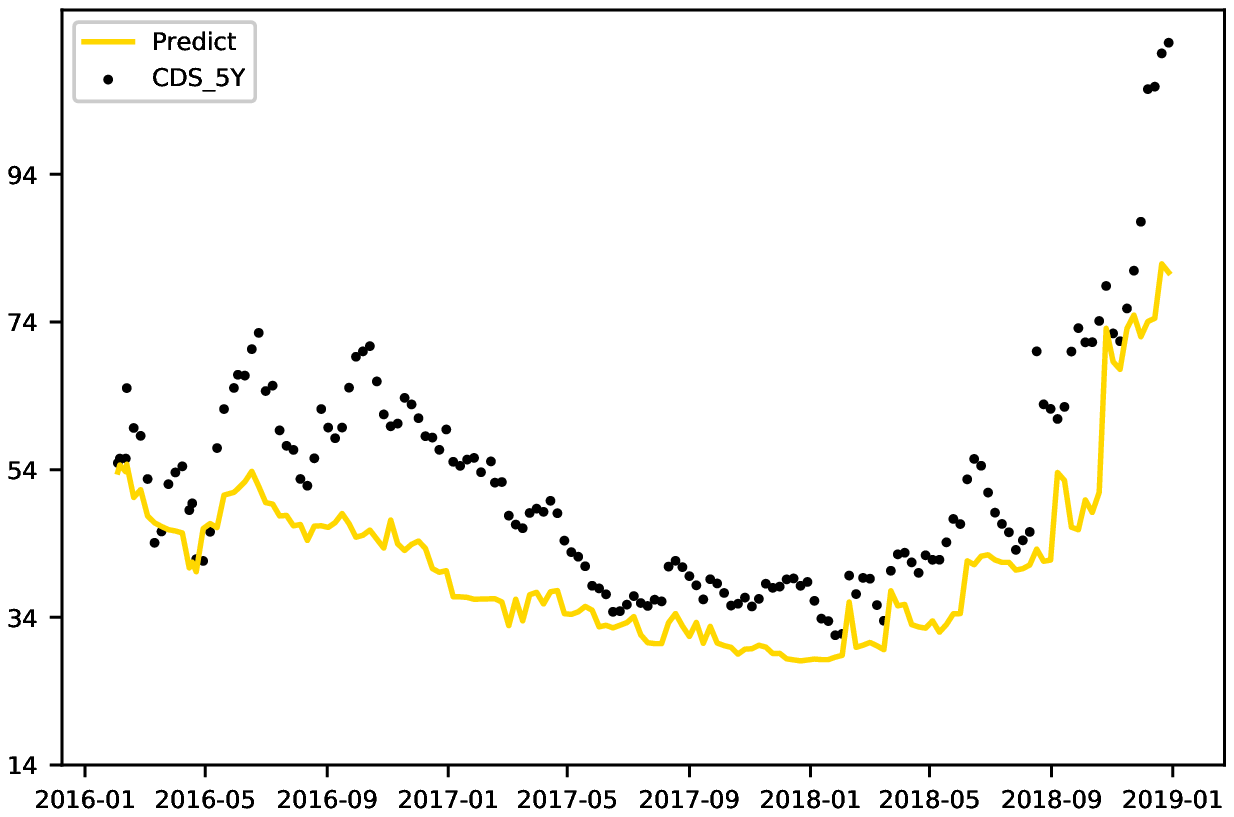}
			\vspace*{-1.0cm}
			\caption*{RF: Bayer}
		\end{subfigure}
	\end{figure}
	\vspace*{-0.6cm}
	\begin{figure}[H]\ContinuedFloat
		\centering
		\advance\leftskip-3cm
		\advance\rightskip-3cm	
		\begin{subfigure}[b]{0.6\textwidth}
			\includegraphics[width=\textwidth]{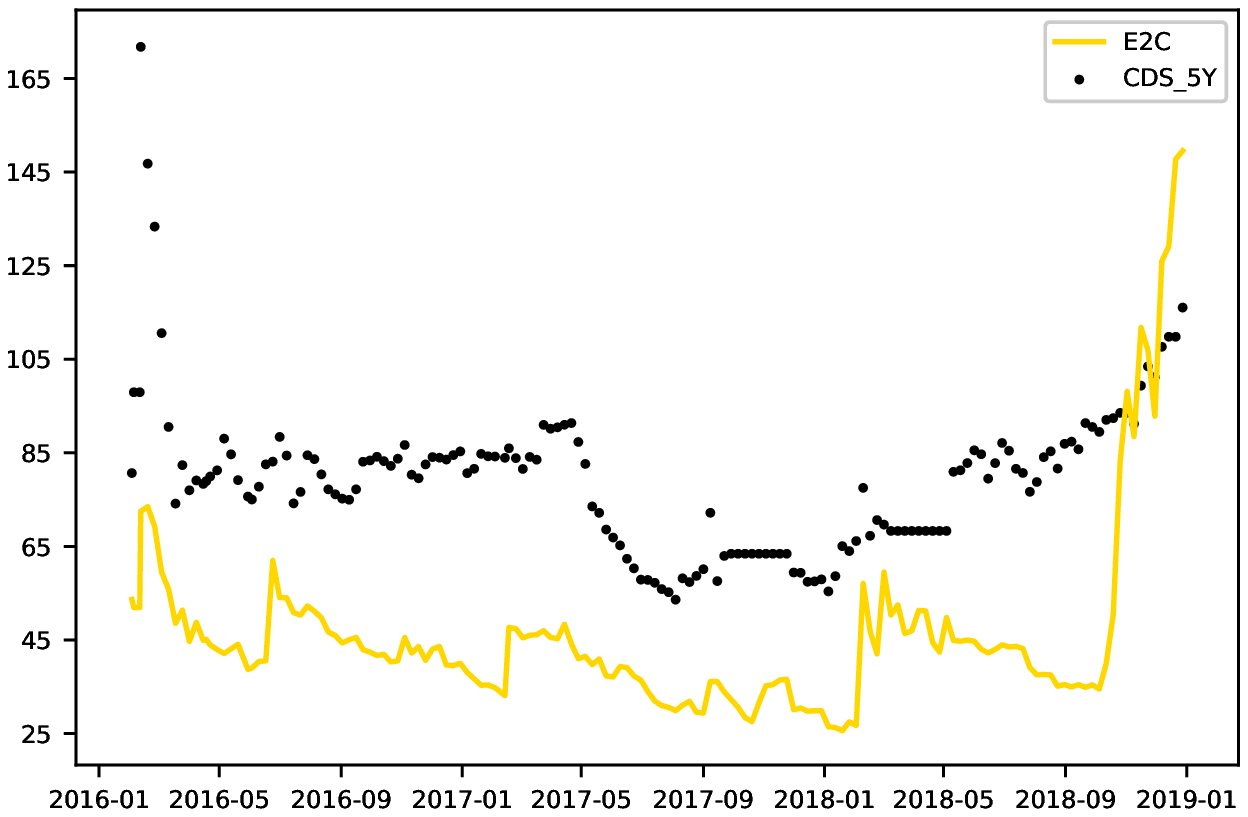}
			\vspace*{-1.0cm}
			\caption*{AIG}
		\end{subfigure}
		\hspace*{-0.0cm}
		\begin{subfigure}[b]{0.6\textwidth}
			\includegraphics[width=\textwidth]{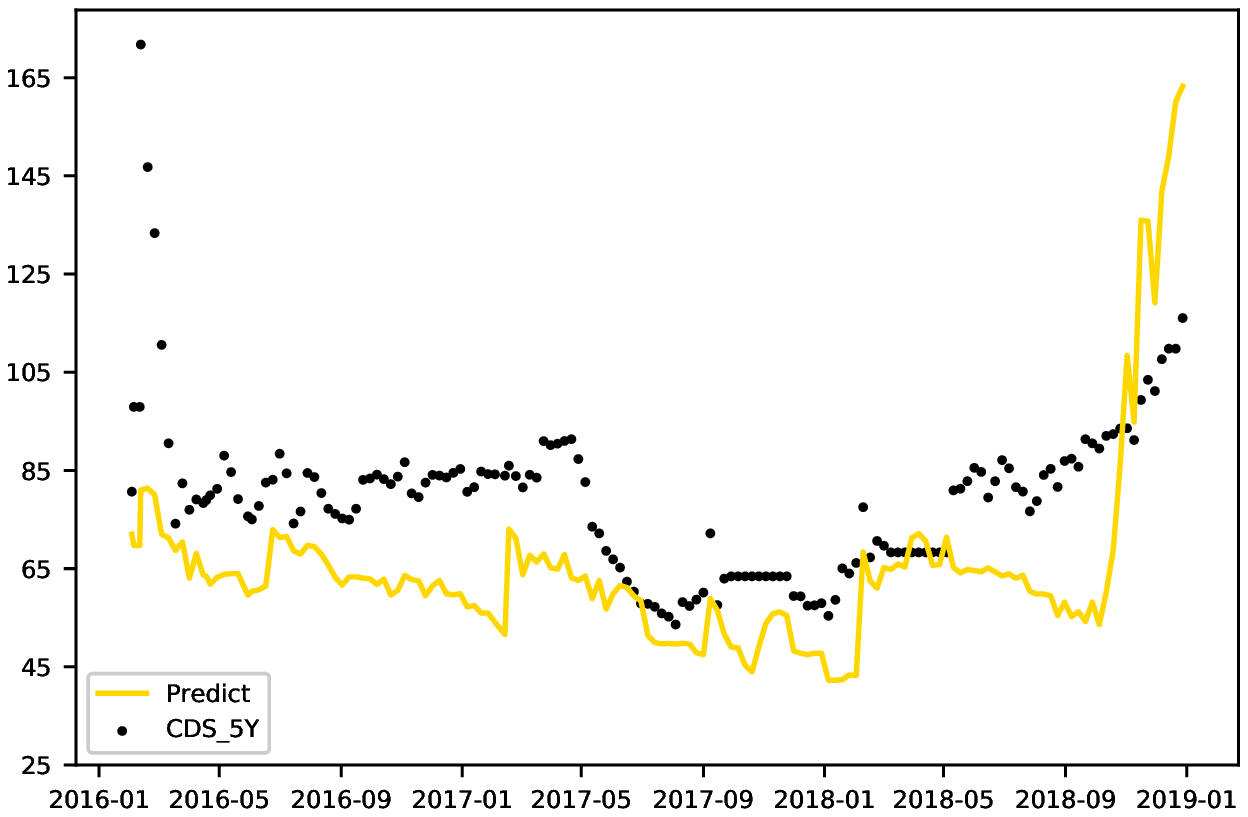}
			\vspace*{-1.0cm}
			\caption*{RF: AIG}
		\end{subfigure}
	\end{figure}
	\vspace*{-0.6cm}
	\begin{figure}[H]\ContinuedFloat
		\centering
		\advance\leftskip-3cm
		\advance\rightskip-3cm	
		\begin{subfigure}[b]{0.6\textwidth}
			\includegraphics[width=\textwidth]{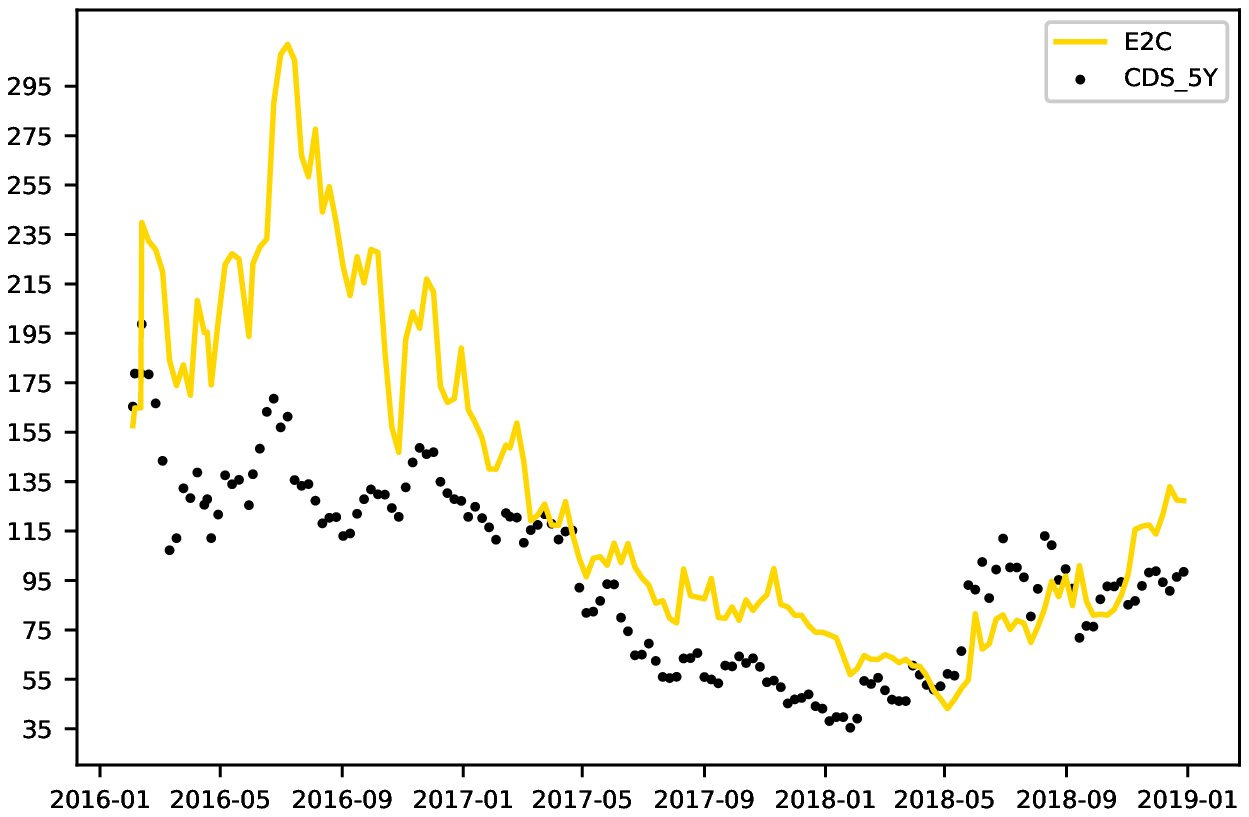}
			\vspace*{-1.0cm}
			\caption*{BBVA}
		\end{subfigure}
		\hspace*{-0.0cm}
		\begin{subfigure}[b]{0.6\textwidth}
			\includegraphics[width=\textwidth]{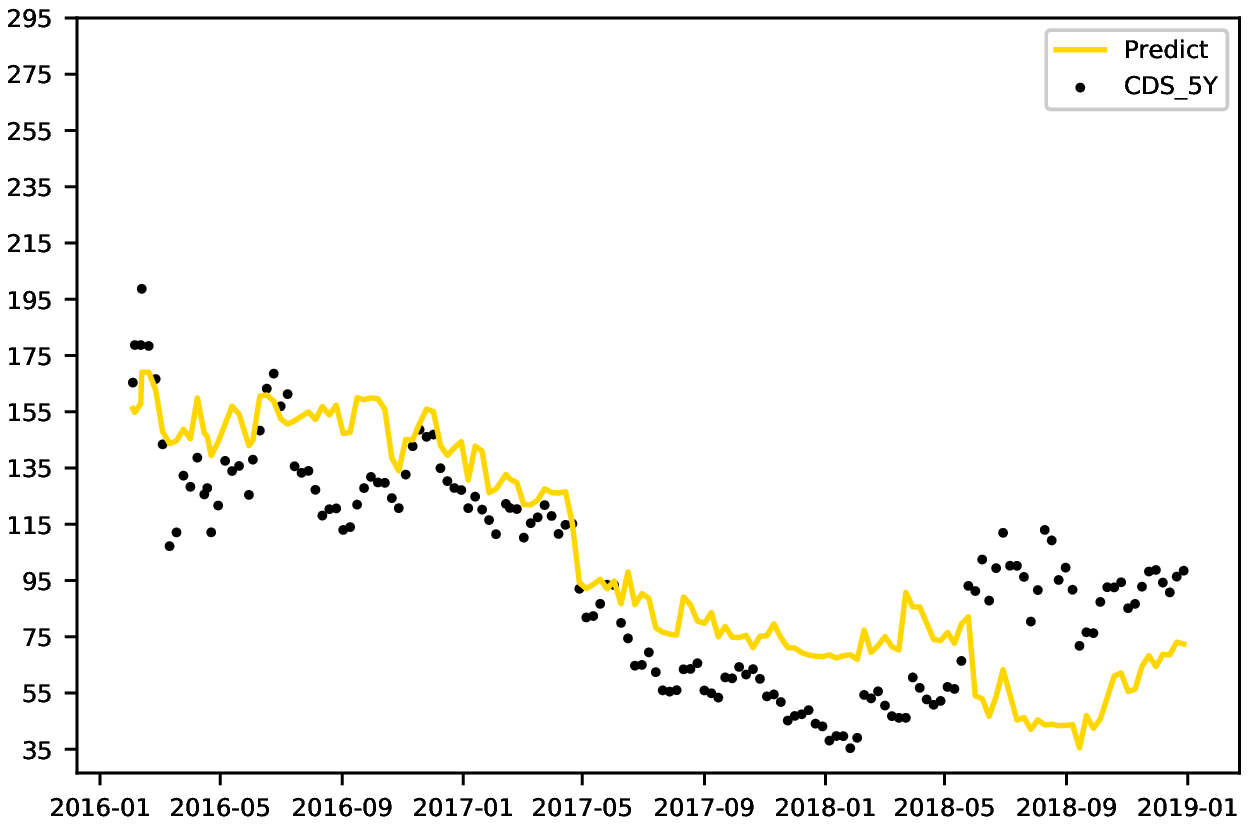}
			\vspace*{-1.0cm}
			\caption*{RF: BBVA}
		\end{subfigure}
	\end{figure}
	\vspace*{-0.6cm}
	\begin{figure}[H]\ContinuedFloat
		\centering
		\advance\leftskip-3cm
		\advance\rightskip-3cm	
		\begin{subfigure}[b]{0.6\textwidth}
			\includegraphics[width=\textwidth]{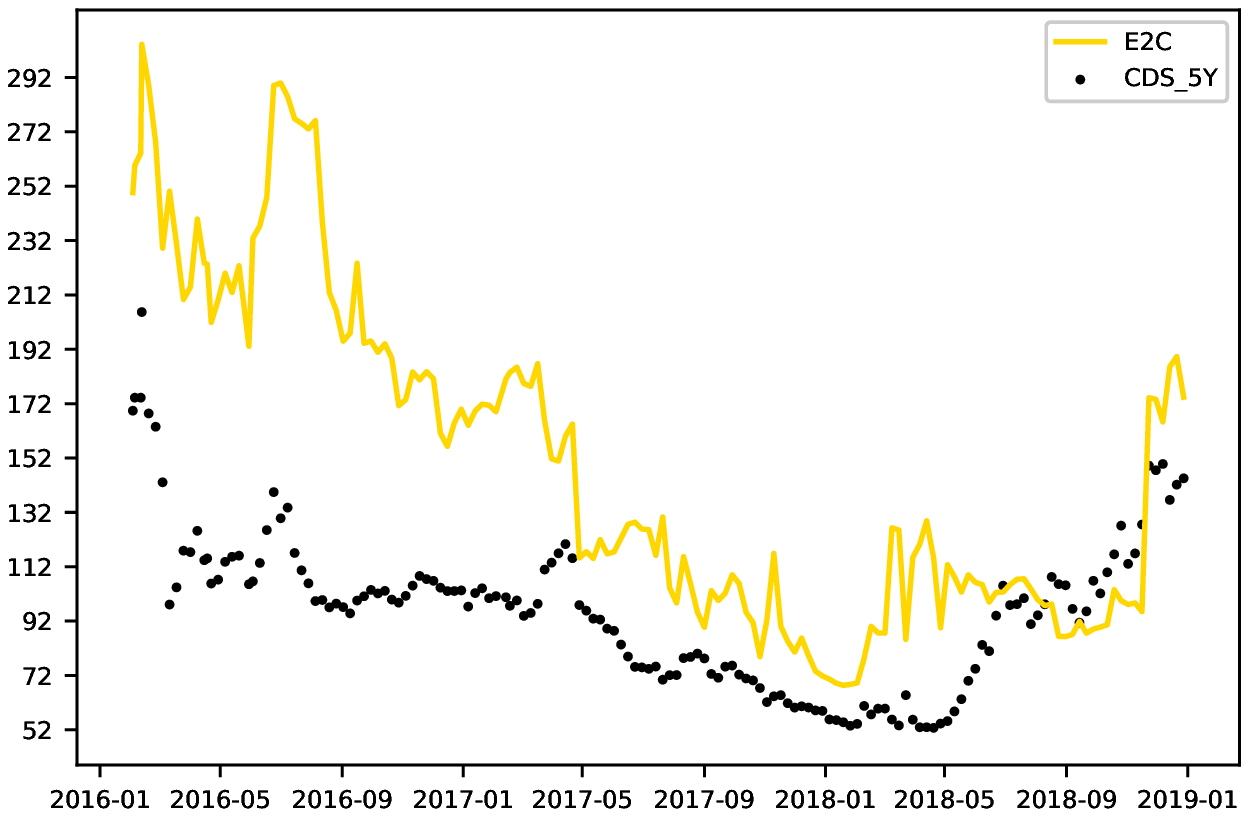}
			\vspace*{-1.0cm}
			\caption*{Renault}
		\end{subfigure}
		\hspace*{-0.0cm}
		\begin{subfigure}[b]{0.6\textwidth}
			\includegraphics[width=\textwidth]{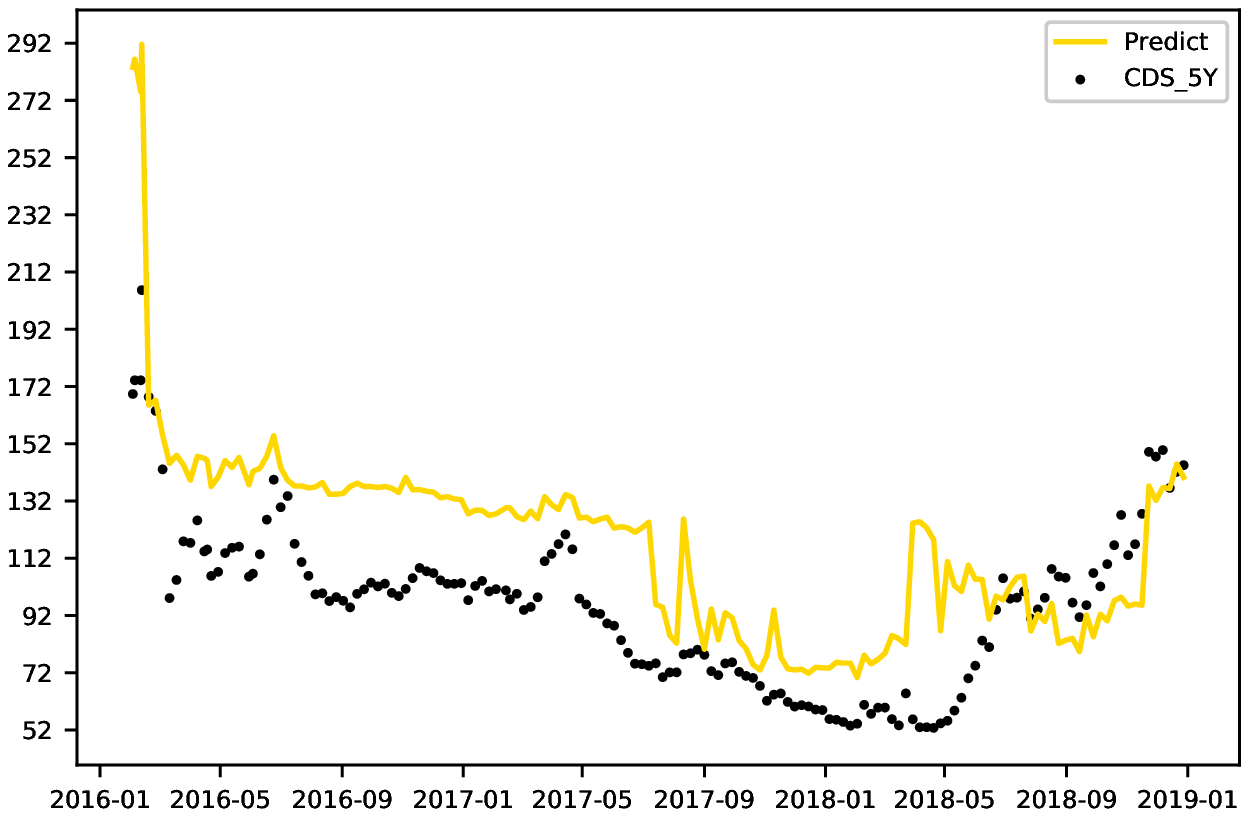}
			\vspace*{-1.0cm}
			\caption*{RF: Renault}
		\end{subfigure}
		\vspace*{+0.2cm}
	\end{figure}
	The graphs on the left-hand side present an already satisfactory approximation of the 5-year CDS using only the E2C equation. However, it is clear on the right-hand-side graphs that the use of a random forest algorithm on a multivariate sample (including the E2C) provides a significantly improved prediction. Training a random forest with these easy-to-access variables is thus a worthwhile process. As robustness check, we ran the random forest algorithm on the same multivariate set-up, removing one sector at a time. Similar levels of accuracy were reached (cf. \nameref{sec:C}, \Cref{tab:10}).
	\subsubsection{Verification of Feature Importance}
\begin{figure}[H]
		\centering
		\advance\leftskip-2cm
		\advance\rightskip-2cm
		\begin{subfigure}[b]{0.6\textwidth}
			\includegraphics[width=9cm, height=6cm]{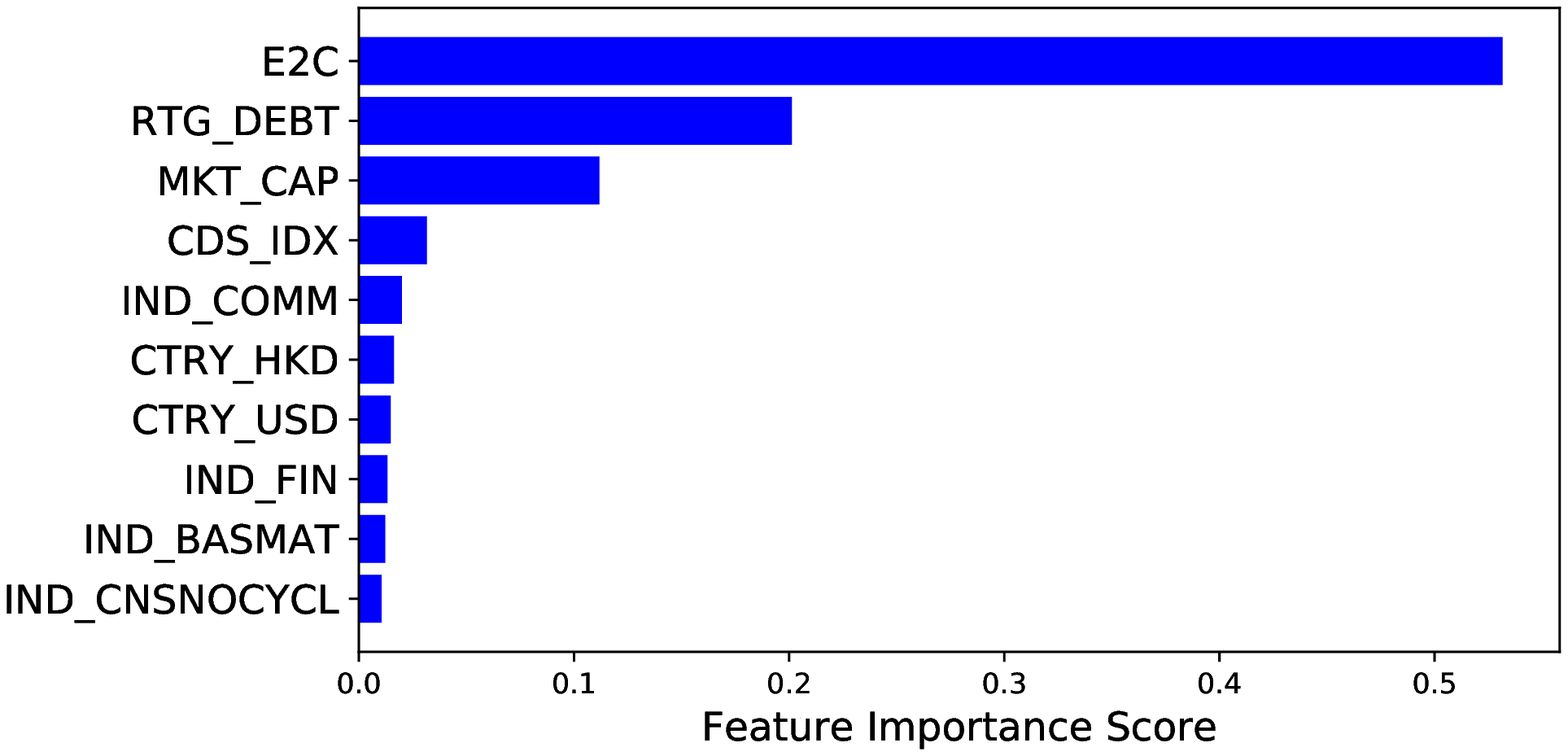}
			\caption*{Feature Importance}
		\end{subfigure}
		\hspace*{-0.8cm}
		\begin{subfigure}[b]{0.6\textwidth}
			\includegraphics[width=9cm, height=6cm]{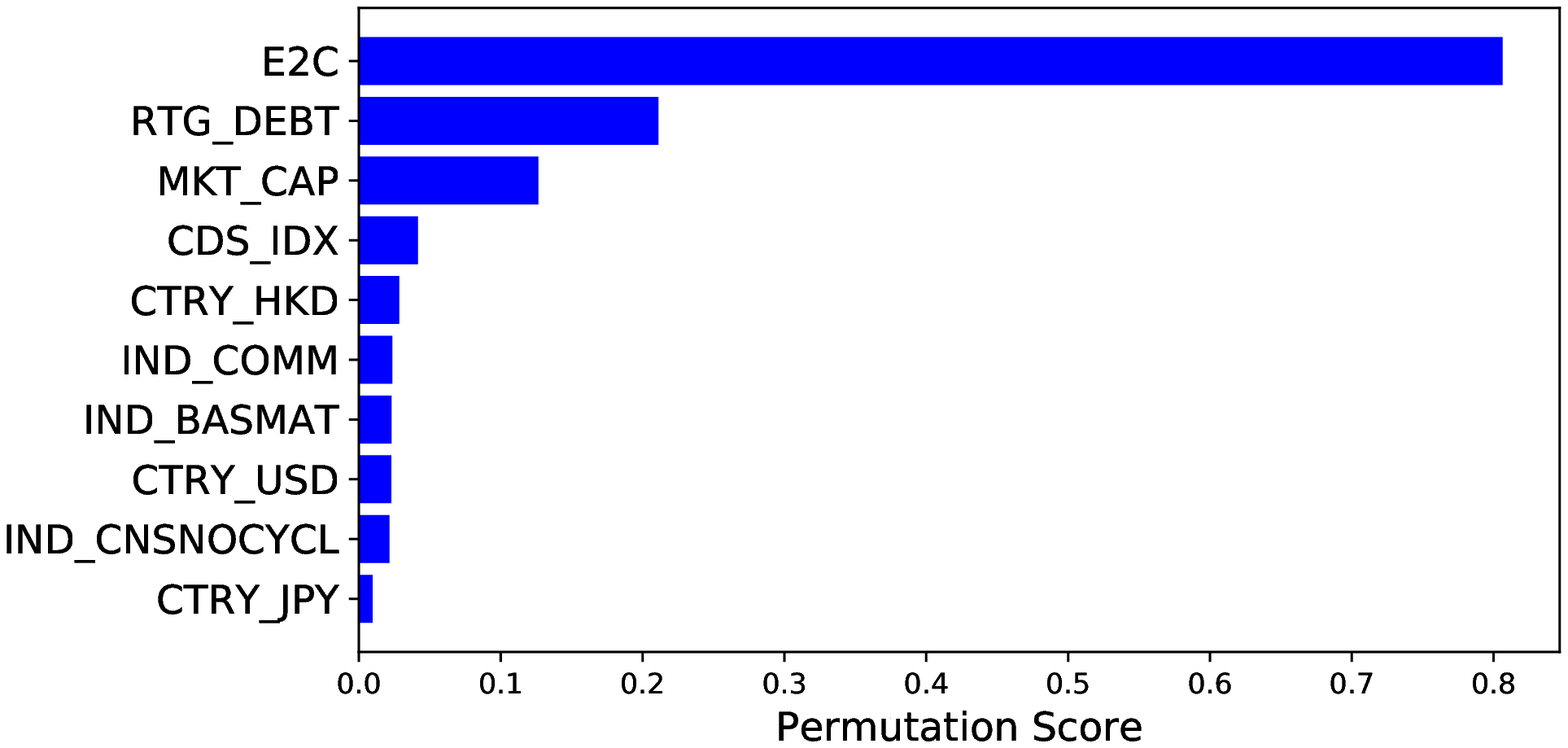}
			\caption*{Feature Importance Permutation}
		\end{subfigure}
		\vspace*{+0.2cm}
		\caption{Features importance assessment}
		\label{fig:8}
	\end{figure}

	The use of a random forest algorithm has been justified above with reference to the results obtained. However, it has an additional property worth noting \textendash \ its transparency which gives access to a quantified measure of each variable's importance. This allows us to assess the individual contribution of the variables (including the E2C) in the improvement obtained with the random forest. The contribution of each feature in predicting the response can be assessed using two methods. Firstly, a method called feature importance, which empirically highlights whether the trees are frequently consulted by a specific feature weighted by the enhancement it brings.  At each node, the improvement brought by a specific selected variable is computed comparing the residual sum of squares before and after splitting. In other words, this method counts, for each tree, the number of times a data sample passes through a node whose decision is based on a specific feature, times the improvement it carries.  Then, the total contribution of the variable is averaged over all trees. Secondly, we measure the variable importance by how much it decreases the model accuracy (averaged over all trees) when this feature is randomly permuted. More specifically, this method computes, for each feature, the mean square error\footnote{As usual, on the out-of-bag samples.} without permutation and compares it to the permuted one. Then, the subtraction is averaged over all random trees and normalized. For instance, for a feature A, we get:
	\begin{align*}
	VI(A) = \dfrac{1}{B}\sum_{b=1}^B\dfrac{R^2_{b} - R^2_{b,Permuted}}{R^2_{b}}
	\end{align*}
	Shuffling a variable is similar in linear regressions to setting a coefficient as null. Consequently, we expect important variables' $R^2$ to be significantly lower in the scrambled cases than in the genuine ones, leading to a high $VI$. In addition, this method does not affect the values' distribution, because variables are just permuted. Moreover, this second metric is useful to distinguish the importance among correlated variables. We display these features' importance measures in the bar charts below, the first method being on the left-hand side. Only the ten most important variables (according to our measures) are shown in \Cref{fig:8}.
	These bar charts confirm that the E2C is by far the main variable contributing to our improved approximation of the CDS\footnote{CreditGrades as substitute for E2C, while providing the same R-squared and being the main feature, is a less dominant variable in this set-up.}. Beyond the E2C, the inclusion of debt ratings and market capitalizations also clearly improves the model. In a sample with a high heterogeneity between companies, it is encouraging to observe that the random forest highlights company-specific variables such as the two listed above.
	\section{Conclusion}
	In this paper, we develop an instrument which is both intuitive and accurate to approximate CDS spreads. It relies on an elementary formula, called E2C, and on random forest regressions learning from a small data set.

	Our first step in reaching this goal is the analysis of the cornerstone concepts driving structural models for Credit Default Swaps. Most of this analysis is closely related to the CreditGrades method, but the specificity of our equation rests on the use of fewer parameters and a more elementary formula. Upon verification, the E2C provides at least as accurate an approximation of the CDS as the CreditGrades model does, and in fact outperforms in certain predictions for firms grouped by debt ratings or industrial sectors.

	Furthermore, we provide a protocol using random forest regressions to correct some remaining issues with accuracy found in all models such as the E2C or CreditGrades. This supervised learning algorithm allows the inclusion of a few additional widely-available financial data as independent variables, to complement the E2C. We detail the steps from the sample pre-processing to the final hyper-parameter setting. The use of the random forest algorithm culminates in an 87.3\% out-of-sample accuracy in our CDS approximations. This method remains simple among non-linear supervised algorithms, and has the added benefit of transparency, which allows an evaluation of the relative importance of our variables. Beyond confirming the impact of debt rating and company size on CDS, this transparency highlights the predominance of the E2C as a reliable predictor among all other variables.
	
	Financial practitioners and researchers have long had to balance precision against transparency in their selection of CDS spreads approximations. The E2C formula we offer in this paper goes a long way toward resolving that tension; as it generally outstrips the accuracy of the CreditGrades model while being more intuitive and straightforward. Furthermore, by conducting random forest regressions on a multivariate universe including our E2C, we obtain an outstandingly reliable and accessible CDS estimation.

	Although our sample covers a large number of companies, industries and countries, we suggest that additional insights may be uncovered with historical data covering a longer time-span. CDS maturity is also not considered in this paper and is left for future research, possibly starting from the exploration of other mathematical ``concentration inequalities'' for the formula's upper bound.
	\bibliography{./Refs/E2C_Biblio}
	\newpage
	\section*{Appendices}
	\paragraph{Appendix A} \label{sec:A}
	\paragraph{Modified ``one-sided'' approach to the Gauss Inequality}\mbox{}\\\\
	A result similar to Gauss's inequality gives, for any random variable $X$ with a decreasing density for negative values:
	\begin{align*}
	{\rm I\!P}(X \leq q) \leq \dfrac{4}{9q^2} \cdot E \left[X^2 \mid X \leq 0 \right] \cdot {\rm I\!P}(X \leq 0)
	\end{align*}
	Then, from the reflection principle for an unbiased stock evolution from the current price:
	\begin{align*}
	{\rm I\!P}(\min S_t \leq 0 \mid t \leq T) = 2 \cdot {\rm I\!P}(S_T \leq 0) = 2 \cdot {\rm I\!P}(S_T - S_0 \leq -S_0)
	\end{align*}
	and
	\begin{align*}
	{\rm I\!P}(S_T \leq S_0) = \frac{1}{2}
	\end{align*}
	Hence,
	\begin{align*}
	{\rm I\!P}(Default \mid t \leq T) \leq \dfrac{4}{9 S_0^2} \cdot E \left[(S_T - S_0)^2 \mid S_T \leq S_0 \right]
	\end{align*}
	Expressing the stock evolution in independent steps: $S_T - S_0 = \sum_{t=1}^{T}dS_t$
	\begin{align*}
	E \left[(\sum_{t=1}^{T}dS_t)^2 \mid S_T \leq S_0 \right] &= E \left[\sum_{t=1}^{T}E\left[dS_t^2 \mid S_t \right] \mid S_T \leq S_0 \right]\\
	&= E \left[\sum_{t=1}^{T} \sigma_{S_t}^2S_t^2 \mid S_T \leq S_0 \right]\\
	&= T \cdot E \left[\sigma_{S}^2S^2 \mid S_T \leq S_0 \right] = T \cdot \delta_{S_T \leq S_0}^2
	\end{align*}
	And this last conditional expectation is proxied by the geometric average of the downside boundary values of a Black-Cox/CreditGrades type model of the firm.
	\newpage
	\paragraph{Debt-per-Share Evaluation}\mbox{}\\\\
	According to CreditGrades, $FinD$ the financial debt amount is determined splitting companies in two categories, banking  and non-banking firms.
	\begin{itemize}
		\item Banking Sector
		\mathleft
		\begin{align*}
		FinD = Long\ Term\ Debt
		\end{align*}
		\item Non-Banking Sector
		\mathleft
		\begin{align*}
		FinD =& Long\ Term\ Debt + Short\ Term\ Debt + 0.5 \cdot (Other\\
		&Long\ Term\ Liabilities + Other\ Short\ Term\ Liabilities) + \\
		&0.4 \cdot Future\ Minimum\ Operating\ Lease\ Obligations
		\end{align*}
	\end{itemize}
	To develop a little further the $FinD$ choice for banks, we highlight that banks deposits are not part of financial leverage, thus excluded from the debt-per-share. In fine, many financial statements items should be assessed on a case-by-case basis\footnote{ST Borrowings, and both Other ST and LT Liabilities contain many other items which in fact should be weighted 0\% into the debt-per-share calculations, ranging from accounts payables and accrued expenses to repos, securities lending of regulated broker-dealer or securities subsidiaries, ...}, leading to a laborious analysis. For the sake of simplification, CreditGrades chose to include in its debt-per-share computations the consolidated Long Term Borrowings. 
	$FinD$, $MinInt$ the minority interests and $PrefEq$ the prefered equities are balance sheet data, thus are given in fundamental report currency. To match with $S_0$ the current stock price and $MktCap_0$ the current market capitalization, we convert them to the stock price currency. Finally, the debt-per-share is defined as follow:
	\begin{align*}
	D=\dfrac{FinD - MinInt}{\dfrac{MktCap_0 + PrefEq}{S_0}}
	\end{align*}
	Note that $MinInt$ is capped at 50\% of $FinD$, $PrefEq$ is capped at 50\% of $MktCap$ and $D$ is floored at 10\% of $S$. Moreover, the weightings of $FinD$ and the cap and floor levels are inspired by CreditGrades \citeyearpar{Finger2002}.
	\newpage
	\paragraph{Appendix B} \label{sec:B}
	\setcounter{section}{2}
	\setcounter{figure}{0}
	\renewcommand\thefigure{\Alph{section}.\arabic{figure}}
	\setcounter{table}{0}
	\renewcommand\thetable{\Alph{section}.\arabic{table}}
	\paragraph{CreditGrades formula}
	\begin{align*}
	{\rm I\!P}(Survival)_{t} &= e^{-h_{CG} \cdot t} = \Phi\left(-\dfrac{A_t}{2} + \dfrac{log(d)}{A_t}\right) - d \cdot \Phi\left(-\dfrac{A_t}{2} - \dfrac{log(d)}{A_t}\right)\\
	where,\\
	d &= \dfrac{S_0 + \bar{L}D}{\bar{L}D} \cdot \exp{\left(\lambda^2\right)}\\
	A^2_t &= \left(\sigma_{S_0} \cdot \dfrac{S_0}{S_0 + \bar{L}D}\right)^2 \cdot t + \lambda^2\\
	\end{align*}
	\paragraph{Descriptive Statistics Tables}\mbox{}\\
	\begin{table}[h]
		\centering
		\advance\leftskip-3cm
		\advance\rightskip-3cm
		\begin{tabular}[t]{|c|c|c|c|} 
			\hline
			Des Stat & CDS\_5Y & E2C & CrdGrd \TBstrut\\
			\hline
			No. of obs. & 47476 & 47476 & 47476\\
			No. of firms & 308 & 308 & 308\\
			No. of periods & 155 & 155 & 155\\
			Mean & 137.4 & 124.5 & 138.7\\
			Std-overall & 230.2 & 236.5 & 268.5\\
			Std-between & 202.2 & 202.8 & 249.4\\
			Std-within & 120.3 & 129.4 & 107.4\\
			Min & 12.9 & 1.4 & 0\\
			q(25\%) & 46 & 26 & 1.1\\
			q(50\%) & 71.6 & 50.6 & 18.6\\
			q(75\%) & 132.3 & 115.5 & 135.6\\
			Max & 4491.1 & 4444.9 & 2246.9\\
			Skew & 7.3 & 5.5 & 2.9\\
			Kurt & 86.3 & 44.1 & 9.2\Bstrut\\
			\hline
		\end{tabular}
		\caption{CDS, E2C and CreditGrades Descriptive Statistics}
		\label{tab:3}
	\end{table}
	\newpage
	\begin{table}[h]
		\centering
		\begin{tabular}{cc|c|c|c|c|c|c|}
			\cline{3-8}
			&        & \multicolumn{2}{c|}{5y CDS} & \multicolumn{2}{c|}{E2C} & \multicolumn{2}{c|}{CrdGrd} \\ \cline{2-8} 
			\multicolumn{1}{c|}{}               & Obs. & mean         & std          & mean        & std        & mean         & std          \\ \hline
			\multicolumn{1}{|c|}{Australia}     & 2170   & 88.9         & 45.3         & 34.4        & 22.6       & 15.1         & 27.6         \\ \hline
			\multicolumn{1}{|c|}{Britain}       & 3402   & 94.8        & 89.4        & 59.6        & 112.8      & 44.8         & 138.8        \\ \hline
			\multicolumn{1}{|c|}{Canada}        & 775    & 318.1          & 275.2        & 376.5       & 518.2      & 373.9        & 458.1        \\ \hline
			\multicolumn{1}{|c|}{Denmark}       & 280    & 88.2         & 49.6           & 112.6       & 70.2       & 277.4        & 220          \\ \hline
			\multicolumn{1}{|c|}{Eurozone}      & 12061   & 103.2        & 208.1        & 102.2       & 139.1      & 119.7        & 194.5        \\ \hline
			\multicolumn{1}{|c|}{Hong Kong}     & 930    & 191.9        & 181.9        & 63.2        & 90.8       & 63.1         & 134.8        \\ \hline
			\multicolumn{1}{|c|}{India}         & 465    & 180        & 47.4         & 326.3       & 193.9      & 433.3        & 251.5        \\ \hline
			\multicolumn{1}{|c|}{Japan}         & 3410   & 50.9         & 36.1         & 110.8       & 101      & 147.4        & 200.6        \\ \hline
			\multicolumn{1}{|c|}{Malaysia}      & 155    & 107.3          & 33.7         & 46.3         & 75        & 46.2         & 102         \\ \hline
			\multicolumn{1}{|c|}{New Zealand}   & 310    & 80.2         & 33.7         & 40        & 31.9       & 26.7         & 36.2           \\ \hline
			\multicolumn{1}{|c|}{Norway}        & 310    & 41.2         & 19.1         & 50        & 36.2       & 24.5         & 47.1         \\ \hline
			\multicolumn{1}{|c|}{Singapore}     & 155    & 57.2         & 9.4          & 8.2         & 4.1        & 0         & 0         \\ \hline
			\multicolumn{1}{|c|}{South Korea}   & 1240    & 73.7         & 22.7         & 90.9        & 57.7       & 180.1        & 235.8        \\ \hline
			\multicolumn{1}{|c|}{Sweden}        & 465    & 86.2         & 39           & 50.1        & 20.3       & 18.1         & 22.5         \\ \hline
			\multicolumn{1}{|c|}{United States} & 21348  & 180.9        & 284.9        & 155.4       & 303      & 167.3        & 326        \\ \hline
		\end{tabular}
		\caption{CDS, E2C and CreditGrades Descriptive Statistics by Country}
		\label{tab:4}
	\end{table}
	\begin{table}[h]
		\centering
		\begin{tabular}[t]{|c|c|c|c|} 
			\hline
			Correl & CDS\_5Y & E2C & CrdGrd\TBstrut\\ 
			\hline
			CDS\_5Y & 100\% & 66.7\% & 62.3\%\Tstrut\\
			E2C & 66.7\% & 100\% & 92.7\%\\
			CrdGrd & 62.3\% & 92.7\% & 100\%\Bstrut\\
			\hline
		\end{tabular}
		\caption{Averaged correlations of CDS, E2C and CreditGrades by Firm}
		\label{tab:5}
	\end{table}
	\begin{table}[H]
		\centering
		\begin{tabular}[t]{|c|c|c|c|} 
			\hline
			Correl & CDS\_5Y & E2C & CrdGrd\TBstrut\\ 
			\hline
			CDS\_5Y & 100\% & 85\% & 76.3\%\Tstrut\\
			E2C & 85\% & 100\% & 89.6\%\\
			CrdGrd & 76.3\% & 89.6\% & 100\%\Bstrut\\
			\hline
		\end{tabular}
		\caption{Averaged correlations of CDS, E2C and CreditGrades by Date}
		\label{tab:6}
	\end{table}
	\newpage
	\begin{table}[h]
		\centering
		\begin{tabular}{cc|c|c|c|c|c|c|}
			\cline{3-8}
			&        & \multicolumn{2}{c|}{RMSE} & \multicolumn{2}{c|}{MAPE} & \multicolumn{2}{c|}{MASE} \\ \cline{2-8} 
			\multicolumn{1}{c|}{}               & Obs. & E2C         & CrdGrd          & E2C        & CrdGrd        & E2C         & CrdGrd          \\ \hline
			\multicolumn{1}{|c|}{A}     & 11396   & 55         & 132         & 0.72        & 1.44       & 16.1         & 31.2         \\ \hline
			\multicolumn{1}{|c|}{BBB}       & 17581   & 61        & 124        & 0.56        & 1.06      & 12.9         & 22.5        \\ \hline
			\multicolumn{1}{|c|}{BB}        & 5360    & 98          & 214        & 0.47       & 0.76      & 9.6        & 15.2        \\ \hline
			\multicolumn{1}{|c|}{B}       & 3637    & 260         & 265           & 0.39       & 0.49       & 6.2        & 6.6          \\ \hline
		\end{tabular}
		\caption{CDS, E2C and CreditGrades Accuracy Metrics by Debt Rating}
		\label{tab:7}
	\end{table}
	\begin{table}[h]
		\centering
		\begin{tabular}{cc|c|c|c|c|c|c|}
			\cline{3-8}
			&        & \multicolumn{2}{c|}{RMSE} & \multicolumn{2}{c|}{MAPE} & \multicolumn{2}{c|}{MASE} \\ \cline{2-8} 
		\multicolumn{1}{c|}{}               & Obs. & E2C         & CrdGrd          & E2C        & CrdGrd        & E2C         & CrdGrd          \\ \hline
		\multicolumn{1}{|c|}{Basic Materials}     & 3285    & 62        & 89        & 0.49        & 0.77       & 6.2         & 9        \\ \hline
		\multicolumn{1}{|c|}{Communication}       & 4907    & 54         & 77           & 0.48       & 0.78       & 9.4        & 12.9          \\ \hline
		\multicolumn{1}{|c|}{Consumer Cyclical}       & 7232   & 70        & 105        & 0.59        & 0.99      & 11.2         & 16.5        \\ \hline
		\multicolumn{1}{|c|}{Consumer Non-Cyclical}        & 5616    & 81          & 108        & 0.53       & 0.88      & 9.9        & 11.7        \\ \hline
		\multicolumn{1}{|c|}{Energy}         & 2227   & 101         & 133         & 0.37       & 0.68      & 4.8        & 5.9        \\ \hline
		\multicolumn{1}{|c|}{Financial}     & 6346   & 127         & 267         & 0.85        & 2.06       & 15.8         & 34.9         \\ \hline
		\multicolumn{1}{|c|}{Industrial}      & 4425   & 51        & 75        & 0.61       & 1.01      & 7.7        & 10.4        \\ \hline
		\multicolumn{1}{|c|}{Utilities}         & 2540    & 77        & 154         & 0.647       & 1.279      & 13.8        & 22.1        \\ \hline
	\end{tabular}
	\caption{CDS, E2C and CreditGrades Accuracy Metrics by Industrial Sector}
	\label{tab:8}
	\end{table}
	\noindent
	RMSE: Root Mean-Square Error\\
	MAPE: Mean Absolute Percentage Error\\
	MASE: Mean Absolute Scaled Error\\
	Consistent with section \ref{sec:TruncMean}, we removed the 10\% top and bottom points.
	\newpage
	\paragraph{Volatility Regime Tests}\mbox{}\\\\
	Model 1: $CDS\_5y_{i,t} = \beta_0 + \beta_1 E2C_{i,t} + \beta_2 VOL_{t} + \beta_3 E2C_{i,t} * VOL_{t} + \epsilon_{i,t}$\\\\
	Model 2: $CDS\_5y_{i,t} = \beta_0 + \beta_1 CrdGrd_{i,t} + \beta_2 VOL_{t} + \beta_3 CrdGrd_{i,t} * VOL_{t} + \epsilon_{i,t}$\\\\
	Where $VOL_{t}$ is a dummy variable, defined as 1 if VIX 3-month volatility index $\geq$ 20.\\\\
	\begin{table}[!htb]
		\begin{minipage}{.5\linewidth}
			\vspace*{-0.0cm}
			\caption*{Model 1}
			\begin{center}
				{
					\centering
					\def\sym#1{\ifmmode^{#1}\else\(^{#1}\)\fi}
					\begin{tabular}{l*{2}{c}}
						\hline\hline
						&\multicolumn{1}{c}{CDS\_5y}&\multicolumn{1}{c}{CDS\_5y}\\
						\hline
						E2C         &       0.8097\sym{***}&       0.8096\sym{***}         \\
						&    (326.36)         &      (289.22)         \\
						VOL      &                     &       1.8482\\
						&                     &    (0.93)         \\
						E2C*VOL      &                     &       -0.0002\\
						&                     &    (-0.04)         \\
						constant      &       36.5921\sym{***}&       36.3731\sym{***}\\
						&     (55.19)         &     (51.13)         \\
						\hline
						\(N\)       &       47476         &       47476         \\
						$R^{2}$  &       0.6917         &       0.6917         \\
						\hline\hline
						\multicolumn{3}{l}{\footnotesize \textit{t} statistics in parentheses}\\
						[-0.5em]
						\multicolumn{3}{l}{\footnotesize \sym{*} \(p<0.05\), \sym{**} \(p<0.01\), \sym{***} \(p<0.001\)}\\
					\end{tabular}
				}	
			\end{center}
		\end{minipage}%
		\begin{minipage}{.5\linewidth}
			\vspace*{-0.0cm}
			\caption*{Model 2}
			\begin{center}
				{
					\centering
					\def\sym#1{\ifmmode^{#1}\else\(^{#1}\)\fi}
					\begin{tabular}{l*{2}{c}}
						\hline\hline
						&\multicolumn{1}{c}{CDS\_5y}&\multicolumn{1}{c}{CDS\_5y}\\
						\hline
						CrdGrd      &    0.6319\sym{***}   &       0.6163\sym{***}\\
						&         (237.69)     &    (212.11)         \\
						VOL      &                     &       -6.3372\sym{*}\\
						&                     &    (-2.58)         \\
						CrdGrd*VOL      &                     &       0.0922\sym{***}\\
						&                     &    (12.83)         \\
						constant      &       49.7583\sym{***} &       50.6751\sym{***}\\
						&     (61.92)         &     (59.21)         \\
						\hline
						\(N\)       &       47476         &       47476         \\
						$R^{2}$  &       0.5434         &       0.5451         \\
						\hline\hline
						\multicolumn{3}{l}{\footnotesize \textit{t} statistics in parentheses}\\
						[-0.5em]
						\multicolumn{3}{l}{\footnotesize \sym{*} \(p<0.05\), \sym{**} \(p<0.01\), \sym{***} \(p<0.001\)}\\
					\end{tabular}
				}	
			\end{center}
		\end{minipage}
		\vspace*{+0.5cm}
		\caption{Goodness-of-Fit of OLS models of volatility regime tests}
		\label{tab:9}
	\end{table}
	\newpage
	\newgeometry{left=3.5cm,right=3.5cm,top=2.3cm,bottom=2.3cm}
	\paragraph{Appendix C} \label{sec:C}
	\setcounter{section}{3}
	\setcounter{figure}{0}
	\renewcommand\thefigure{\Alph{section}.\arabic{figure}}
	\setcounter{table}{0}
	\renewcommand\thetable{\Alph{section}.\arabic{table}}
	\paragraph{Random Forests Robustness Tests}\mbox{}\\
	\begin{table}[h]
		\centering
		\begin{tabular}{cc|c|c|c|c|}
			\cline{3-6}
			&        & \multicolumn{2}{c|}{In-Sample} & \multicolumn{2}{c|}{Out-of-Sample} \\ \cline{1-6} 
			\multicolumn{1}{|c|}{Removed Sector}               & Remaining Obs. & mean (\%)         & std (\%)          & mean (\%)        & std (\%)          \\ \hline
			\multicolumn{1}{|c|}{Basic Materials}     & 41753    & 99.3        & 0.06        & 84        & 7.16        \\ \hline
			\multicolumn{1}{|c|}{Communication}       & 39990    & 99.2         & 0.07           & 87       & 3.55          \\ \hline
			\multicolumn{1}{|c|}{Consumer Cyclical}       & 36795   & 99.4        & 0.05        & 85.6        & 5.26        \\ \hline
			\multicolumn{1}{|c|}{Consumer Non-Cyclical}        & 39129    & 99.2          & 0.08        & 85.1       & 6.51        \\ \hline
			\multicolumn{1}{|c|}{Energy}         & 43245   & 99.3         & 0.07         & 86.9       & 4.54        \\ \hline
			\multicolumn{1}{|c|}{Financial}     & 37791   & 99.3         & 0.07         & 87.5        & 4.33         \\ \hline
			\multicolumn{1}{|c|}{Industrial}      & 40328   & 99.3        & 0.06        & 86       & 4.91        \\ \hline
			\multicolumn{1}{|c|}{Utilities}         & 42780    & 99.2        & 0.07         & 85.9       & 4.37        \\ \hline
		\end{tabular}
		\caption{Random forests robustness tests with removal of industrial sectors}
		\label{tab:10}
	\end{table}
\end{document}